\documentclass[%
 aip,
 cha,
 amsmath,amssymb,
 reprint,%
]{revtex4-1}

\usepackage{graphicx}% Include figure files
\usepackage{dcolumn}% Align table columns on decimal point
\usepackage{bm}% bold math
\usepackage[utf8]{inputenc}
\usepackage[T1]{fontenc}
\usepackage{mathptmx}

\usepackage{tabularx}
\usepackage{multirow}
\usepackage{amsthm}
\usepackage{color}
\usepackage[utf8]{inputenc}
\usepackage[T1]{fontenc}
\usepackage{textcomp}
\usepackage{gensymb}

\newtheoremstyle{query}%
{}{}%space above/below
{\color{red}}%body style
{}%heading indent
{\sffamily\bfseries}{:}{12pt}%heading style/punctuation/space after
{}% head spec
\theoremstyle{query}

\newcommand{\diff}{\mathrm{d}}%This is a better way to get the ``differential d'': the previous way puts an unwanted space after the d.
\newcommand{\bff }{}

\begin{document}

\preprint{AIP/123-QED}

\title[
%Sample title
]{Can we use linear response theory to assess geoengineering strategies?}
% Force line breaks with \\

\author{Tam\'as B\'odai}
\email{bodai@pusan.ac.kr}
\affiliation{Center for Climate Physics, Institute for Basic Science, Busan, 46241, Republic of Korea}
\affiliation{Pusan National University, Busan, 46241, Republic of Korea}
%\altaffiliation[]{Center for Climate Physics, Institute for Basic Science, Busan, Republic of Korea, 46241}
%Lines break automatically or can be forced with \\

\author{Valerio Lucarini}%
 %\email{Second.Author@institution.edu.}
\affiliation{ 
Centre for the Mathematics of Planet Earth, University of Reading, RG6 6AX, UK}
\affiliation{
Department of Mathematics and Statistics, University of Reading, RG6 6AX, UK%\\This line break forced with \textbackslash\textbackslash
}
\affiliation{
CEN, Meteorological Institute, University of Hamburg, 20144, Germany}

\author{Frank Lunkeit}
 %\homepage{http://www.Second.institution.edu/~Charlie.Author.}
\affiliation{
CEN, Meteorological Institute, University of Hamburg, 20144, Germany}%

\date{\today}% It is always \today, today,
             %  but any date may be explicitly specified

\begin{abstract}
{Geoengineering can control only some climatic variables but not others, resulting in side-effects.} We investigate in an intermediate-complexity climate model the applicability of linear response theory (LRT) to the assessment of a geoengineering method. This application of LRT is twofold. First, our objective (O1) is to assess only the \textit{best} possible geoengineering scenario by looking for a suitable modulation of solar forcing that can {cancel} out or otherwise modulate a climate change signal that would result from a rise in carbon dioxide concentration [CO$_2$] alone. Here we consider only the cancellation of the expected global mean surface air temperature $\Delta\langle[T_s]\rangle$. It is in fact a straightforward inverse problem for this solar forcing, and, considering an infinite time period, we use LRT to provide the solution in the frequency domain in closed form as $f_s(\omega) = (\Delta\langle[T_s]\rangle(\omega) - \chi_{g}(\omega)f_{g}(\omega))/\chi_s(\omega)$, where the $\chi$'s are linear susceptibilities. We  provide procedures suitable for numerical implementation that apply to \textit{finite} time periods too. {Second, to be able to utilize LRT to quantify side-effects, the response with respect to uncontrolled observables, such as regional averages $\langle T_s\rangle$, must be approximately \textit{linear}. Therefore, our objective (O2) here is to assess the linearity of the response.}
We find that under geoengineering {in the sense of (O1)}, i.e., under combined greenhouse and {required} solar forcing, the asymptotic response $\Delta\langle[T_s]\rangle$ is actually not zero. {This turns out not to be  due to nonlinearity of the response under geoengineering, but} rather a consequence of inaccurate determination of the linear susceptibilities $\chi$. {The error} is in fact due to a significant quadratic nonlinearity of the response under system identification achieved by a forced experiment. This nonlinear contribution can be easily removed, which results in much better estimates of the linear susceptibility, and, in turn, in a fivefold reduction in $\Delta\langle[T_s]\rangle$ under geoengineering practice. This correction dramatically improves also the agreement of the spatial patterns of the predicted linear and the true model responses. However, considering (O2), {such an agreement is not perfect and is worse in the case of the precipitation patterns as opposed to surface temperature. Some evidence suggests that it could be due to a greater degree of nonlinearity in the case of  precipitation.}

\end{abstract}

\maketitle

\begin{quotation}
Geoengineering strategies with the aim of %ameliorating 
mitigating climate change are  receiving increasing attention,\cite{Synthesis_report,NAP18805,NAP18988,Geoeng_book,1748-9326-9-1-014001,doi:10.1002/asl.316,Kravitz_blog,esd-10-453-2019,Stilgoe:2019,Collomb:2019} not only because of their potential to solve one of the greatest  challenges faced by modern society, but also because of the great risk that such an unprecedented endeavor entails. {Here we would like to advocate that the study of climate change in general, and geoengineering in particular, would benefit from} 
response theory~\cite{Kubo:1966,Ruelle:2009} and the theory of nonautonomous dynamical systems.~\cite{Sell_I,Sell_II,PhysRevA.41.784,Crauel1994,Crauel1997,Arnold:1998,KRbook:2011,CLRbook:2013} These mathematical tools were  introduced into climate science many years ago,\cite{Leith:1975,Bell:1980,Nicolis1985} but only recently have they started to really gain traction.~\cite{GRL:GRL18205,GB:2007,Kirk-Davidoff:2009,MGY:2010,npg-20-239-2013,npg-18-7-2011,doi:10.1029/2010GL045208,Ragone2016,Lucarini2017,Herein2015,Herein2017,BT:2012,DBT:2015,PhysRevE.94.022214} The first application of response theory to the study and efficient assessment of geoengineering in particular was by Kravitz and MacMartin. \cite{acp-16-15789-2016} They  assessed the linearity of the response, %It concerns our point (O2), 
but regarding global averages only. However,  regional temperature responses to radiative forcing can be nonlinear,\cite{doi:10.1002/2015JD023901,doi:10.1029/180GM09,Good2015,Lucarini2017} and 
{there has been an indication~\cite{doi:10.1002/2015JD023901} that}
they can be nonlinear in the case of geoengineering too. 
% VL:
We show that it is possible to describe in a concise and general way the response of the climate system to two or more forcings with given time-dependent modulations. In particular---and this is the case of interest in geoengineering---if a forcing is given, one can arrange the time modulation of $N$ other forcings in such a way as to achieve a desired time-dependent change for $N$ climatic observables of interest. The pitfall of this approach is that (a) the response of any other observable is, in principle, uncontrolled and (b) nonlinearities can become more and more relevant as forcings are added to the system. This indicates that there are some fundamental caveats in the setup of geoengineering strategies.
\end{quotation}

% \begin{aq}
% Note that in addition to editing the text for language, I've also made a number of formatting changes to bring the paper into the correct style for AIP journals, for example in the tables, where AIP strongly discourages the use of vertical rules and (too many) internal horizontal ones, and to the reference citations, which should generally follow punctuation. There are also a few adjustments to LaTeX coding to comply with Revtex guidelines, notably that explicit centering commands need not, and indeed should not, be used inside figure and table environments (although I must admit I've yet to come across these causing any problems!). As explained in a comment in the .tex file preamble, I've amended the \textbackslash diff command so that it does not produce an unwanted space after the ``d'' (there should be a space after the d$x$ etc. as a whole, but this is put in by hand). Finally, please see the comments about the repeated footnote in Appendices~\ref{sec:discrete_time} and \ref{sec:circ_conv}.
% \end{aq}

\section{Introduction: Using response theory to formulate geoengineering strategies and outcomes}

First we summarize briefly the existing mathematical tools (Sec.~\ref{sec:elements}) that 
will provide us with the framework to describe %a TB: reverted to 'the'
the geoengineering problem as an inverse problem (O1) (Sec.~\ref{sec:geoeng_problem}). We will then elucidate the utility of this inverse problem approach and compare it with the alternative control problem and other approaches (Sec.~\ref{sec:utility}), 
and we will subsequently provide the context for the need to assess geoengineering strategies (O2) (Sec.~\ref{sec:this_paper}).

\subsection{Elements of response theory}\label{sec:elements}

In \textit{nonautonomous dissipative dynamical systems}, like the climate system, given in the form
\begin{equation}\label{eq:ode}
 \dot{x}=F(x)+\epsilon g(x,t),%g(x)f(t) %24.07.18.
\end{equation}
the \textit{response} of the system to an external forcing $g(x,t)$ can be \textit{unambiguously} defined in terms of the so-called \textit{snapshot attractor}~\cite{PhysRevA.41.784} of the system, and the natural probability distribution or the measure $\mu(x,t)$ supported by it. This applies also to chaotic systems, when the snapshot attractor is a fractal object. Both the attractor and the measure are \textit{unique} objects; they are defined by an \textit{ensemble} of trajectories initialized in the \textit{infinite} past. The time dependence of the snapshot attractor, also called a pullback attractor,~\cite{Crauel1994,Arnold:1998,Chekroun20111685} and its measure give what is often termed  the ``forced response,'' and their geometrical details at any instant describe (statistical aspects of) the \textit{internal variability} in a conceptually sound sense.~\cite{DBT:2015}

For a scalar observable $\Psi(x)$ too, the (forced) response is uniquely given by a projection of the measure. {Response theory~\cite{Risken,Abramov2008,Ruelle:2009} asserts that the most basic ensemble-based statistics, the mean $\langle\Psi\rangle(t)=\int \diff x\,\Psi(x)\mu(x)(t)$ can be decomposed into linear ($j=1$) and nonlinear ($j>1$) contributions:
\begin{equation}\label{eq:perturbative}
 \Delta\langle\Psi\rangle(t) = \langle\Psi\rangle(t) - \langle\Psi\rangle_0 = \sum_{j=1}^{\infty} \epsilon^j\langle\Psi\rangle^{(j)}(t),
\end{equation} 
where the first-order, i.e., linear, term can be obtained as 
\begin{widetext}
\begin{equation}\label{eq:first_order_term_time_full}
 \langle\Psi\rangle^{(1)}(t) = \int \diff x\,\Psi(x)\int^{\infty}_{-\infty}\diff\tau\,(\exp[(t-\tau)L_F(x)][L_g(x,\tau)\bar{\mu}(x)])(x,t,\tau),
\end{equation} 
\end{widetext}
where $\bar{\mu}(x)$ is the natural invariant measure of the autonomous system ($g=0$), and the operators are defined as $L_F\mu = -\mathrm{div}(F\mu)$ and $L_g\bar{\mu} = -\mathrm{div}(g\bar{\mu})$. %, in the notation of~\cite{Abramov2008}.
In (\ref{eq:perturbative}), $\langle\Psi\rangle_0$ is the unperturbed ($\epsilon=0$) expectation, and the series converges only if the forcing $\epsilon g(x,t)$ is small enough. 
If the forcing depends on time in a multiplicative fashion, % quoted from Risken
$g(x,t)=g(x)f(t)$, then we can write 
\begin{equation}\label{eq:first_order_term_time}
 \langle\Psi\rangle^{(1)}(t) = G_{\Psi}^{(1)}(t) \ast f(t) = \int^{\infty}_{-\infty}\diff\tau\, G_{\Psi}^{(1)}(\tau)f(t-\tau),
\end{equation} 
where the \textit{Green's function} is implied by Eqs.~(\ref{eq:first_order_term_time_full}) and (\ref{eq:first_order_term_time}) to be 
\begin{equation}\label{eq:Greens_fun_def}
 %G_{\Psi}^{(1)}(t)  =\int\bar{\mu}(dx)\Theta(t) g\cdot\nabla(S^t_0\Psi(x))
 G_{\Psi}^{(1)}(t)  =\int \diff x\,\Psi(x)(\exp[tL_F(x)][L_g(x)\bar{\mu}(x)])(x,t). % 14.02.18. following  Climate_snapshot_attractor_and_response.pptx and (Abramov & Majda 2008).
\end{equation} 
Note that the higher-order terms $\langle\Psi\rangle^{(j)}$ can be expressed as multiple \textit{convolution integrals} involving multitime Green's functions.\cite{Lucarini2017}
}
Taking the Fourier transform (FT) of Eq.~(\ref{eq:first_order_term_time}), we have, via the convolution theorem,\cite{Katznelson:1976} a response formula in the frequency domain:
\begin{equation}\label{eq:first_order_term_freq}
 \langle\Psi\rangle^{(1)}(\omega) = \chi_{\Psi}^{(1)}(\omega)f(\omega),
\end{equation} 
where $\chi_{\Psi}^{(1)}(\omega)= \mathrm{FT}[G_{\Psi}^{(1)}(t)] $ is called the linear \textit{susceptibility}.

\subsection{The geoengineering problem}\label{sec:geoeng_problem}

It has been proposed~\cite{NAP18988} that the effect of greenhouse forcing can be mitigated by applying another external forcing to the Earth system by some geoengineering means that has, in a way, an opposing effect. There are various types of  forcing  that can achieve this, but here we will consider those---generically referred to as ``solar-radiation management'' (SRM)~\cite{RMA:2010,RRIKM:2012}---that can be modeled by a modulation of the solar constant. We will call this simply the ``solar forcing.'' Clearly, these are means that modulate the shortwave incoming radiation. Readily proposed geoengineering methods include a fleet of reflective satellites of large Sun-facing surface area put into orbit around the Earth, aerosols sprayed into the atmosphere, artificially generated clouds, etc. A modulated solar constant model represents these geoengineering scenarios with various degree of approximation, {not necessarily a good approximation.\cite{1748-9326-9-1-014001}}

Formally, the geoengineering problem involves a forced/nonautonomous system, where at least two terms contribute to the forcing. For simplicity, to start with, we consider the case of only two forcing terms,  both of which are additive; that is, the dynamical system of  interest is
\begin{equation}\label{eq:ode2}
 \dot{x}=F(x)+\epsilon (g_g(x)f_g(t) + g_s(x)f_s(t)),
\end{equation}
where the subscripts already indicate  the physical meanings of the forcings, $g$ for ``greenhouse'' and $s$ for ``solar.''  Except for the need for a convergent series in Eq.~(\ref{eq:perturbative}), an arbitrary value can be assigned to the ``small'' parameter $\epsilon$, and in order to obtain a result in the uncomplicated form of Eq.~(\ref{eq:needed_forcing_freq}), we choose the same $\epsilon$ for both forcing components. {Equation~(\ref{eq:first_order_term_time}) implies that t}he first-order contribution $\langle\Psi_{\varSigma}\rangle^{(1)}(t)$ to the \textit{total response} $\Delta\langle\Psi_{\varSigma}\rangle$ under combined forcing, i.e., geoengineering (where the subscript in $\Psi_{\varSigma}$ is to indicate the presence of multiple forcings), can be written as the superposition of first-order contributions to respective responses to the two forcings in two separate scenarios when these forcings are acting alone. Formally, this is expressed as
\begin{equation}\label{eq:tot_resp_time}
 \langle\Psi_{\varSigma}\rangle^{(1)}(t) = G_{\Psi,g}^{(1)}(t) \ast f_g(t) + G_{\Psi,s}^{(1)}(t) \ast f_s(t),
\end{equation} 
The FT of this equation is 
\begin{equation}\label{eq:tot_resp_freq}
 \langle\Psi_{\varSigma}\rangle^{(1)}(\omega) = \chi_{\Psi,g}(\omega) f_g(\omega) + \chi_{\Psi,s}(\omega) f_s(\omega).
\end{equation} 
{Note that the nonlinear response is more complicated with multiple forcings present: it is not just a sum of multiple convolution integrals~\cite{Lucarini2017} as in the case of a single forcing scenario.}

If the ``forward'' problem is the prediction of the response to a given forcing, then the \textit{inverse} problem of ``predicting'' the necessary forcing for a desired response, being our objective (O1), seems to be well defined in view of the above equations. To a linear approximation, the necessary or \textit{required} forcing is
\begin{equation}\label{eq:needed_forcing_freq}
 f_s(\omega) \approx \frac{\Delta\langle\Psi_{\varSigma}\rangle(\omega) - \chi_{\Psi,{g}}(\omega)f_{g}(\omega)}{\chi_{\Psi,s}(\omega)}.
\end{equation} 
In the above, $\epsilon=1$ is taken. 

The case of two forcings, one greenhouse and one geoengineering, can be generalized to $N+1$ forcings when we desire to control $N$ climate observables $\bm{\Psi}^T = (\Psi_1,\dots,\Psi_N)$ by modulating $N$ geoengineering forcings $\bm{f}_s^T = (f_{s1},\dots,f_{sN})$. With these, generalizing Eq.~(\ref{eq:tot_resp_freq}), we can write in matrix form~\cite{Lucarini2018}
\begin{equation}\label{eq:tot_resp_freq_gen}
 \langle\bm{\Psi}_{\varSigma}\rangle^{(1)}(\omega) = \bm{\chi}_{\bm{\Psi},g}(\omega) f_g(\omega) + \bm{\chi}_{\bm{\Psi},s}(\omega)\cdot \bm{f}_s(\omega),
\end{equation} 
This equation can be inverted to give the vector of geoengineering forcings $\bm{f}_s$: 
\begin{equation}\label{eq:needed_forcing_freq_gen}
 \bm{f}_s(\omega) = \bm{\chi}_{\bm{\Psi},s}^{-1}(\omega) \cdot (\langle\bm{\Psi}_{\varSigma}\rangle^{(1)}(\omega) - \bm{\chi}_{\bm{\Psi},g}(\omega) f_g(\omega)), 
\end{equation} 
provided the inverse of the matrix $\bm{\chi}_{\bm{\Psi},s}$ exists. The problem of static response is considered by Lu \textit{et al.},\cite{Lu_etal:2019} who take the $N$ geoengineering forcings as those acting at $N$ different gridpoints of a climate model and look for forcing fields to which the climate system is most susceptible. 

\subsection{The utility of the inverse problem and its alternatives}\label{sec:utility}

In Appendix~\ref{sec:inverse_problem}, we provide a procedure to solve the inverse problem with $N=2$ using discrete and finite time series. That situation can be interpreted as a control problem, which is in fact a rather special type of \textit{optimal} control. This way, the required forcing can be predetermined and need not be updated during its application. In a different approach in, for example, Refs.~\onlinecite{RMA:2010,RRIKM:2012}, the solar forcing was constructed on the basis of some models of how much radiative forcing a sudden change of some greenhouse gas concentration or the stratospheric optical depth would yield. In addition, these authors created a scenario ensemble of SRMs, and selected the most effective SRMs. {The latter assessment strategy is clearly rather inefficient and inaccurate, and that would still be the case had the ensemble been generated using response theory.} {We note that MacMartin \textit{et al.}\cite{MacMartin20140134} proposed for the first time to solve an inverse problem as a ``design problem'' for geoengineering. They invert the analytical solution of a conceptual model for the global average surface temperature, the parameters of which conceptual model are inferred via fitting it to Earth System Model simulation data. Our method is more generic in that it does not require analytic inversion or the use of a simplified model, and it can also consider any observable to be controlled, not just the global average temperature.}

The inverse problem would have a \textit{direct} practical relevance were we to have $f_g(t)$ a given, as assumed. However, this is clearly not the case: predicting  greenhouse gas emissions is an extremely complicated task, since it is determined among others by \textit{social} processes, for which we do not have good models. Nevertheless, efforts are underway~\cite{Rogelj2018,GIDDEN2018187} (see also \url{https://crescendoproject.eu/research/theme-4/}). The current standard practice to tackle this challenge, as reflected by the IPCC reports,\cite{Synthesis_report} is to consider half a dozen so-called ``methodologically constructed'' twenty-first century emission scenarios. This way, instead of climate predictions, one produces so-called climate \textit{projections} belonging to hypothetical future emission scenarios. Therefore, the solution to our inverse problem has a rather \textit{indirect} practical relevance; {the inverse problem approach would allow us} to carry out 
\textit{scenario analyzes}. {The reader can find elsewhere~\cite{6858658,MacMartin2014,MacMartin20140134,esd-7-469-2016}  descriptions and analysis of a} \textit{feedback} control problem of \textit{direct} practical relevance, when the solar forcing is being determined in real time with the use of some controller, adapting to a progressing greenhouse forcing, trying to realize the desired response. Note that with feedback control, in a scenario analysis setting, a new simulation needs to be run for each emission scenario, {making it very inefficient for an extensive assessment exercise. Note also that for feedback control, what can be observed as a reference (the basis of feedback) is not the forced response in terms of an ensemble, but only a single realization. Therefore, what the feedback and open-loop control strategies would respectively realize---the climatology in terms of say ensemble means, on the one hand, and the internal variability, on the other---could be very different. They are expected to be more significantly different the stronger the internal variability exhibited by the observable $\Psi$ chosen to be controlled, e.g., in the case of local vs global average quantities. In an extreme case, one can consider a geoengineering method \textit{designed}~\cite{Kravitz_blog} to extinguish the oscillatory El-Ni\~no phenomenon, presumably in order to prevent floods or droughts that are part of the internal variability of Earth's climate.}

We point out that in, for example, Eq.~(\ref{eq:needed_forcing_freq}), we write $\Psi$ to denote a generic observable. This means that we can \textit{choose} a particular (scalar) observable that we desire to evolve in a particular way. With reference to the classic term ``global warming,'' in contrast to ``climate change,'' we will attempt to enforce  cancellation of the global average surface air temperature (Sec.~\ref{sec:surf_temp}). With the increasingly wide-ranging analyzes of climate change scenarios, however, it is clear that ``climate change'' should have a comprehensive meaning, and not just be a synonym for ``global warming''.\cite{Conway:2008} In fact, physical quantities other than temperature could have a greater social or ecological impact.\cite{Synthesis_report} Therefore, unlike in the present work, in practice, we might want to choose some variable other than global temperature to control. Beside the \textit{physical type} of the observable quantity, we can make arbitrary choices with respect to the \textit{spatial scale} of the quantity, such as local or regional averages (Sec.~\ref{sec:res_spatial}), zonal averages (Appendix~\ref{sec:res_zonal}), global averages (Sec.~\ref{sec:res_global}), etc. 

\subsection{This paper}\label{sec:this_paper}

\begin{figure}  %[t!]
    %\begin{center}
        \includegraphics[width=1\linewidth]{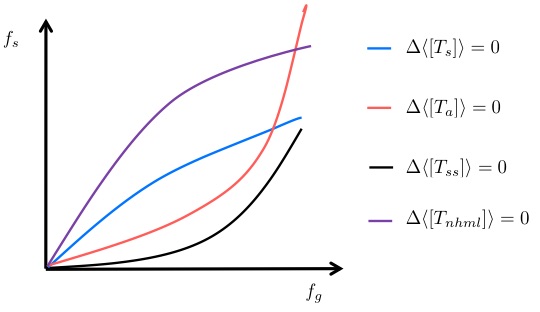}
    %\end{center}
    \caption{\label{fig:isolines} A cartoon of hypothetical isolines in the plane of greenhouse and solar forcings $f_g$--$f_s$ for various observables: globally averaged surface air temperature ($\Delta\langle[T_s]\rangle=0$), globally averaged atmospheric temperature ($\Delta\langle[T_a]\rangle=0$), averaged sea surface temperature ($\Delta\langle[T_{ss}]\rangle=0$), and surface air temperature averaged at the midlatitudes of the Northern hemisphere ($\Delta\langle[T_{nhml}]\rangle=0$). The diagram is a reproduction of Fig.~5 of Ref.~\onlinecite{Lucarini:2013}. } 
\end{figure}

We now turn to motivating the need for comprehensive assessment of geoengineering scenarios. Once an observable $\Psi$ is chosen to evolve in a particular way, {which determines $f_s(t)$ according to Eq.~(\ref{eq:needed_forcing_freq})}, the evolution of any other observable $\Phi$ will be \textit{a given}---the solution of a \textit{forward} problem formally identical to (\ref{eq:tot_resp_freq}): 
%
% \begin{equation}
%  \langle\Phi_{\varSigma}\rangle^{(1)}(\omega) = \chi_{\Phi,g}(\omega) f_g(\omega) + \chi_{\Phi,s}(\omega) f_s(\omega),
% \end{equation} 
\begin{equation}
 \langle\Phi_{\varSigma}\rangle^{(1)}(t) = G_{\Phi,g}^{(1)}(t) \ast f_g(t) + G_{\Phi,s}^{(1)}(t) \ast f_s(t),
\end{equation} 
yet with an $f_s$ given by Eq.~(\ref{eq:needed_forcing_freq}). {Note that} $\langle\Phi_{\varSigma}\rangle^{(1)}(t)\neq\langle\Psi_{\varSigma}\rangle^{(1)}(t)$ when $G_{\Phi,g}(t)\neq G_{\Psi,g}(t)$ and/or $G_{\Phi,s}(t)\neq G_{\Psi,s}(t)$, which is the generic case. Regarding the desire for cancellation, $\Delta\langle\Psi_{\varSigma}\rangle=0$, we can frame~\cite{Lucarini:2013} geoengineering---considering for simplicity only quasistatically slow changes of $f_g(t)$---as a confinement to the 0 isoline of $\Delta\langle\Psi_{\varSigma}\rangle$ over the plane of $f_g$ and $f_s$. In general, this isoline is different for different observables $\Phi\neq\Psi$; i.e., under a linear response, these straight isolines fan out from the origin of the $f_g$--$f_s$ plane. This is illustrated in Fig.~\ref{fig:isolines}, where the curvature of the isolines for larger values of $f_g$ and $f_s$ reflect also the more general situation of nonlinear responses. It is a straightforward implication that when the system is confined to one isoline, it cannot be confined to the different isolines of other variables $\Phi_i$; i.e., (unwanted) changes $\Delta\langle\Phi_{i,\varSigma}\rangle\neq0$ will ensue. In other words: the proposed geoengineering method will provide just a partial solution at best. While one aspect of the problem is solved, other aspects can be neglected, or even changed to the worse, possibly with catastrophic consequences.\footnote{Furthermore, we note that, as  is often acknowledged, ``no-one is living under the average climate.'' Although some live closer than others. That is, while the primary problem can be solved for some, even that will not be solved for others. Therefore, the debate on climate engineering is unlikely to %have fewer political overtones and a weaker motive 
be less political than the climate debate itself. {As Alan Robock put it 10 years ago,\cite{Robock1166} ``If geoengineering worked, whose hand would be on the thermostat? How could the world agree on an optimal climate?''}} {A long list of studies have to date addressed the issue of side effects; see, e.g., Refs.~\onlinecite{doi:10.1029/1999GL006086,RMA:2010,RRIKM:2012,1748-9326-9-1-014001,MacMartin20140134,doi:10.1002/jgrd.50856,acp-16-15789-2016,MacMartin20160454}.} This possibility is the main \textit{motivation} of our present investigation {too, concerning in particular the question (O2) of whether linear response theory can provide an efficient tool to map out and quantify accurately the various side effects of a variety of scenarios. A comprehensive assessment would consider a variety of geoengineering scenarios, emission scenarios, Earth System Models, and possibly other things, for which  the efficiency of computation is crucial.} In this study, having enforced (approximately, to various degrees) a cancellation of global average surface air temperature, $\Delta\langle\Psi_{\varSigma}\rangle=\Delta\langle[T_{s,\varSigma}]\rangle\approx0$, we will \textit{diagnose} unwanted changes, i.e., the total response, in terms of
\begin{itemize}
 \item $\Phi = [T_s]_{\lambda}$: zonal averages (Appendix~\ref{sec:res_zonal});
 \item $\Phi = T_s$: regional averages of the surface temperature (Sec.~\ref{sec:res_spatial});
 \item $\Phi = T_{tr}$: regional averages of the temperature near the tropopause (Sec.~\ref{sec:res_spatial}); 
 \item $\Phi = [P_{y}]$ and $P_{y}$: the annual precipitation (Sec.~\ref{sec:res_precip}).
\end{itemize}
Note that we denote spatial averaging by square brackets, subscripted by the spatial variable(s) with respect to which we average over its whole range, e.g., longitudes $\lambda$ for zonal averages. However, for global averaging we drop the subscripting altogether (instead of writing, e.g., $[T_s]_{\lambda,\mu}$). {Some of these observables have been considered in a number of studies,\cite{RMA:2010,RRIKM:2012,1748-9326-9-1-014001,doi:10.1002/jgrd.50856,acp-16-15789-2016,MacMartin20160454} and our results are mostly consistent with the published ones; however, as our novel contribution (O2), we will also investigate carefully whether these responses can be predicted by linear response theory.}

{The premise of our objective as set out above is that of Robock,\cite{Robock1166} recently quoted in a blog by Kravitz,\cite{Kravitz_blog} in which blog he asks the questions whether ``there is only one thermostat'' and whether ``the climate can be optimized regionally.'' Regarding the second  of these, Ban-Weiss and Caldeira~\cite{1748-9326-5-3-034009} and MacMartin \textit{et al.}~\cite{MacMartin2012} applied different spatial patterns to reduce side effects to surface temperature. This idea is actually already covered by our framework  with $N=2$, in which the function $g_s(x)$ in Eq.~(\ref{eq:ode2}) needs to be specified accordingly. Regarding the first question, Kravitz \textit{et al.}\cite{esd-7-469-2016,doi:10.1002/2017JD026874} proposed that one names \textit{multiple} objectives and looks accordingly  for multiple suitable ``control knobs'' for the climate. They employed a feedback control. The alternative---the inverse problem approach (O1) that we propose here---is shown above to be straightforward to generalize to the multiple-objective--multiple-control-knob situation, simply through the possibility of solving the linear matrix equation~(\ref{eq:tot_resp_freq_gen}) by inverting for $\bm{f}_s$. With regard to side effects, however, whatever way we construct the geoengineering forcing, the situation is hardly different from the single-objective--single-control-knob situation: there will be objectives that we could inadvertently miss from the list, or objectives that are not convenient to include, and then we need to assess the side effects in terms of the corresponding uncontrolled observables---or, indeed, assess the climatology as comprehensively as possible or as is desired.
Regarding the assessability of side effects using response theory (O2), even if nonlinear response formulae are available, feasibility might be hampered by an increasing number of objectives or control forcings.}

We point out that {in the Planet Simulator intermediate-complexity GCM (PlaSim in short),}~\cite{Fraedrich:2012}  the greenhouse and solar forcings have been found to be approximately equivalent in terms of the stationary response of the global average surface air temperature~\cite{BOSCHI20131724} insomuch that its isolines are parallel straight lines (even if there is a curvature of the surface). This was found to be the case in rather extensive ranges of the forcings, 90--2880~ppm and 1200--1500~W\,m$^{-2}$, respectively. That is, any curvature of the blue line as shown in Fig.~\ref{fig:isolines} occurs outside of the said ranges. However, {in the context of geoengineering, the concern is} whether these forcings are equivalent in the same sense in terms of other variables ($\Phi$) too, {as discussed. We will demonstrate in PlaSim that with regard to  regional averages $T_s$, the correspondence of forcings is still remarkable, but there is nevertheless a residual response with a nontrivial pattern under geoengineering. Furthermore, our analysis 
hints that (O2) this residual response might not be so linear, and less so for precipitation, which goes beyond the findings of MacMartin and Kravitz,\cite{acp-16-15789-2016} who demonstrate clearly the linearity  only of the global average response under geoengineering, but average the linear predictions of spatial patterns over nine models. On the other hand, Cao \textit{et al.}~\cite{doi:10.1002/2015JD023901} do  indicate that the local response for several observables under geoengineering might be nonlinear by comparing the prediction of a linear regression model with simulation results for the HadCM3L model. However, we point out that it is possible that the local susceptibilities represented in the simple linear model were inaccurately estimated from model simulation data, and so further analysis would be needed to attribute the said mismatch to nonlinearity. This is what we attempt here.}

%TO HERE

This work follows Ragone \textit{et al.}~\cite{Ragone2016} and Lucarini \textit{et al.}~\cite{Lucarini2017} The latter  demonstrated that it is convenient for response theory to predict spatial patterns too, which, as outlined above, form the basis for one of the types of diagnostics that we pursue in order to assess the success of the geoengineering method. In both of these papers, the authors used
PlaSim~\cite{Fraedrich:2012} to demonstrate the power of their methodology, but with slightly differing setups of the model. Here we adopt the setup of Lucarini  \textit{et al.}\cite{Lucarini2017} featuring meridional ocean heat transport. The present work also builds on Gritsun and Lucarini \cite{GRITSUN201762} in adopting a simple technique to obtain a better estimate of the linear susceptibility, which was independently discovered by Liu \textit{et al.}~\cite{doi:10.1175/JCLI-D-17-0462.1}  {Clearly, a better susceptibility estimate would be useful in making a linear prediction only if the actual response were linear. Under [CO$_2$]-doubling, Ragone \textit{et al.}~\cite{Ragone2016} and Lucarini \textit{et al.}~\cite{Lucarini2017} found a nonlinear response of the \textit{global average}  $\Delta\langle[T_s]\rangle$, and so no linear prediction would be productive in that case; however, under geoengineering, the total response is aimed to be much smaller, and so in principle the response may be linear. This is found to be approximately the case in PlaSim, and, so, as one of the main contributions of this work (O1), by improving the susceptibility estimates, we can improve greatly on our prediction of a solar forcing $f_s(t)$ required for cancellation, $\Delta\langle\Psi_{\varSigma}\rangle(t)=0$.}

The structure of the remainder of this paper is as follows. Next, in Sec.~\ref{sec:forcing_scenarios}, we detail our methodology, listing a set of experiments performed to establish the response characteristics of the climate model and to assess nonlinearities, among other things. Then, in Sec.~\ref{sec:results}, we provide results, first pertaining to objective (O1) about the success of the primary objective of geoengineering, namely, the cancellation (Sec.~\ref{sec:res_global}), and then our diagnostics of other observables (Secs.~\ref{sec:res_spatial} and \ref{sec:res_precip}). Finally, in Sec.~\ref{sec:improved}, in terms of the stationary climate only (O1), we outline an improved method for obtaining the required solar forcing for cancellation, and also (O2) {analyze our improved diagnostics with respect to the linearity of the response.} In Sec.~\ref{sec:summary} we summarize our results and give our perspective for worthwhile future work. In a series of appendices we give details of the notation and algorithm for spectral analysis in discrete time (Appendix~\ref{sec:discrete_time}), the way we obtain the Green's functions (Appendix~\ref{sec:obtain_greens}), our novel solution method to the discrete- and finite-time inverse problem for a required solar forcing (Appendix~\ref{sec:inverse_problem}), and the circular convolution theorem (Appendix~\ref{sec:circ_conv}), and we relegate  the diagnostics of zonal averages to Appendix~\ref{sec:res_zonal}.

\section{Methodology: Forcing scenarios}\label{sec:forcing_scenarios}

The form of the forcing signal $f_{g}$ due to changes in the carbon dioxide concentration $[{\rm CO}_2]$, for which we want to solve the geoengineering inverse problem, is a ramp that was used by Lucarini \textit{et al.}~\cite{Lucarini2017} This is a standard forcing type, also used for the CMIP6 DECK (Diagnostic, Evaluation and Characterization of Klima) protocols.\cite{gmd-9-4019-2016} More precisely, it is not a time-continuous ramp, for the reason detailed in Appendix~\ref{sec:obtain_greens}, but  $[{\rm CO}_2]$, and so $f_{g}$, is kept constant for one year after each incremental increase. The $[{\rm CO}_2][n+1]-[{\rm CO}_2][n]$ increment is a (small) \textit{fraction} of the current value $[{\rm CO}_2][n]$, and therefore increasing in a superlinear fashion with time $[n]$, but, owing to the logarithmic dependence of the radiative forcing on the $[{\rm CO}_2]$ concentration,\cite{JGRD:JGRD51883} it realizes a linear radiative forcing signal\footnote{\label{foot:greenhouse_forcing} This is meant  in a loose sense, because, strictly speaking, the realized radiative greenhouse forcing (which we do not even try to define here) must not be considered as an external forcing. The external forcing is indeed the $[{\rm CO}_2]$ concentration. A logarithmic scaling of this signal, however, makes no difference insomuch as a causal Green's functions exist between this scaled variable and well-behaved observables. The scaling is intuitive and standard practice, and we will allow ourselves to refer to $\ln([{\rm CO}_2]/[{\rm CO}_2]_0)$ as the radiative greenhouse forcing.} $f_{g}[n]$ (see Appendix~\ref{sec:discrete_time} for the notation for a discrete time series),  i.e., a constant-in-time ($n$) radiative forcing increment $f_{g}[n+1]-f_{g}[n]$. Hence the name ``ramp.'' Such a form of the %(radiative) not sure why i'm supposed to delete this
forcing signal is useful in diagnosing or interpreting results. For example, if the response characteristic to solar forcing $f_s$ is similar to that of $f_{g}$, then the required solar forcing to cancel global change would also be approximately ramp-like. 

Note, however, that a linearity of the response characteristic to any forcing is {usually} checked by a comparison of the linear prediction with the truth in terms of a model simulation subject to the same forcing. {The latter we will refer to as ``reference'' in the following, instead of ``truth.''} Beside the nonlinearity, another factor that gives rise to a discrepancy is a statistical error due to the finite ensemble size. However, the latter has a very distinct feature that can be visually told apart easily from the contribution of nonlinearity. We reiterate that by applying a staircase-like forcing, we guarantee that the said discrepancy has no contribution due to  calculations being performed in discrete time. 

We point out that, at asymptotic times, there is no discrepancy at all, because of the way we estimate the Green's function (Appendix~\ref{sec:obtain_greens}): the discrepancy emerges \textit{transiently} only. Its all-time maximum  is a useful intuitive measure of nonlinearity in the examined regime. In any case, the larger the response, the greater is the nonlinear contribution to it, and so---in the context of system identification---the more inaccurate our estimates of the susceptibilities (Appendix~\ref{sec:obtain_greens}) become. Therefore, beside our \textit{base scenario} of (overall) doubling $[{\rm CO}_2]$, we will also check if we can obtain a more accurate (and so useful for the geoengineering problem) estimate of the Green's function using a \textit{weaker} identification forcing. In particular, we apply a $[{\rm CO}_2]$ change that results in \textit{half} of the (overall) radiative forcing change of that by doubling $[{\rm CO}_2]$. (This is realized by $[{\rm CO}_2]_{\infty}/[{\rm CO}_2]_0=\sqrt{2}$, according to the above-mentioned logarithmic law.\cite{JGRD:JGRD51883}) Note that in the case of this weaker forcing, irrespective of the different plateau level, the increments of the $[{\rm CO}_2]$ changes realize the same 1\%/yr relative change. 

We refer the reader to Table~\ref{tab:forcing_scenarios} for an overview of the various identification and test forcing scenarios that we used in the present study. Among them, we have CQ2 defined by 0.1\%/yr relative changes, which makes it a much slower change than the base scenario. The response to such a slow ramp forcing should be ramp-like as long as the linear term in Eq.~(\ref{eq:perturbative}) dominates over the nonlinear ones. This forcing scenario will therefore provide us another reference in interpreting other results with respect to linearity.

In Table~\ref{tab:forcing_scenarios}, we have not indicated the plateau level of the solar forcing $f_s$ used in conjunction with $f_{g}$. We chose this level such that the response asymptotically in terms of the global average surface air temperature is the same but of opposite sign as that due to the corresponding $f_{g}$. This level can be easily determined to a good approximation by an iterative procedure. Beside those in Table~\ref{tab:forcing_scenarios}, we will introduce a few more forcing scenarios in Sec.~\ref{sec:improved}: some forcing scenarios that will give improved results, and other forcing scenarios that aid the interpretation of our results. The abrupt stepwise forcing scenarios, CS1, SS1, CS2, SS2, are employed to numerically determine the Green's functions as detailed in Appendix~\ref{sec:obtain_greens}.

\begin{table}
\caption{Sets of simulation data specified by the forcing. Each data set is codenamed by a three-character code, the first character coding the quantity in which the forcing is presented (C for $[{\rm CO}_2]$ and S for solar irradiance), the second character coding the ``form'' of the forcing signal (S for step, R for ramp, and Q for slow ramp), and the third character coding the plateau level of the (corresponding---see  text) greenhouse forcing (2 for $[{\rm CO}_2]_{\infty}/[{\rm CO}_2]_0=2$ and 1 for $[{\rm CO}_2]_{\infty}/[{\rm CO}_2]_0=\sqrt{2}$). The CS2 and CR2 data sets are preexisting to the present study,\cite{Lucarini2017} consisting of 200 ensemble members. All new data sets listed here consist of 20 ensemble members each, except for CQ2, which consists of 10. %WAIT, WE ALSO HAD SS2I AND BR2C, SEE EMAIL 06.10.2016.
}\label{tab:forcing_scenarios}
  %\begin{center}
    %\begin{tabularx}{0.4\textwidth}{c | *5{>{\Centering}c} | c} 
    \renewcommand{\arraystretch}{1.3}
    \begin{ruledtabular}\begin{tabular}{llcccccc}      
      %\hline%\toprule
   &Form: &\multicolumn{2}{c}{Step} & \multicolumn{2}{c}{Ramp} & Slow ramp  \\
       Forcing&Plateau:& $2$ & $\sqrt{2}$ & 2 & $\sqrt{2}$ & 2 \\%\midrule
       \hline
      $[{\rm CO}_2]$ && CS2 & CS1 & CR2 & CR1 & CQ2   \\
      Solar          && SS2 & SS1 & SR2 & SR1        \\
    Combined    &    &     &     & BR2 & BR1 &       \\
     %Quantity        &     &     &     &     &     &  \\
      %\hline
    %   %\bottomrule
    \end{tabular}\end{ruledtabular}
  %\end{center}
\end{table}

\section{Results}\label{sec:results}

\subsection{Surface air temperature}\label{sec:surf_temp}

\subsubsection{Global average}\label{sec:res_global}

\begin{figure*}  %[t!]
   % \begin{center}
        \begin{tabular}{cc}
            \includegraphics[width=0.5\linewidth]{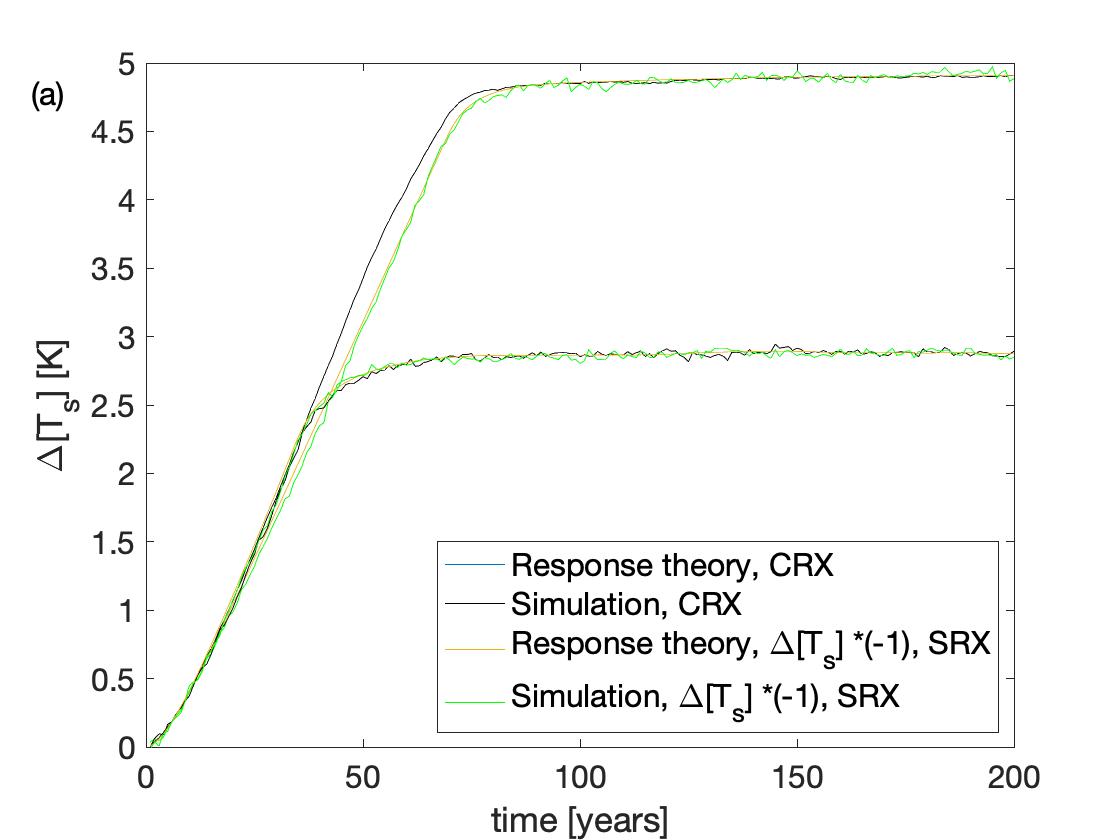} 
            \includegraphics[width=0.5\linewidth]{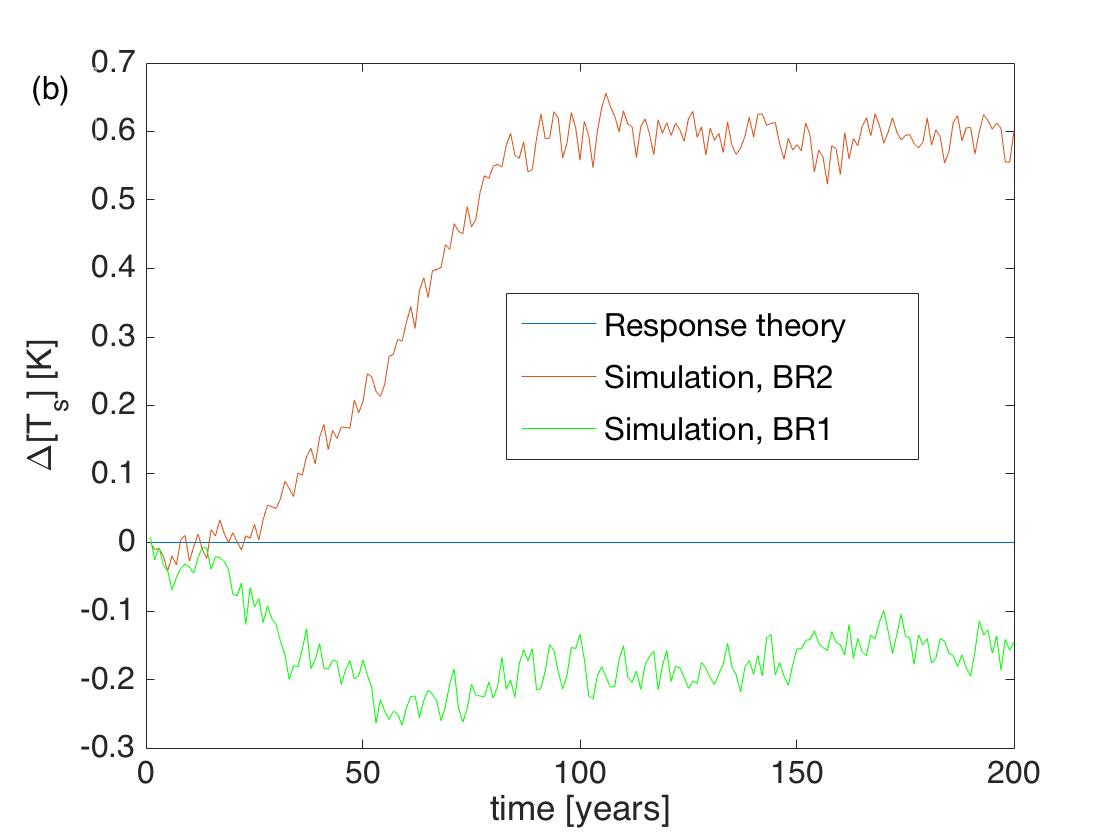}
        \end{tabular}
        \caption{\label{fig:resp2ramp} Predicted and true surface air temperature responses to ramp-like forcings. Forcing scenarios are (a) CR1, CR2, SR1, SR2 and (b) BR1, BR2. {\bff Note that in the keys, e.g. in ``CRX'', ``X'' can stand for either ``1'' or ``2'', and, intuitively, these scenarios can be told apart in the diagram by the greater/smaller response levels.} {Note also that in  (a), the two yellow curves perfectly cover the corresponding {\bff light} blue ones, because $f_s$ is calculated to cancel global warming at all times.}  
        }
    %\end{center}
\end{figure*}

The global average surface air temperature is the variable with respect to which we \textit{prescribe} the cancellation. We do \textit{not} consider any other variable in this role throughout the present study. Having predicted the solar forcings (SR1, SR2) required to produce no total response used in combination with prescribed $[{\rm CO}_2]$ forcings (CR1, CR2) adopting the methodology described in Appendix~\ref{sec:inverse_problem}, we plot the predicted linear responses in Fig.~\ref{fig:resp2ramp}(a). Clearly, these predictions can be viewed either as components of the \textit{predicted} total response (BR1, BR2), or the predicted response in separate scenarios (CR1, CR2, SR1,  SR2). Alongside these predictions, we plot the true response in the scenarios when the forcings are applied separately, i.e., the responses evaluated by direct numerical simulations (CR1, CR2, SR1, SR2). The comparison of prediction and {reference} reveals that (i) the response to stronger forcing is more nonlinear in the case of greenhouse forcing (CR2) in comparison with solar forcing (SR2) and  (ii) with a weaker identification (CS1, SS1) and test forcing (CR1, SR1), the linear prediction for CR1 is much better than that for CR2, while SR1 is seemingly as good as SR2. For the scenarios of combined forcing (BR1, BR2), only the true response is nontrivial if nonlinear, which is displayed in Fig.~\ref{fig:resp2ramp}(b). Indeed, because of the nonlinearity, the total asymptotic response is nonzero. {(Note that the fluctuations at asymptotic time are due to the finite ensemble size.)} It is visibly nonzero even with the weaker forcings.  However, it is just about 10\% of that with greenhouse forcing solely, even in the case of the stronger forcings.

% \begin{figure}  %[t!]
%     \begin{center}
%         \begin{tabular}{cc}
%             \includegraphics[width=0.5\linewidth]{08.jpg} 
%             \includegraphics[width=0.5\linewidth]{09.jpg}
%             \\
%             \includegraphics[width=0.5\linewidth]{39.jpg} 
%             \includegraphics[width=0.5\linewidth]{40.jpg}
%         \end{tabular}
%         \caption{\label{fig:resp2ramp_CQ2} True response under a very slow ramp forcing, CQ2.  The global mean is shown in panel (a), in panel (b) each colored curve belongs to a gridpoint, and in panel (c)/(d) the colored curves show the zonal averages at different latitudes on the Northern/Southern hemisphere. }
%     \end{center}
% \end{figure}

\begin{figure}  %[t!]
    %\begin{center}
	\includegraphics[width=1\linewidth]{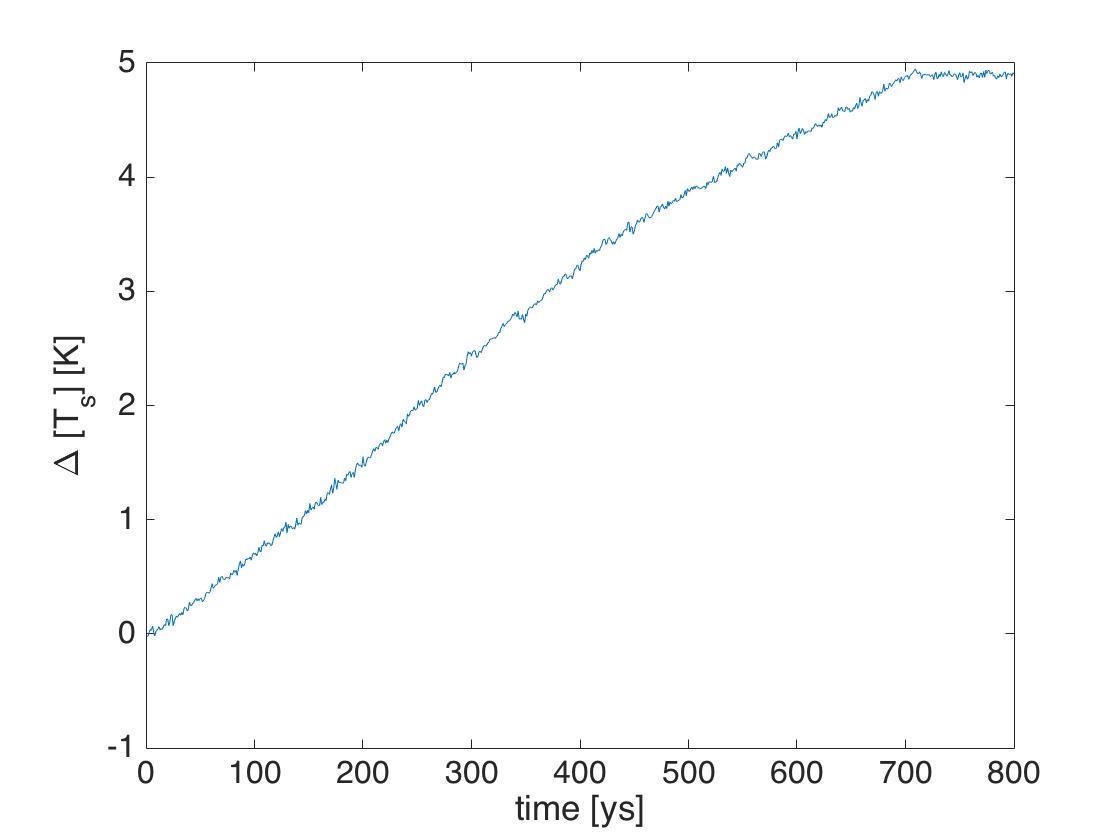}
        \caption{\label{fig:resp2ramp_CQ2} True response of the global mean surface temperature under a very slow ramp forcing, CQ2.}
   % \end{center}
\end{figure}

The pronounced nonlinearity [point (i) above] shows up also in other experiments. With a very slow forcing CQ2, we registered the response as shown in Fig.~\ref{fig:resp2ramp_CQ2}. Although the rate of forcing is unchanged throughout the time period of almost 700 years, the response switches to a slower rate between 400 and 500 years, or, between 3 and 4~K changes in the temperature.\footnote{A slower rate of change of the response to a slow forcing translates to a smaller static susceptibility (at $\omega=0$), i.e., sensitivity.} The placement of this change in the rate, compared with the asymptotic temperature change of almost 3~K upon the weaker CR1 forcing seen in Fig.~\ref{fig:resp2ramp}(a), is in good agreement with the observation of a much more closely linear response to that weaker forcing as compared with CR2. A crude indicator of non/linearity can be extracted from the CQ2 experiment, but also from comparing the asymptotic/stationary responses (denoted by a subscript $\infty$) in the XX1 and XX2 experiments, as {the following ratio}:
% Remeber that i asked Frank about the solar forcing levels in an email! ... I see data in email sent on 14.April 2017.
%
%\begin{equation}\label{eq:simple_nonlin}
% \rho = \frac{\frac{\Delta \langle[T_s]_{\infty,2}\rangle}{\Delta \langle[T_s]_{\infty,1}\rangle}}{\frac{f_{\infty,2}}{f_{\infty,1}}} = \frac{\frac{\Delta \langle[T_s]_{\infty,2}\rangle}{f_{\infty,2}}}{\frac{\Delta \langle[T_s]_{\infty,1}\rangle}{f_{\infty,1}}}.  %(\Delta \langle[T_s]_{\infty,2}\rangle/\Delta \langle[T_s]_{\infty,1}\rangle)/(f_{\infty,2}/f_{\infty,1}).
%\end{equation} 
%
\begin{equation}\label{eq:simple_nonlin}
 \rho = \frac{ {\Delta \langle[T_s]_{\infty,2}\rangle}/{\Delta \langle[T_s]_{\infty,1}\rangle}}{{f_{\infty,2}}/{f_{\infty,1}}} = \frac{ {\Delta \langle[T_s]_{\infty,2}\rangle}/{f_{\infty,2}}}{{\Delta \langle[T_s]_{\infty,1}\rangle}/{f_{\infty,1}}}.  \end{equation} 
{(Note that we write an ``X'' in place of one of the possible characters in the scenario identification code when it does not matter which of the possible characters is written there.)} This value is $\rho=0.99$ with solar forcing and 0.85 with greenhouse forcing, in agreement with what the comparison of predicted and true responses, seen in Fig.~\ref{fig:resp2ramp}(a), allowed us to conclude above.

\subsubsection{Spatial pattern}\label{sec:res_spatial}

\begin{figure*}  %[t!]
    %\begin{center}
        \begin{tabular}{cc}
            \includegraphics[width=0.5\linewidth]{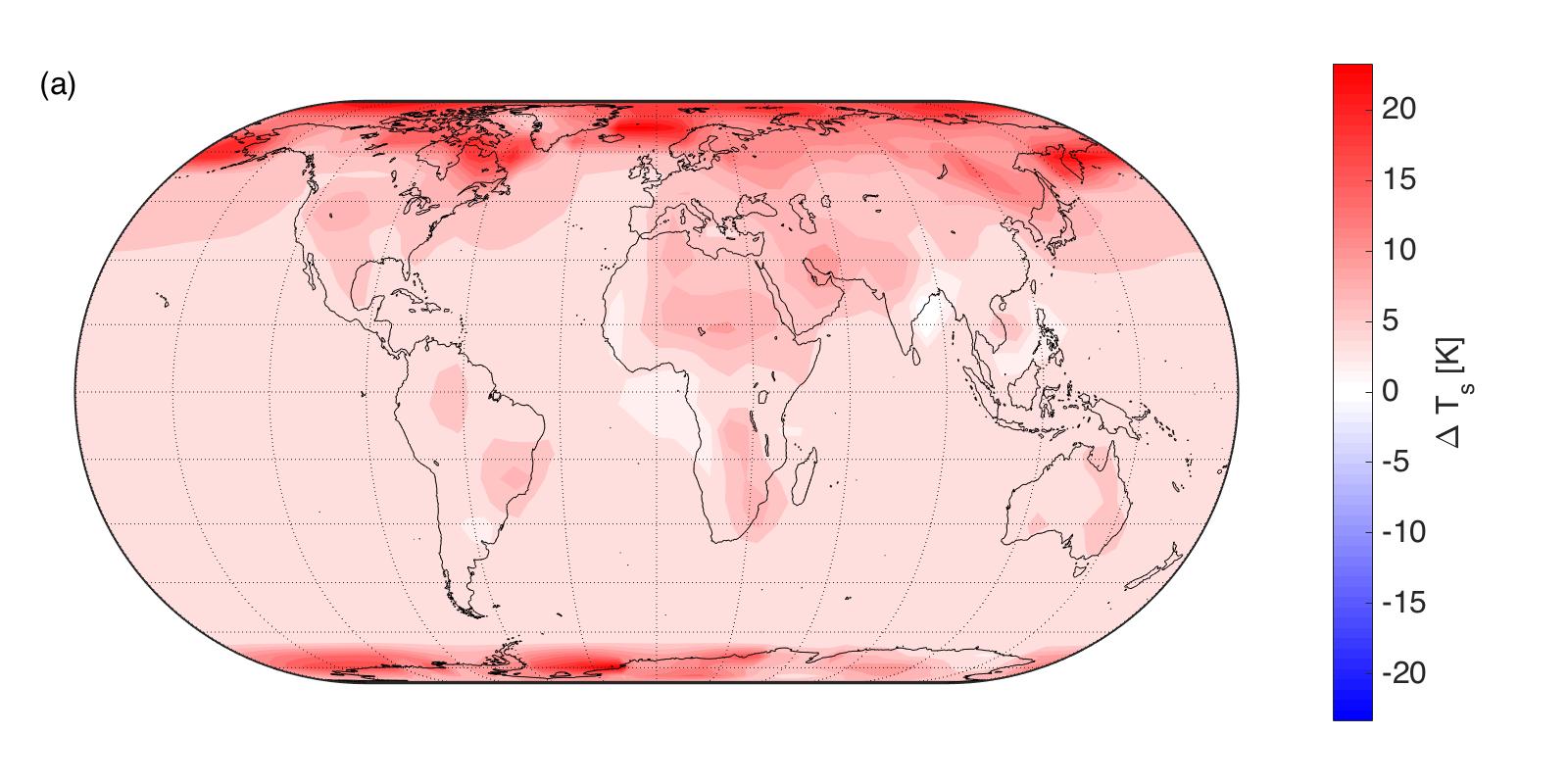} 
            \includegraphics[width=0.5\linewidth]{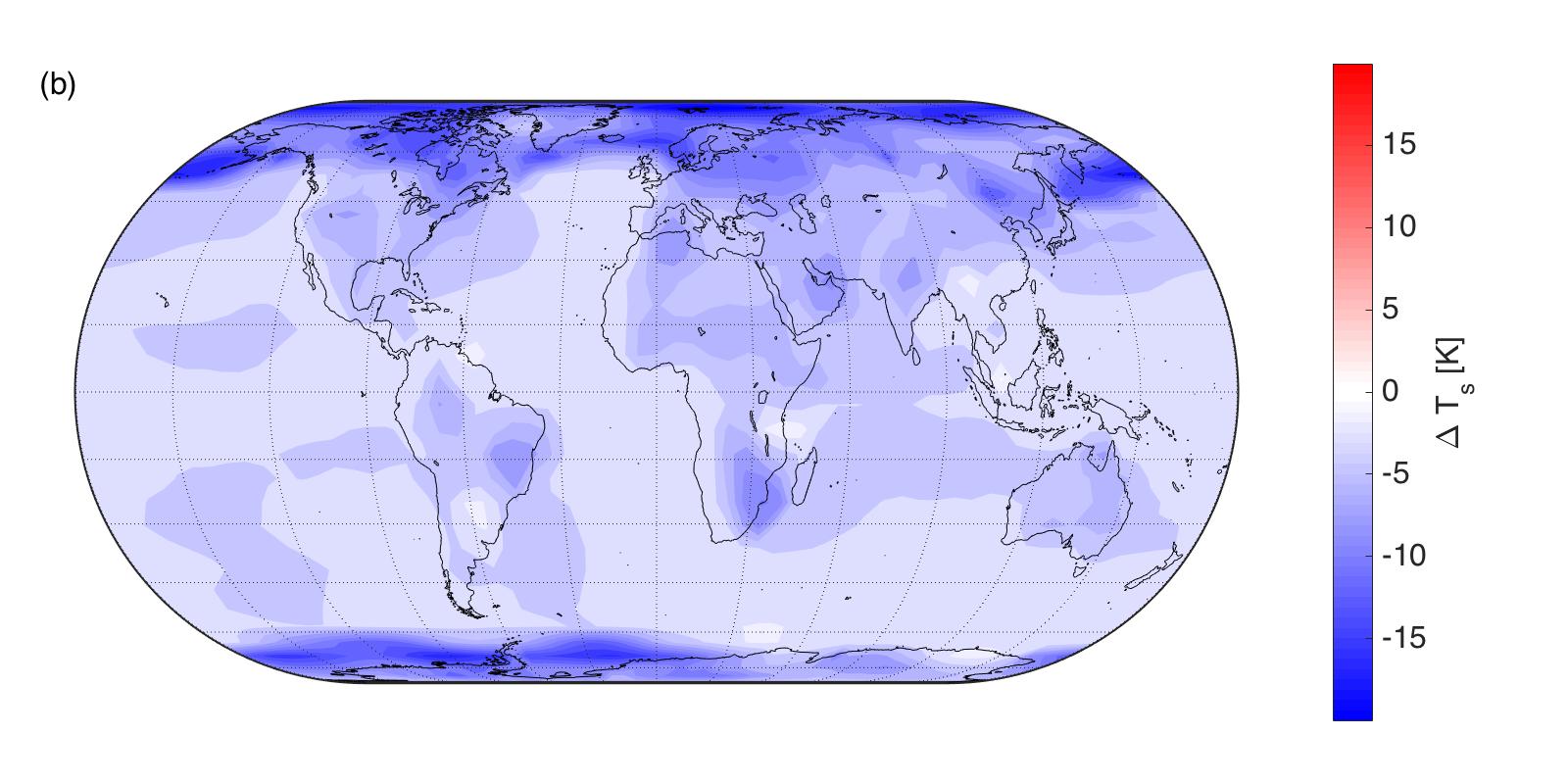} \\
            \includegraphics[width=0.5\linewidth]{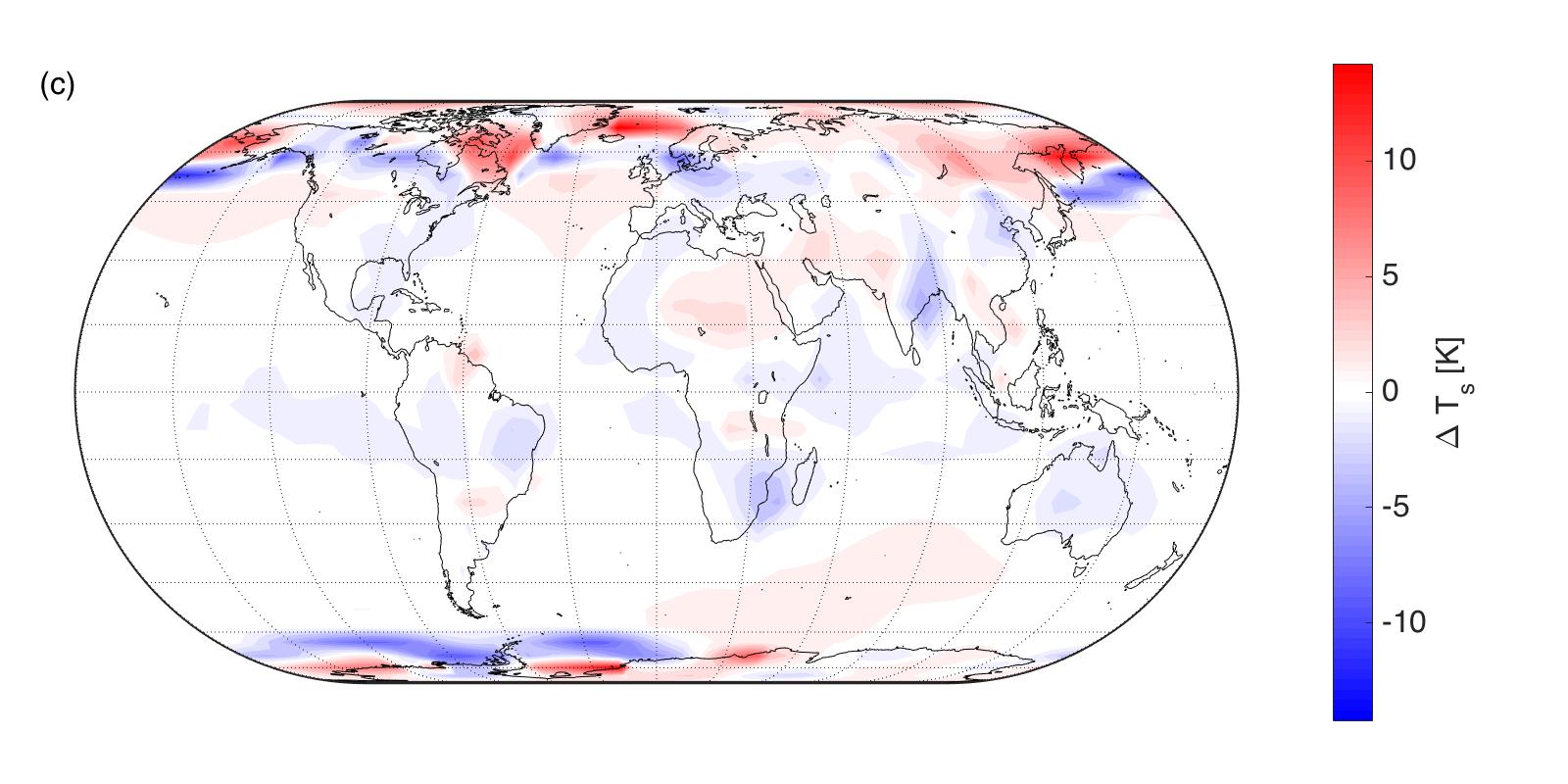} 
            \includegraphics[width=0.5\linewidth]{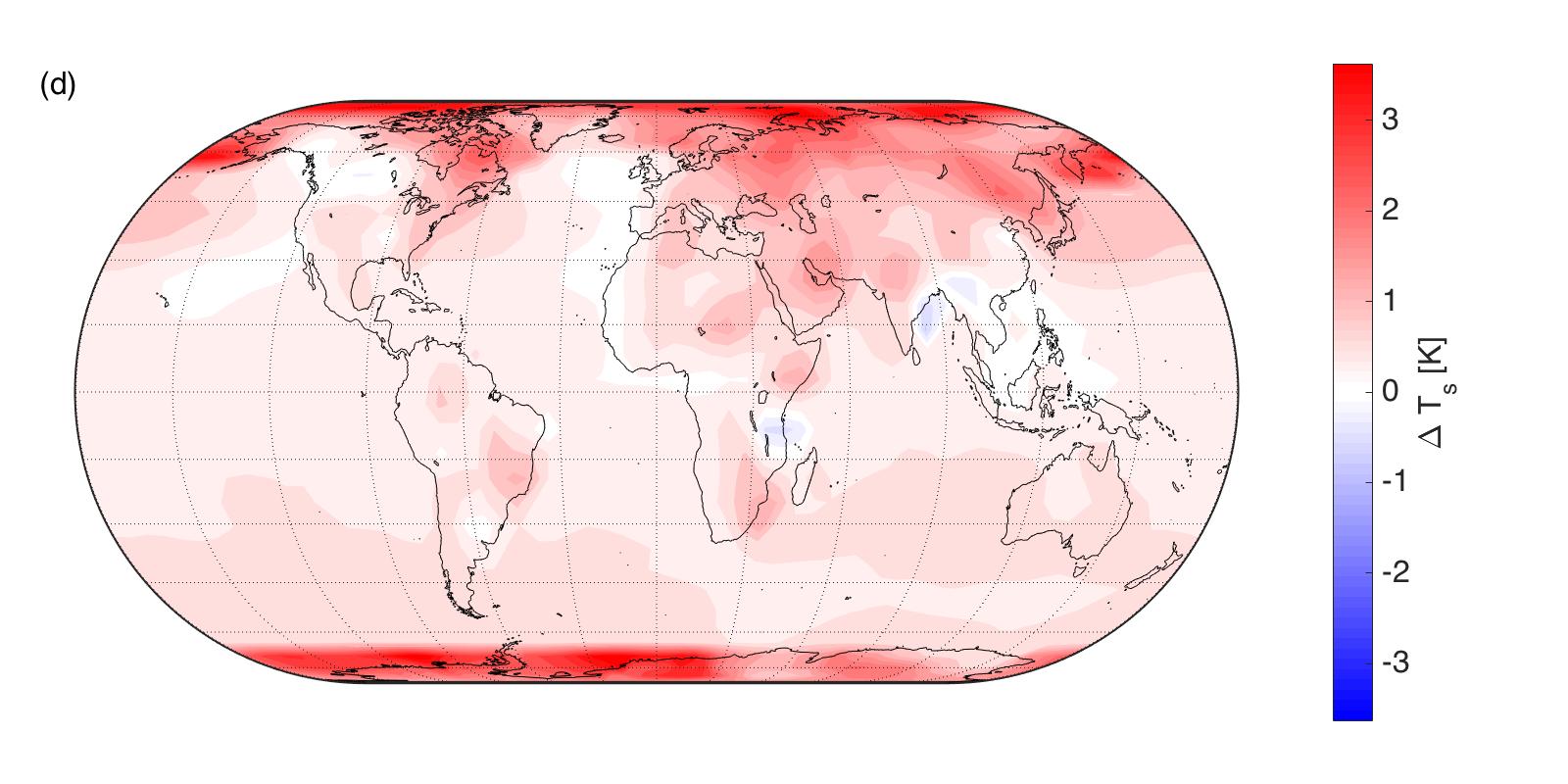} \\
            %\hspace{0.5\linewidth}
            % 23.02.2018.
            \includegraphics[width=0.5\linewidth]{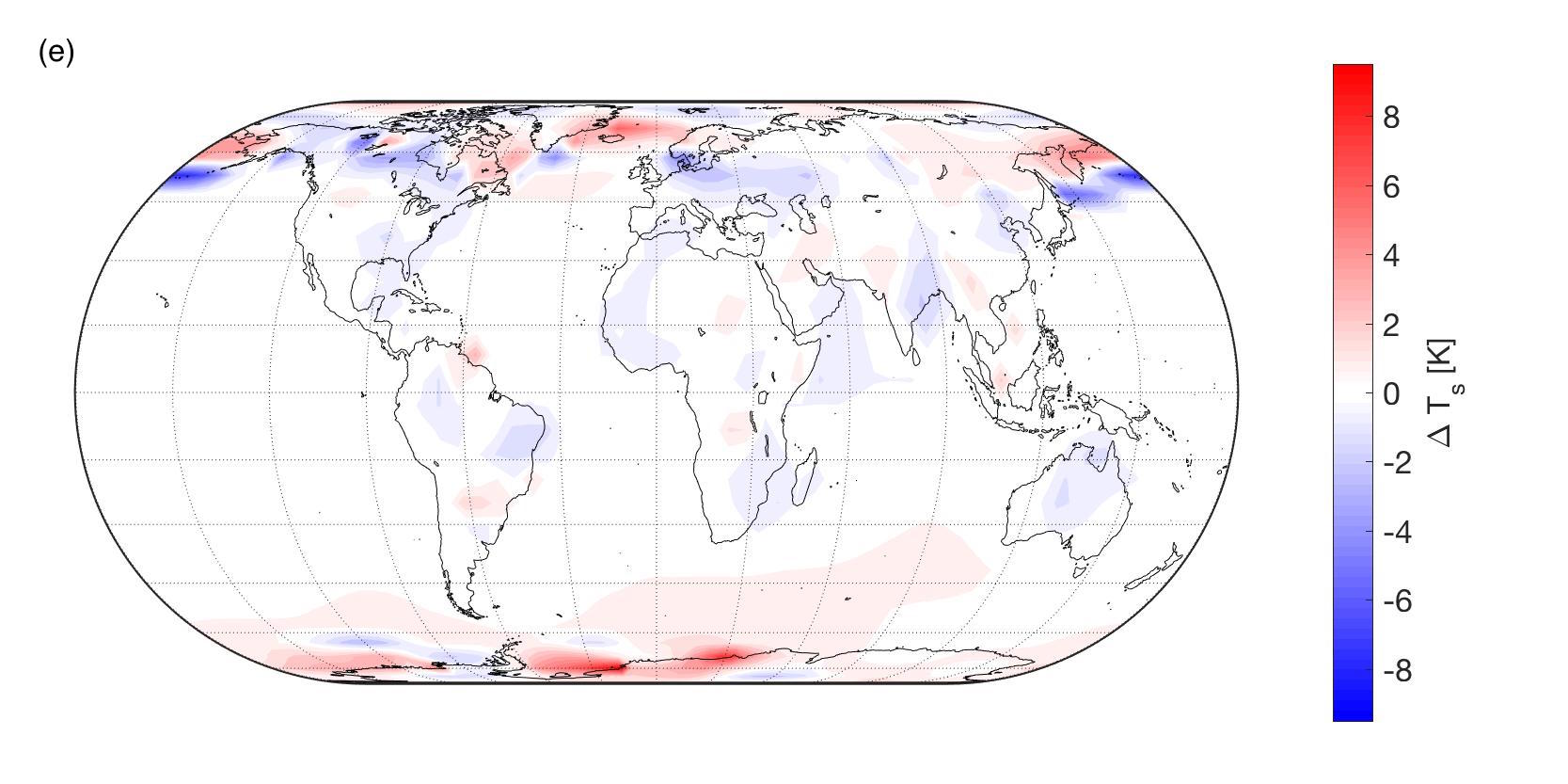}
            \includegraphics[width=0.5\linewidth]{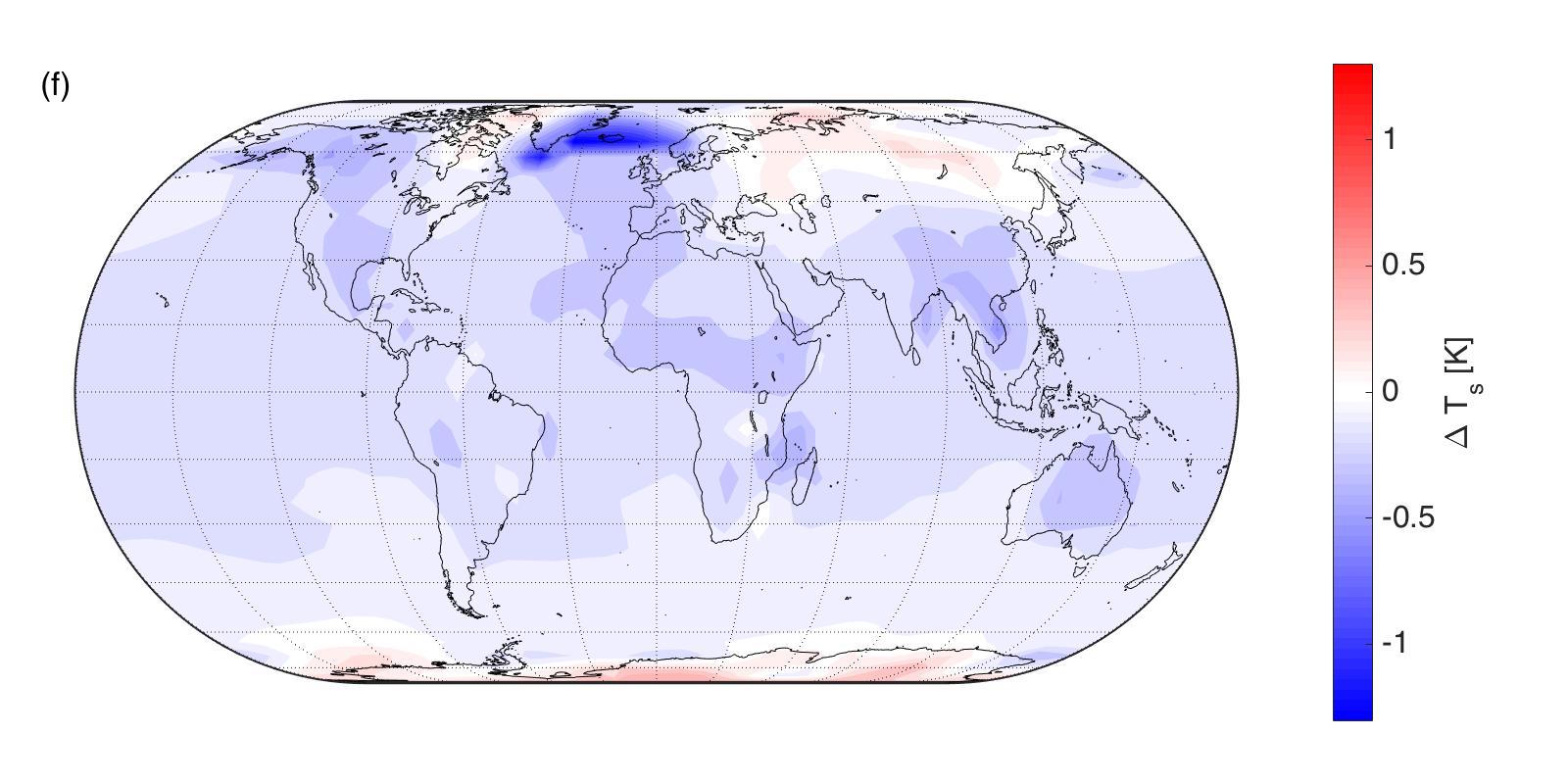}  
        \end{tabular}
        \caption{\label{fig:ecs_spatpattern} Spatial variation of the stationary climate in terms of the surface air temperature belonging to different forcing levels specified by plateaus of forcings collected in Table~\ref{tab:forcing_scenarios}: (a) CX2; (b) SX2; (c) BX2; (d) BX2; (e) BX1; (f) BX1. All diagrams picture the {reference}, except for (c) and (e), which show the linear predictions. Note the different ranges of the temperature for the color bars. 
        }
    %\end{center}
\end{figure*}

We continue with the \textit{diagnostics} of the geoengineering method in view of \textit{uncontrolled} observables. We begin by looking at the observable of the same physical quantity, the surface air temperature, but of a spatial scale other than the global average. We would like to map out the spatial variation of the total response. A comprehensive view of the spatial variations of the response is given by the distribution over the 2D surface, computing the response at each gridpoint separately, as done by Lucarini \textit{et al.}~\cite{Lucarini2017} Similarly to zonal averages (Appendix~\ref{sec:res_zonal}), the response patterns to greenhouse and solar forcings are very similar in the stationary climate regimes; see Figs.~\ref{fig:ecs_spatpattern}(a) and \ref{fig:ecs_spatpattern}(b) for the strong forcings CR2 and SR2, respectively. 
{(See Ref.~\onlinecite{doi:10.1029/2005JD005776} for such a comparison in a complex model.)
The patterns in Figs.~\ref{fig:ecs_spatpattern}(a) and \ref{fig:ecs_spatpattern}(b)} are misaligned slightly, which results in nonzero predicted total responses (BR2) of opposite sign in neighboring regions. This is shown in Fig.~\ref{fig:ecs_spatpattern}(c). The picture for the weaker forcings, CR1, SR1 (not shown), and BR1 [Fig.~\ref{fig:ecs_spatpattern}(e)], is similar.

\begin{figure*}  %[t!]
   % \begin{center}
        \begin{tabular}{cc}
            \includegraphics[width=0.5\linewidth]{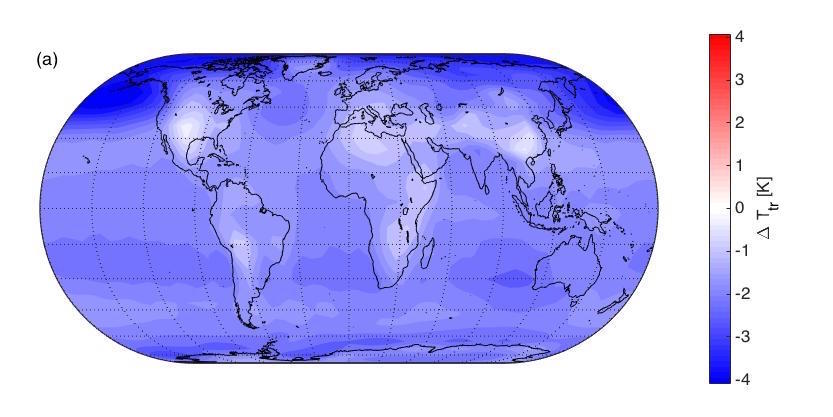} 
            \includegraphics[width=0.5\linewidth]{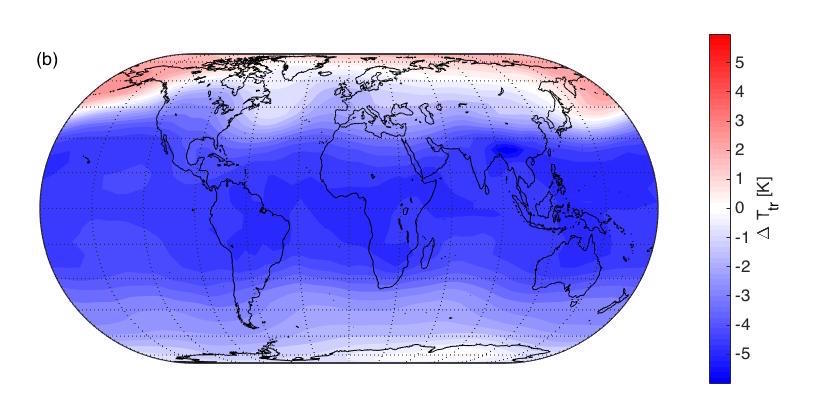} 
        \end{tabular}
        \caption{\label{fig:resp2ramp_spatpattern_aloft} True spatial variation of the stationary climate in terms of the air temperature in the topmost model layer, nearest to the tropopause: (a) BX2; (b) SX2. 
        }
    %\end{center}
\end{figure*}

Unsurprisingly, large predicted residual total responses occur where the response is large to either greenhouse or solar forcing alone. However, the predicted total response turns out to be grossly erroneous; the {reference} regarding the surface air temperature, shown in Figs.~\ref{fig:ecs_spatpattern}(d) for BR2 and \ref{fig:ecs_spatpattern}(f) for BR1, shows that \textit{significant cancellation is achieved even locally}. [We note that the overwhelmingly red and blue colors in Figs.~\ref{fig:ecs_spatpattern}(d) and \ref{fig:ecs_spatpattern}(f), respectively, are consistent with the signs of the true residual total global change shown in Fig.~\ref{fig:resp2ramp}(b).] However, looking at the temperatures at the highest model level, nearest the tropopause, 
the response under combined forcing [BX2, Fig.~\ref{fig:resp2ramp_spatpattern_aloft}(a)] is comparable in magnitude to the response under, say, solar forcing alone [SX2, Fig.~\ref{fig:resp2ramp_spatpattern_aloft}(b)]---nothing like the situation on the surface as seen
in Figs.~\ref{fig:ecs_spatpattern}(b) and \ref{fig:ecs_spatpattern}(f).

\subsection{Annual precipitation}\label{sec:res_precip}

\begin{figure*}  %[t!]
    %\begin{center}
        \begin{tabular}{cc}
            \includegraphics[width=0.5\linewidth]{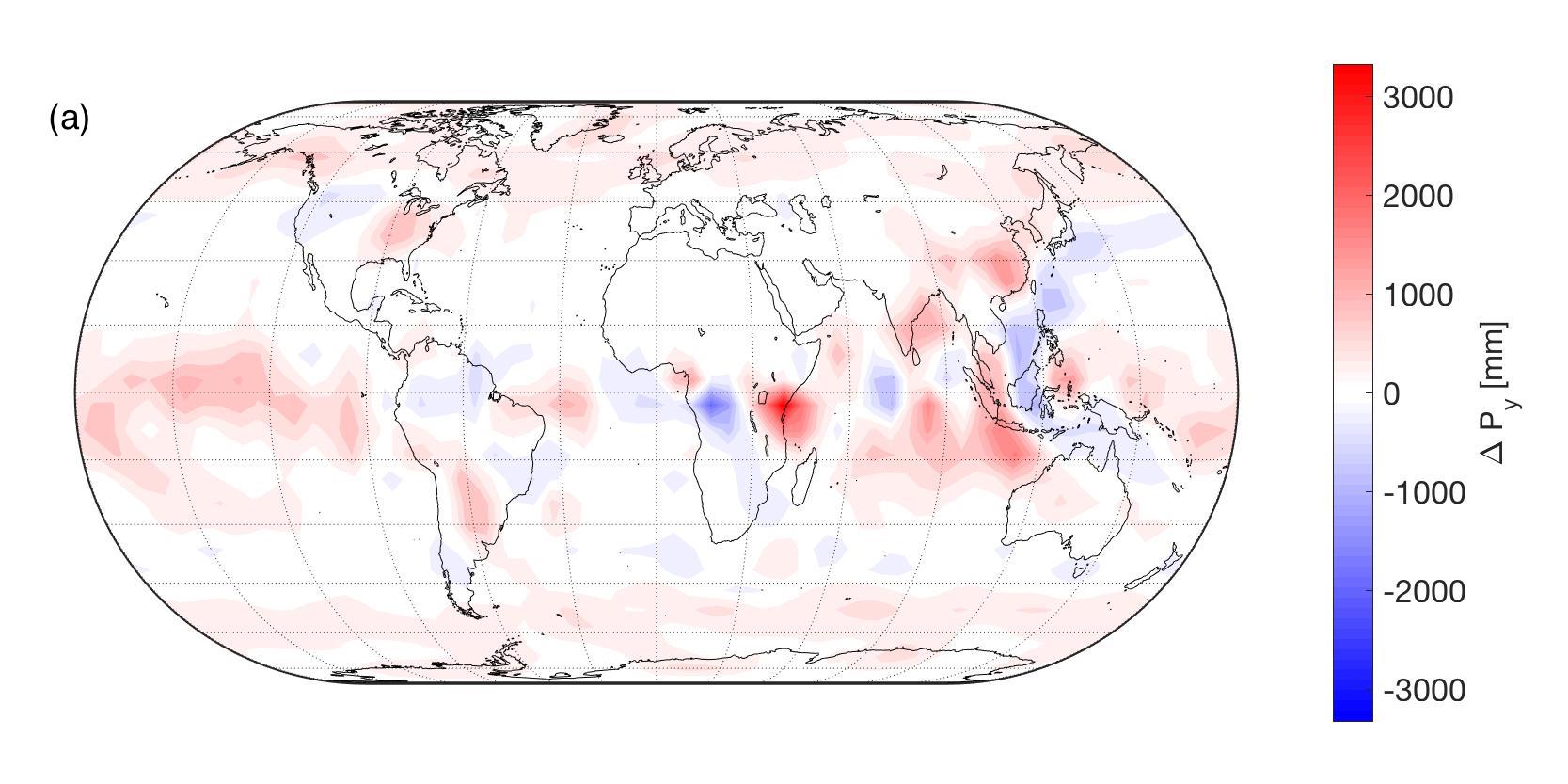} 
            \includegraphics[width=0.5\linewidth]{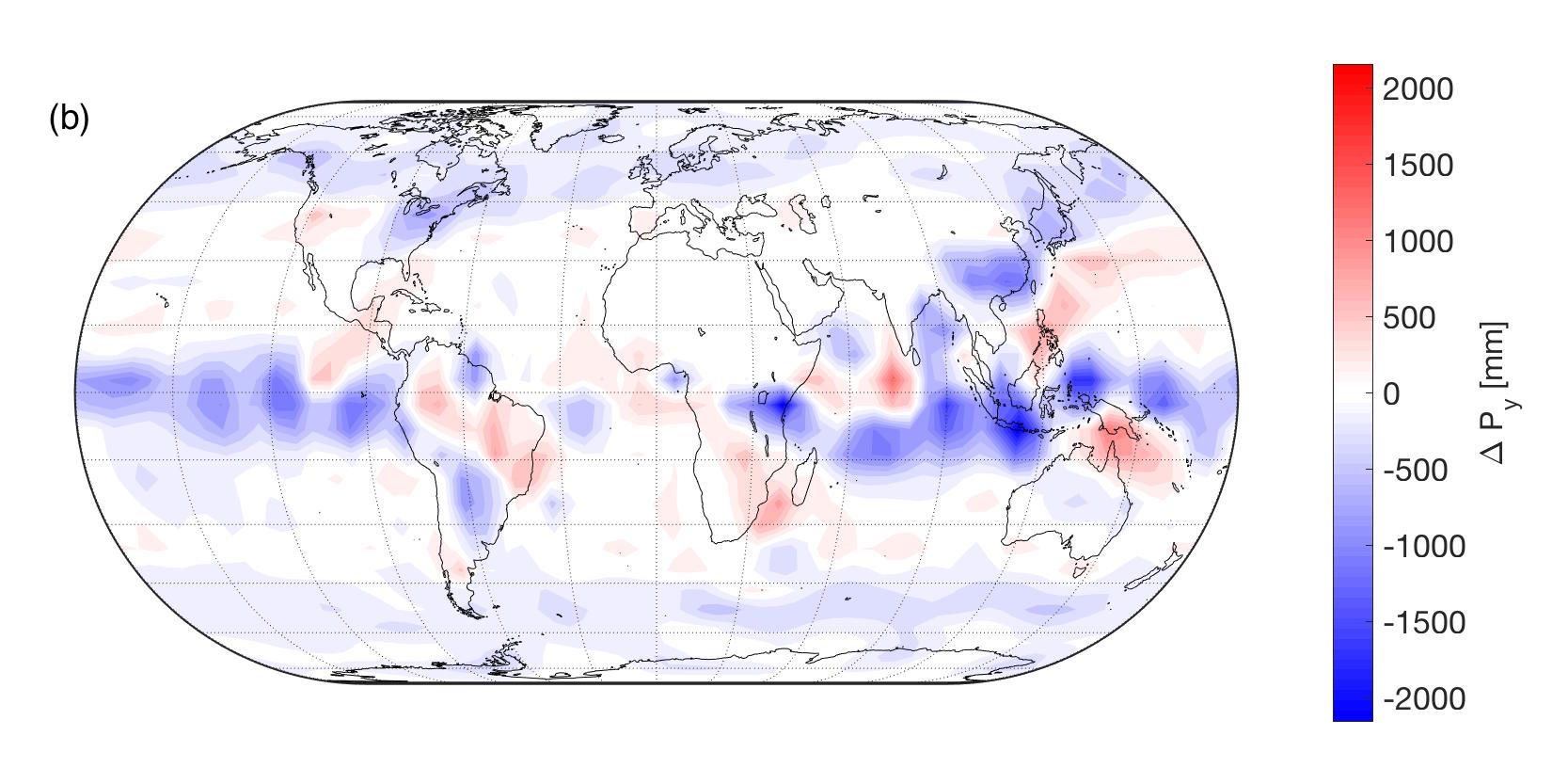} \\
            \includegraphics[width=0.5\linewidth]{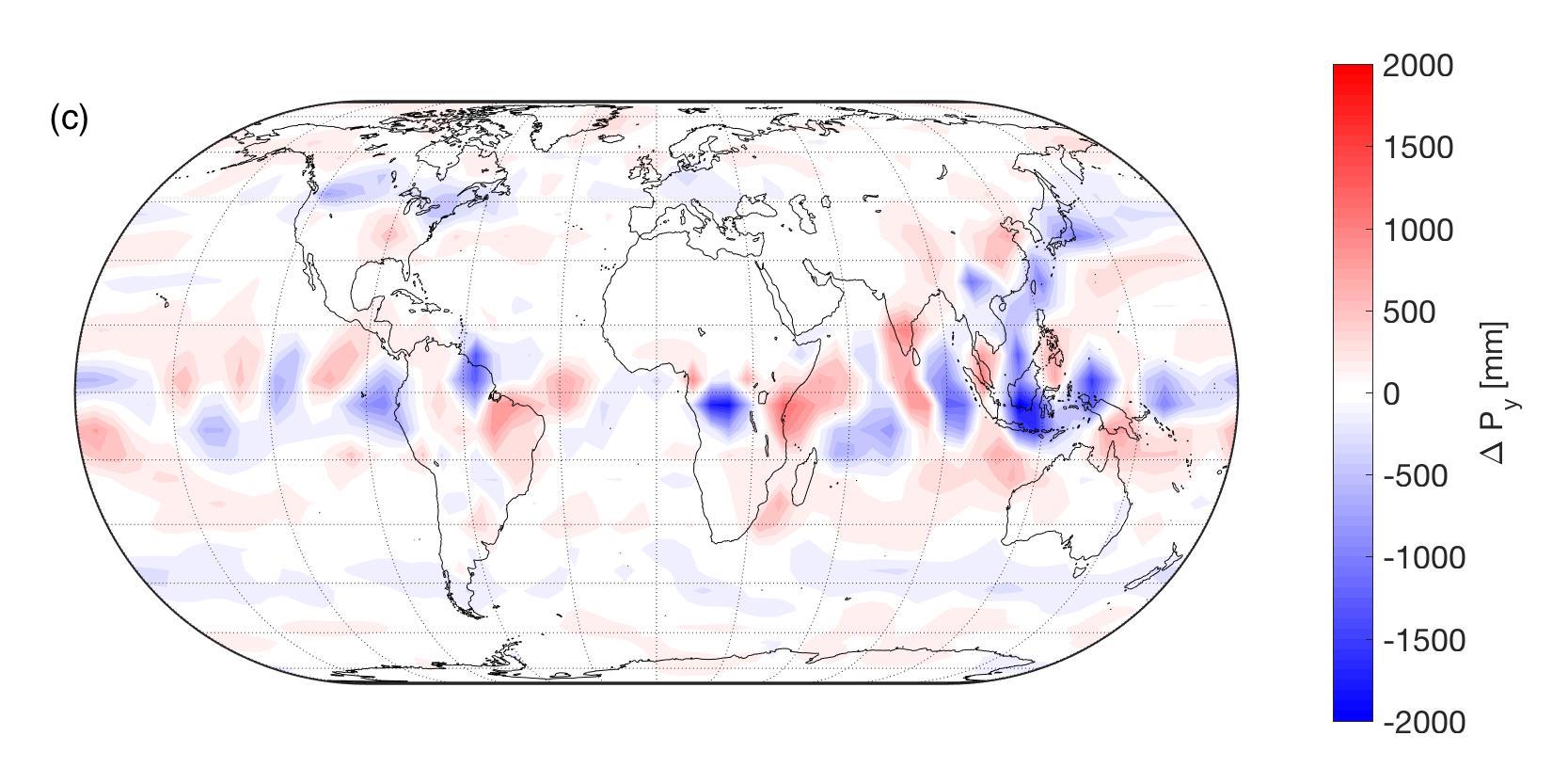} 
            \includegraphics[width=0.5\linewidth]{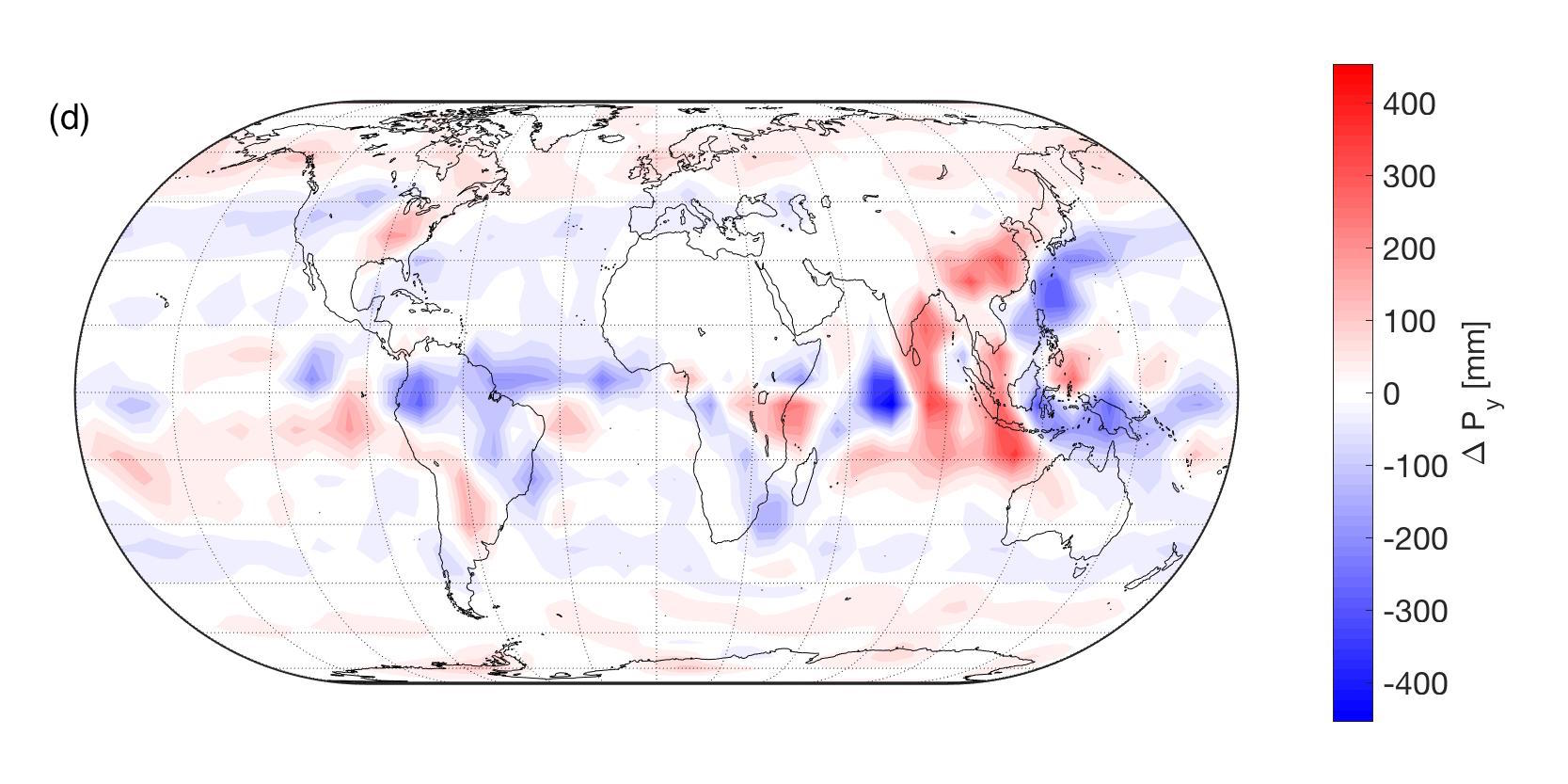} 
            \\
            %\hspace{0.5\linewidth}
            \includegraphics[width=0.5\linewidth]{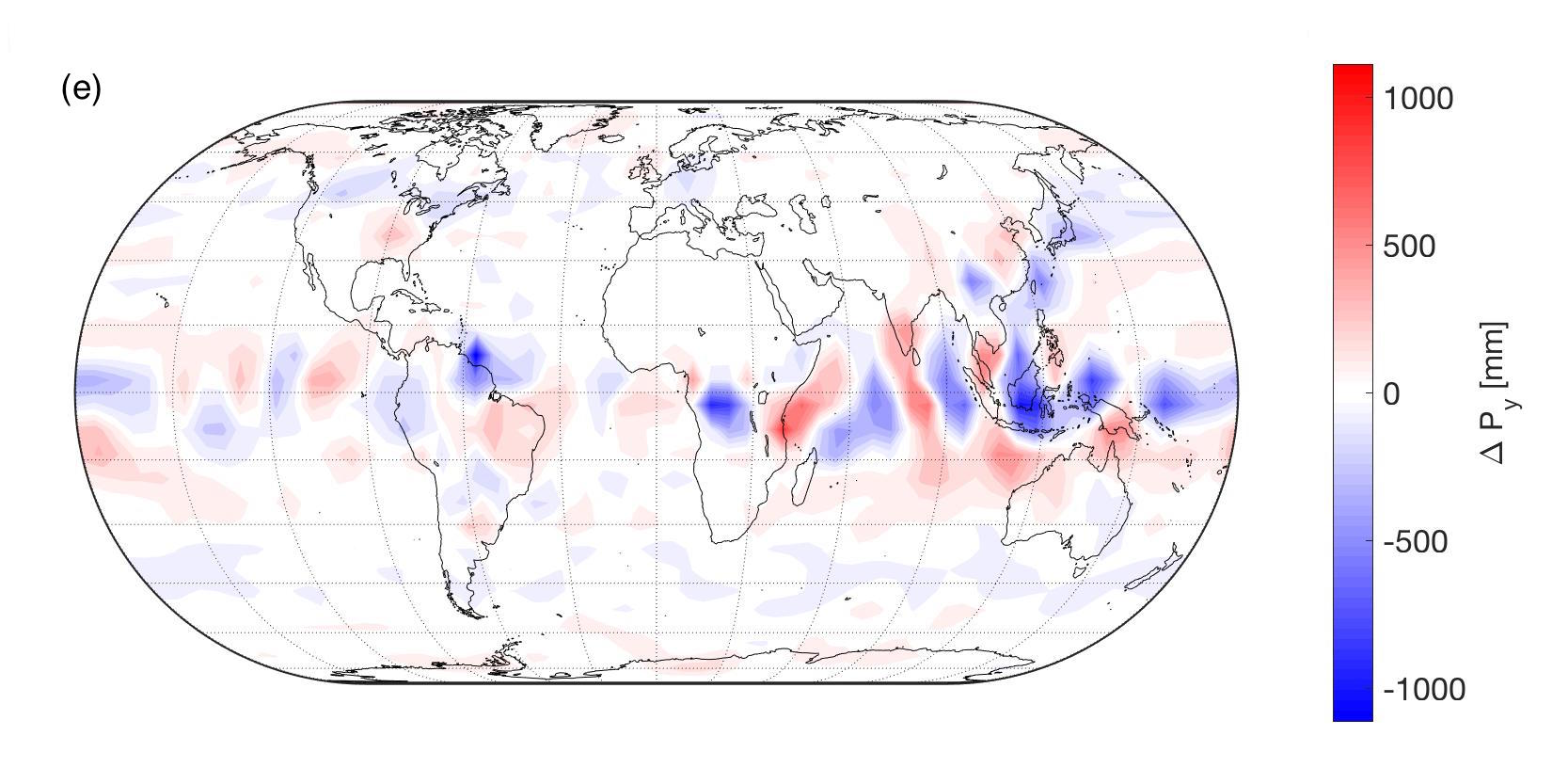}
            \includegraphics[width=0.5\linewidth]{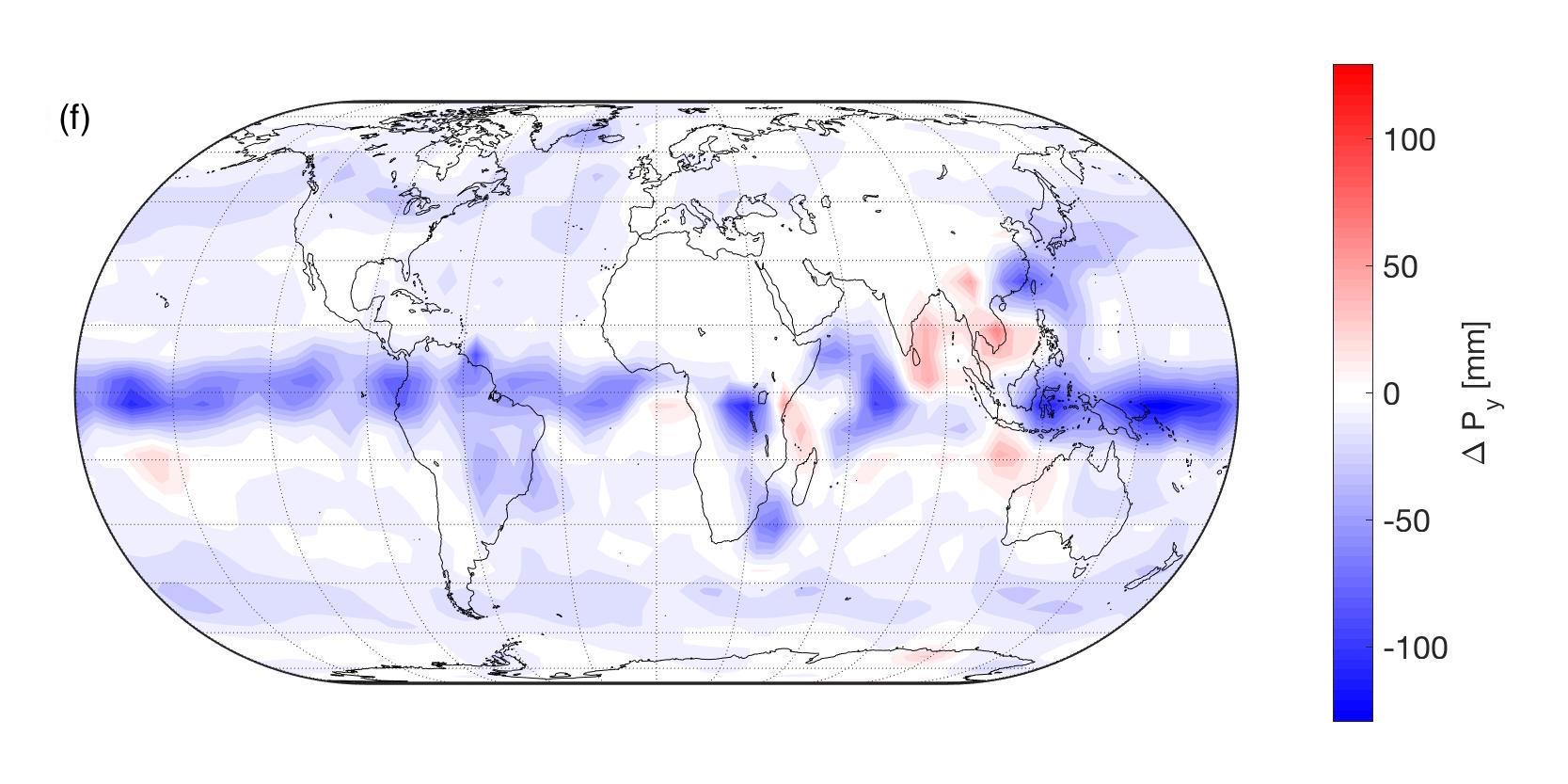}  
        \end{tabular}
        \caption{\label{fig:ecs_spatpattern_precip} Same as Fig.~\ref{fig:ecs_spatpattern}, but for the annual precipitation.}
    %\end{center}
\end{figure*}

\begin{figure*}  %[t!]
   % \begin{center}
        \begin{tabular}{cc}
            \includegraphics[width=0.5\linewidth]{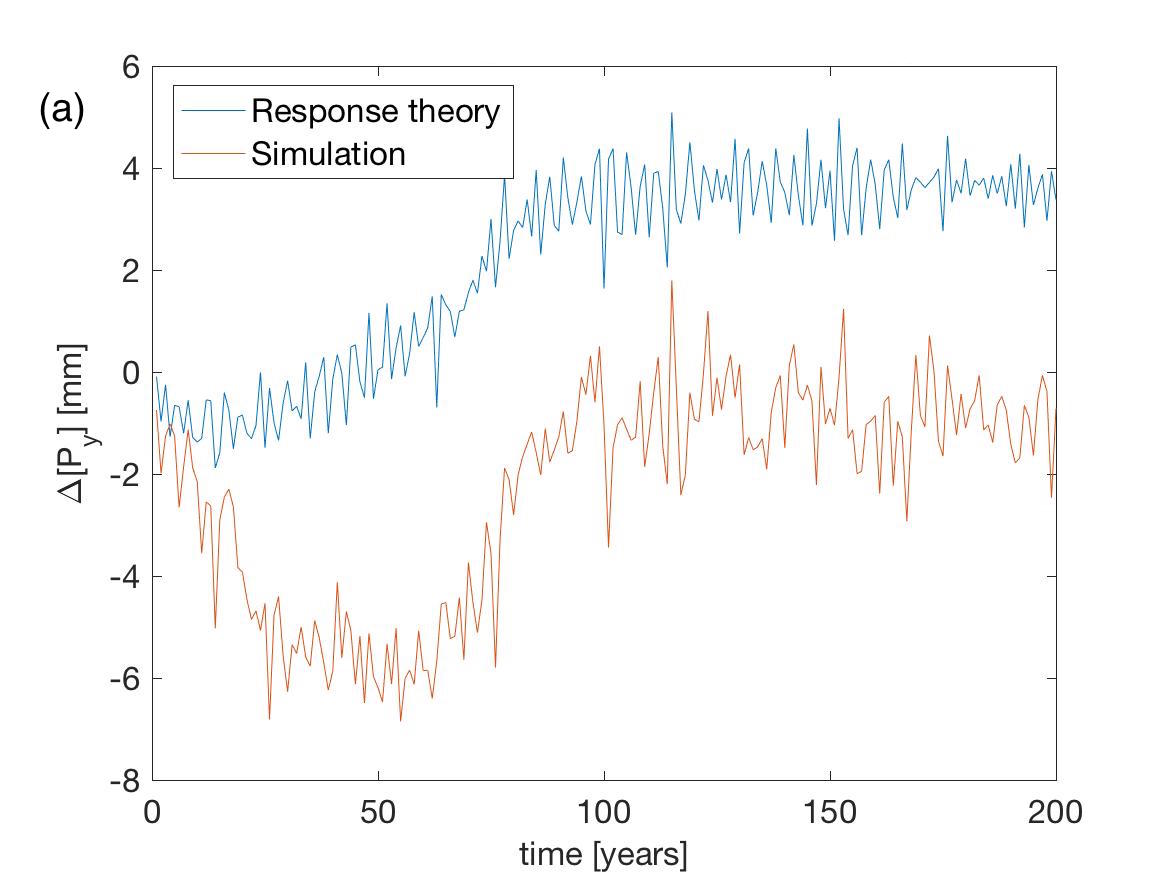} 
            \includegraphics[width=0.5\linewidth]{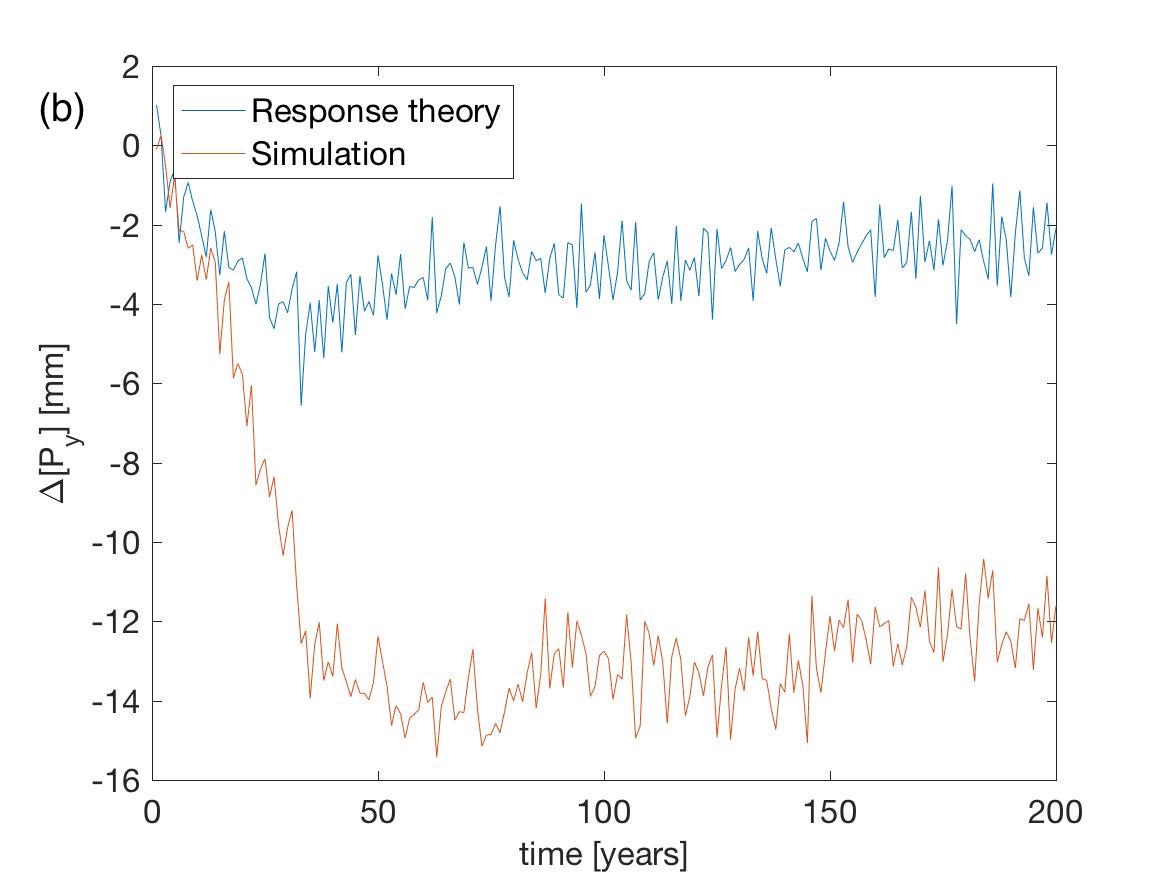} 
        \end{tabular}
        \caption{\label{fig:resp2ramp_combine_globave_precip} Same as Fig.~\ref{fig:resp2ramp}{(b)}, but for the annual precipitation {and showing separately the cases of (a) BR2 and (b) BR1}. 
        }
   % \end{center}
\end{figure*}

Here we present results for another diagnostic observable, the annual precipitation $P_y$. 
In terms of the spatial patterns of the response, very similar conclusions can be drawn for the precipitation as for the surface air temperature, and these conclusions are supported by the set of diagrams in Fig.~\ref{fig:ecs_spatpattern_precip}. The difference is that  
the largest responses are observed at equatorial regions, {and it is not clear what mechanism causes this}. Most importantly, \textit{significant cancellation is actually achieved as opposed to the much less favorable linear prediction}. This is so even if the solar forcing used is the same as before, i.e., one determined with the aim of canceling global warming (not wettening; {in the same spirit as Fig.~4 of Ref.~\onlinecite{acp-16-15789-2016}}). The significant cancellation clearly suggests that the response characteristics of $P_y$ to greenhouse and to solar forcing, say in terms of the respective Green's functions, are very similar, as with the corresponding Green's functions of $T_s$. Nevertheless, there are differences too, as seen in Fig.~\ref{fig:resp2ramp_combine_globave_precip}, such as the more obvious nonmonotonicity of the evolution, despite a monotonic forcing. Furthermore, the linear prediction for the total response in the stationary climate is nonzero, which is clearly because we are looking at an uncontrolled observable. This linear prediction, however, is quite ``unreliable,'' as can be expected from
{the mismatch of the true and predicted} spatial patterns. Otherwise, both the predicted and the true total {global mean} responses to combined forcing look rather negligible compared with the responses to the greenhouse or solar forcings acting separately; see also Table~\ref{tab:precip_global_change_equilib}. Interestingly, the transient responses (not shown) have similar qualities to those of the temperature: nonlinearity is most obvious for CR2 as opposed to CR1, SR1, and SR2.

\begin{table}
\caption{Global average stationary climatology of the annual precipitation belonging to different forcing levels.}\label{tab:precip_global_change_equilib}
  %\begin{center}  
  \renewcommand{\arraystretch}{1.3}
   \begin{ruledtabular} \begin{tabular}{lcccc}
      %\toprule
       Forcing           & CX1 & CX2 & SX1 & SX2 \tabularnewline %\hline%\midrule
       $\Delta \langle[P_y]\rangle_{\infty}$ (mm) & 74  & 124 & $-$71  & $-$121  
      %\bottomrule
    \end{tabular}\end{ruledtabular}
  %\end{center}
\end{table}
% Data from MATLAB/Response_theory/Clim_eng/New_set/Results/2016.09.20/1/51.jpg

We note that equatorial drying under a similar geoengineering scenario has also been reported by others.\cite{1748-9326-9-1-014001,acp-16-15789-2016} However, in {these studies} a quadrupling of [CO$_2$] was considered. We point out that it does {seem to} matter what levels of change we consider: under [CO$_2$]-doubling, we  {actually find  less drying} than in the case of the $\sqrt{2}$-fold [CO$_2$] increase. {This finding can, however, have different reasons. One candidate is that the response under combined forcing is nonlinear; another  is that (assuming that the response under combined forcing is approximately linear) the required solar forcing was determined inaccurately [which already resulted  in a residual response as seen in Fig.~\ref{fig:resp2ramp}(b)]. {Note that in Refs.~\onlinecite{1748-9326-9-1-014001,acp-16-15789-2016}, an exact cancellation of global mean surface temperature was achieved in the stationary climate, like, for example, in the G1 GeoMIP experiment. Given this, Fig.~4 of Ref.~\onlinecite{acp-16-15789-2016} indicates that the response of the global mean is approximately linear in most CMIP5 models considered, at least up to a certain forcing level that was actually lower than [CO$_2$]-doubling. In the following, we  
{indicate} that both of {the said} effects play a role, i.e., nonlinearity is {likely} also present in our case; however, it {might} not be the dominant component.}} {Drying while global average surface temperature was maintained in a model was reported also by Ricke \textit{et al.}~\cite{RMA:2010,RRIKM:2012}}

\section{Improved methodology and results}\label{sec:improved}

\subsection{Achieving a cancellation of the global mean surface temperature change}\label{sec:improved_I}

\begin{figure*}  %[t!]
    %\begin{center}
        \begin{tabular}{cc}
            \includegraphics[width=0.5\linewidth]{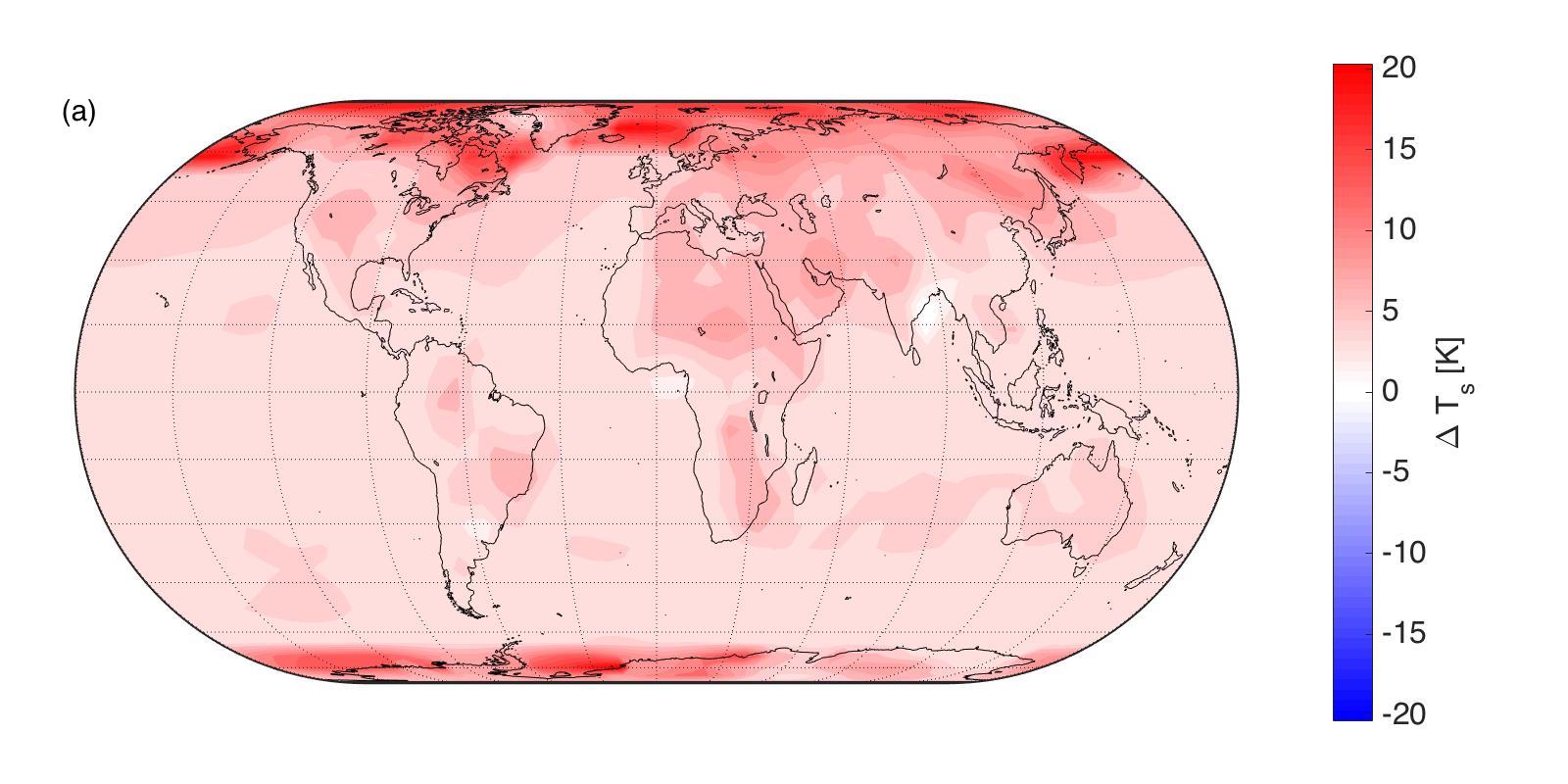} 
            \includegraphics[width=0.5\linewidth]{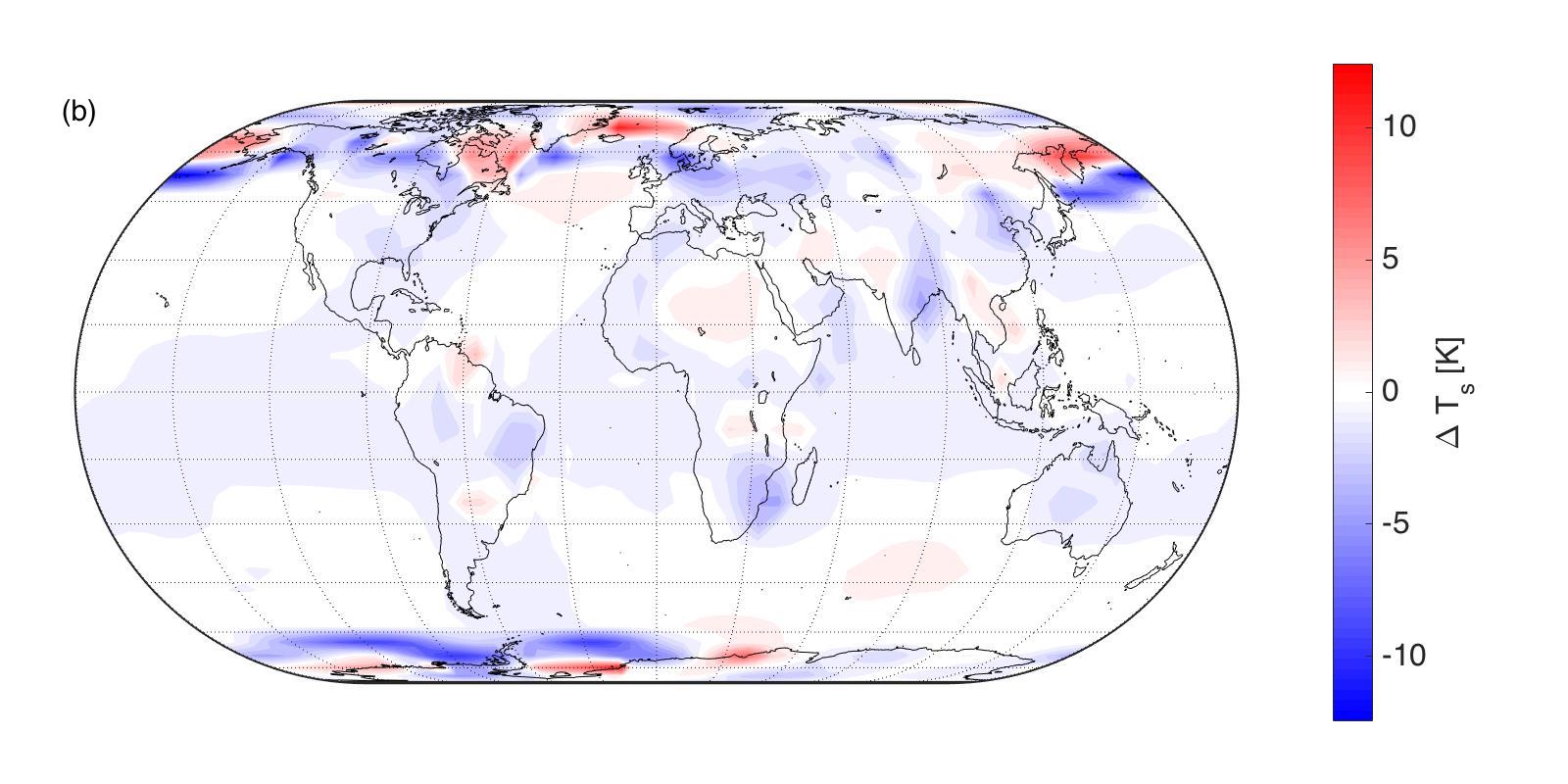} 
        \end{tabular}
        \caption{\label{fig:ecs_spatpattern_inv} Spatial variation of the stationary climate in terms of the air temperature: (a) true response under SS2I; (b) predicted response under combined forcings used for SS2 and SS2I amounting to no forcing. 
        }
   % \end{center}
\end{figure*}

This subsection pertains to our objective (O1). The very close resemblance of the patterns seen in Figs.~\ref{fig:ecs_spatpattern}(a) and \ref{fig:ecs_spatpattern}(b) hints that the effect of a changing $[{\rm CO}_2]$ on the radiative forcing shaping the surface air temperature is very similar to that due to a changing solar strength. However, with these data, we are not properly informed about just how similar, because, for example, the CR2 and SR2 forcings act in \textit{opposite} directions and, owing to nonlinearities, they do not have to have the same effect even if the effects due to forcings acting in the same direction are indistinguishable. Therefore, we produced just that missing simulation: complementing SS2, for which the applied solar forcing is a step of equal magnitude but opposite sign. For this forcing, the stationary climate 
is shown in Fig.~\ref{fig:ecs_spatpattern_inv}(a), to be referred to as SS2I. It is virtually indistinguishable from the pattern resulting for CS2, seen in Fig.~\ref{fig:ecs_spatpattern}(a), including a lack of such misalignment as the comparison of Figs.~\ref{fig:ecs_spatpattern}(a) and \ref{fig:ecs_spatpattern}(b)   revealed. {This goes beyond the report on the (approximate) ``equivalence'' of greenhouse and solar forcings with respect to (asymptotic in time) \textit{global average} surface temperature;\cite{BOSCHI20131724} this is extended now to \textit{regional averages}, i.e., spatial patterns, of that variable with a remarkable degree of approximation. {[Just how close this equivalence is} will be indicated by Fig.~\ref{fig:ecs_spatpattern_cancel}(a).] 
} 

The superposition of the stationary climates for SS2 and SS2I, displayed in Fig.~\ref{fig:ecs_spatpattern_inv}(b), is in turn almost indistinguishable from the asymptotic total response to combined BR2 forcing, seen in Fig.~\ref{fig:ecs_spatpattern}(c). 
By inspection of Eq.~(\ref{eq:perturbative}), this pattern turns out to be created by even-order nonlinear perturbative terms of the response.
The selection of the even-order terms takes exactly the superposition of the responses from two experiments where the forcings are equal but of opposite sign: $\varepsilon_1=-\varepsilon_2$.

Instead of eliminating the even-order terms by superposition, we can retain only the odd-order terms by subtraction. We proceed in this direction, assuming that the third-order and higher odd-order terms make a negligible contribution. This way, we attempt to improve on the results for the linear susceptibility---and so ultimately on our prediction of the required solar forcing needed for canceling global warming. This is done with the aim of making an advance regarding our objective (O1). We can then apply this forcing in a new experiment coded as BR2C (``C'' for ``cancel''). For this experiment, we can utilize (although we will not examine the transient\footnote{The precise treatment of the transient proceeds by solving the same inverse problem as outlined in Appendix~\ref{sec:inverse_problem}, centered around Eq.~(\ref{eq:tot_resp_iter}), except that the impulse responses in that equation, e.g., $\tilde{h}_{\Psi,{g}}$, need to be produced as an average from two simulations each, as also done by Gritsun and Lucarini.\cite{GRITSUN201762}}) our finding that the response characteristics to greenhouse and solar, i.e., short-wave and long-wave radiative, forcings are very similar, which would allow the application of a solar forcing that is a simple straight ramp, just like $\log([{\rm CO}_2]/[{\rm CO}_2]_0)(t)$, having the same length before the plateau. {(This should be the rationale behind the G2 experiments of GeoMIP.)} That is, what we improve on here is only the \textit{level} of the plateau. It is rather straightforward to obtain the following equations for this level $f_{\infty,BR2C,s}$:
\begin{align}
 \chi_{[T_s],\infty,s} &= \frac{|\Delta\langle[T_s]\rangle_{\infty,SS2}|+|\Delta\langle[T_s]\rangle_{\infty,SS2I}|}{2|f_{\infty,SS2}|}, \\
 \chi_{[T_s],\infty,g}&= \frac{|\Delta\langle[T_s]\rangle_{\infty,CS2}|+|\Delta\langle[T_s]\rangle_{\infty,CS2I}|}{2|f_{\infty,CS2}|}, \\
 |\Delta\langle[T_s]\rangle_{\infty,BR2C}| &= \chi_{[T_s],\infty,s}|f_{\infty,BR2C,s}| - \chi_{[T_s],\infty,g}|f_{\infty,BR2C,g}| \label{eq:tot_resp_noniter},\\
 |\Delta\langle[T_s]\rangle_{\infty,BR2C}| &= 0.
\end{align} 
The subscripts  $\infty$ refer to the asymptotic, stationary climate regime; other subscripts refer to the experiment, i.e., forcing scenario. Observe that we need data from a new experiment, CS2I, where the ``I'' indicates an experiment related to CS2 analogously to the relation of SS2I to SS2. Since we are interested in the stationary climate regime only, owing to ergodicity we can produce just a single long trajectory instead of an ensemble. The result of this is $\Delta\langle[T_s]\rangle_{\infty,CS2I}=-5.11$~K (while we already have $\Delta\langle[T_s]\rangle_{\infty,SS2I}=4.36$~K, and, from Fig.~\ref{fig:resp2step},  $\Delta\langle[T_s]\rangle_{\infty,SS2}=-\Delta\langle[T_s]\rangle_{\infty,CS2}=-4.90$~K). Having  $|f_{\infty,BR2C,g}|=|f_{\infty,CS2}|$, we can express the sought-for forcing in relative terms based on the temperature data only, such as
\begin{equation}\label{eq:improved_pred}
 \frac{|f_{\infty,BR2C,s}|}{|f_{\infty,SS2}|} = \frac{|\Delta\langle[T_s]\rangle_{\infty,CS2}|+|\Delta\langle[T_s]\rangle_{\infty,CS2I}|}{|\Delta\langle[T_s]\rangle_{\infty,SS2}|+|\Delta\langle[T_s]\rangle_{\infty,SS2I}|} = 1.08. 
\end{equation} 
In fact, we carried out the BR2C experiment independently: \textit{iteratively} determining a solar forcing that cancels to a very good approximation the total response (similarly to how the level for, e.g., SS2 was determined observing the result of CS2). This forcing in the above relative terms was found to be 1.11, % based on data from FL's email on 14.04.17.
agreeing well with our prediction of 1.08. % I have the data for this in FL's email sent on 20.04.17.

Given that our prediction is smaller than the forcing actually needed  for cancellation, we can predict an \textit{upper bound} on the actual total response to our predicted forcing by substituting the actually needed value $|f_{\infty,BR2C,s}|/|f_{\infty,SS2}|=1.11$  into Eq.~(\ref{eq:tot_resp_noniter}) {(assuming that the response under combined forcing is linear).} This gives $\Delta\langle[T_s]\rangle_{\infty,BR2C} < 0.134$~K. Considering that the total residual response with the original naive methodology (Appendices~\ref{sec:obtain_greens} and \ref{sec:inverse_problem}) was 0.6~K, this means that with the improved methodology we managed to reduce the total response almost to  \textit{one-fifth} or even less of the said first result. (Of course, the exact reduction can be easily determined by an extra simulation, which we have not run.) In fact, some residual total response even with the improved method could be expected, since the simple measure of nonlinearity (\ref{eq:simple_nonlin}) indicated that linearity is much more violated by an increasing radiative forcing as opposed to a decreasing one. This suggests that the third-order \textit{odd} perturbative term is not very small relative to the second-order one---contrary to the assumption of our improved methodology. {Another source of error could be a nonlinear component of the response under combined forcing.} This is what we examine next.

\subsection{Uncontrolled response and its (non)linearity}\label{sec:improved_II}

\begin{figure*}  %[t!]
    %\begin{center}
        \begin{tabular}{cc}
            \includegraphics[width=0.5\linewidth]{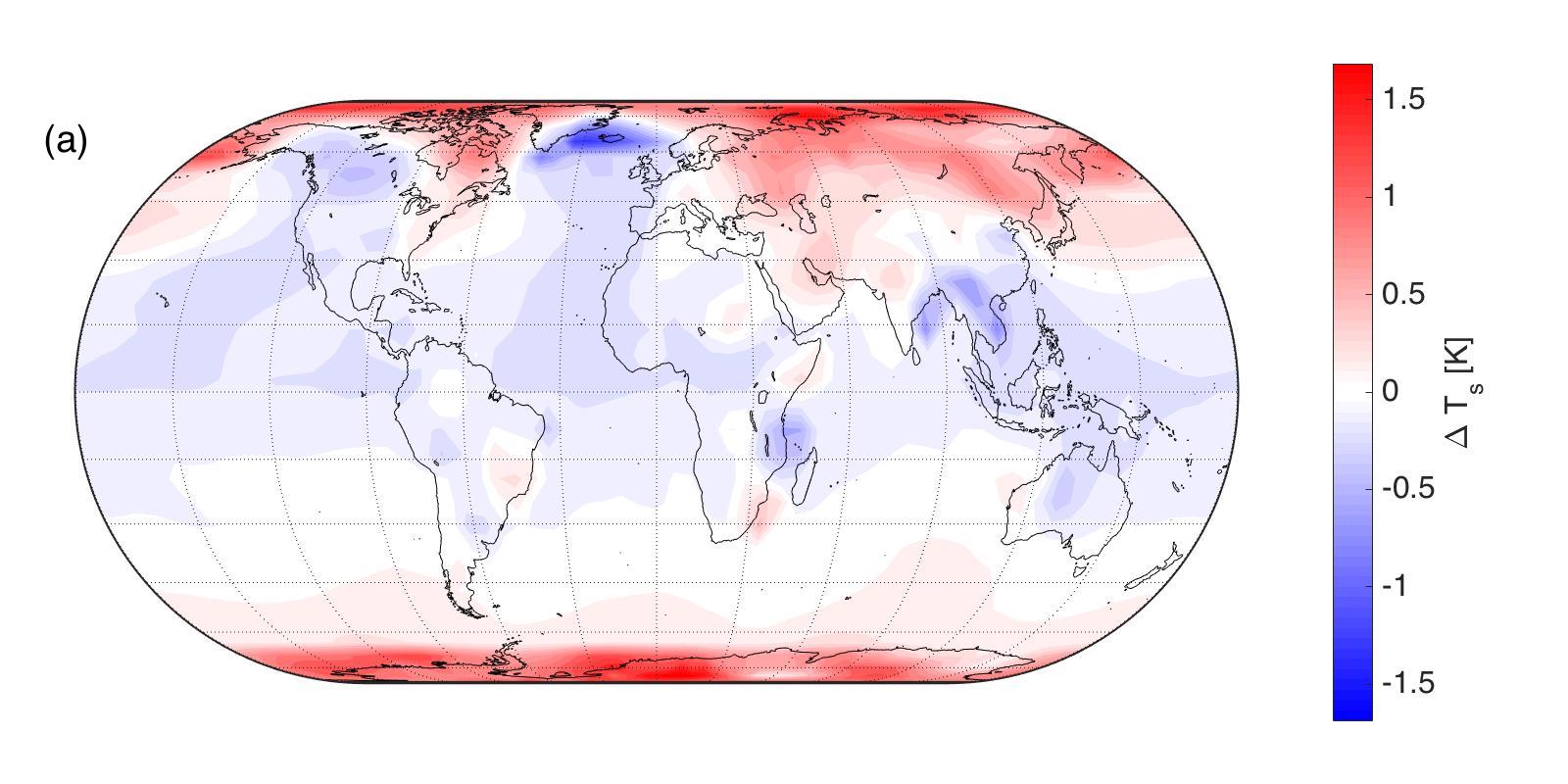} 
            \includegraphics[width=0.5\linewidth]{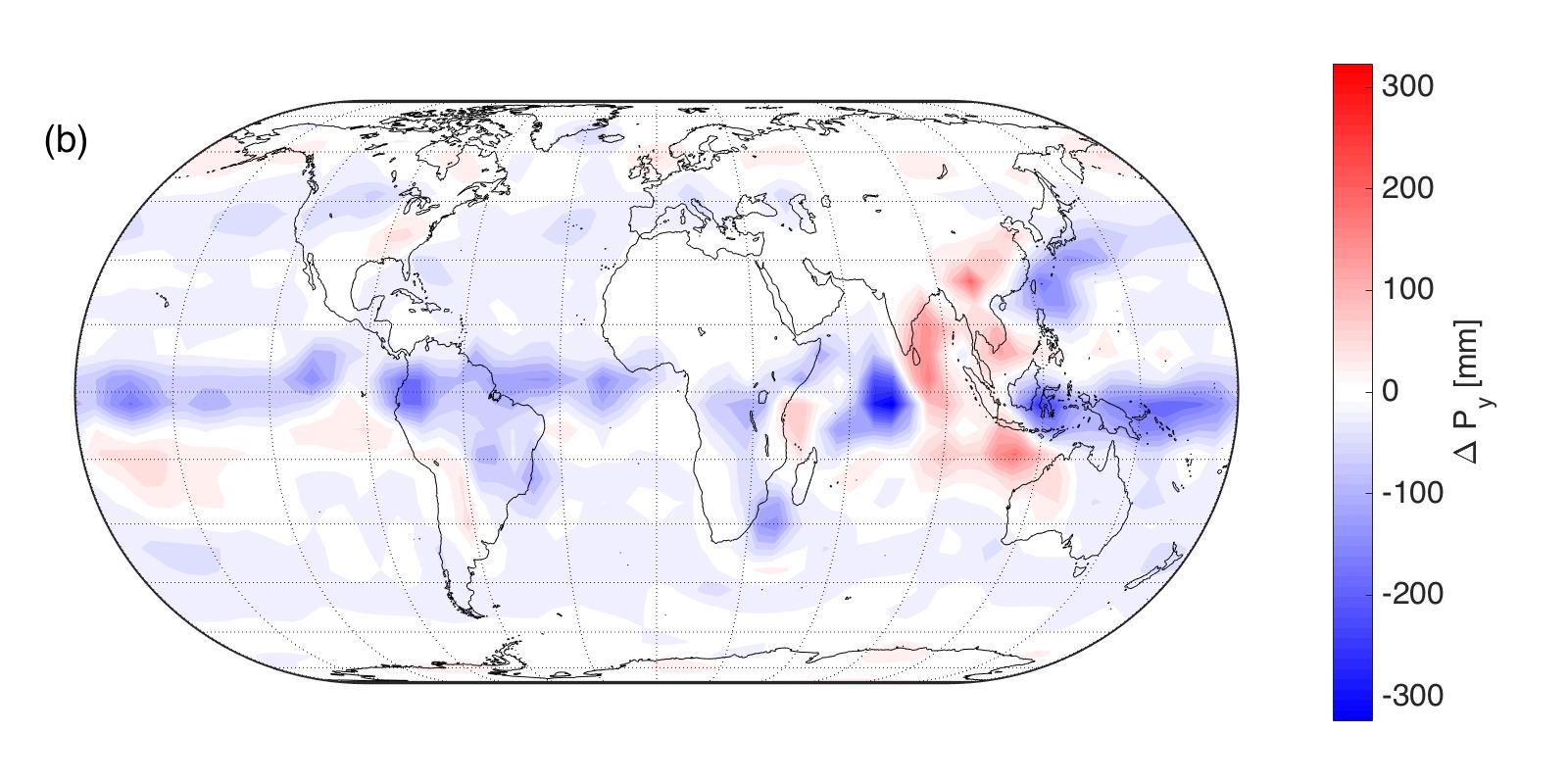} \\
            \includegraphics[width=0.5\linewidth]{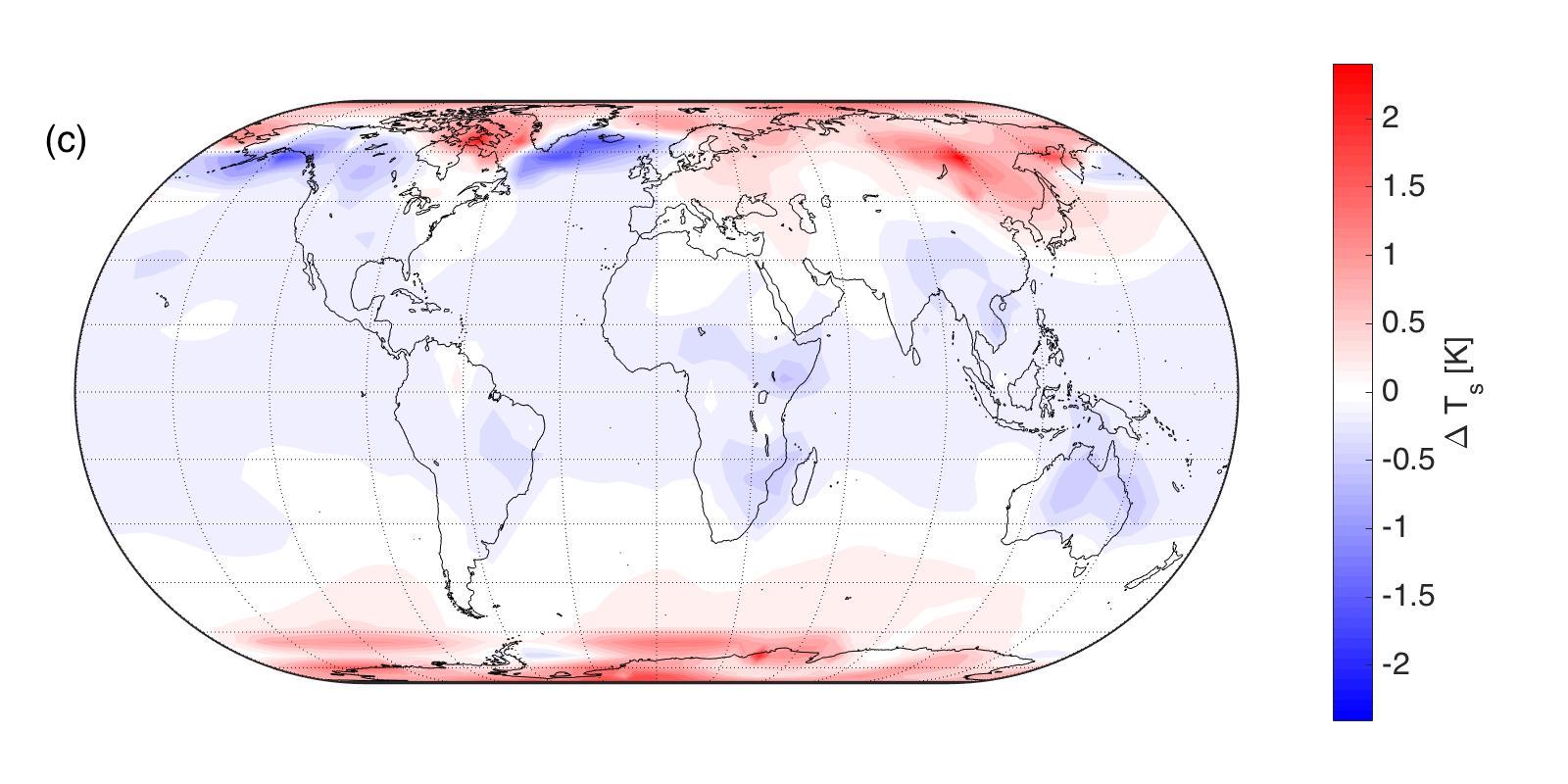}
            \includegraphics[width=0.5\linewidth]{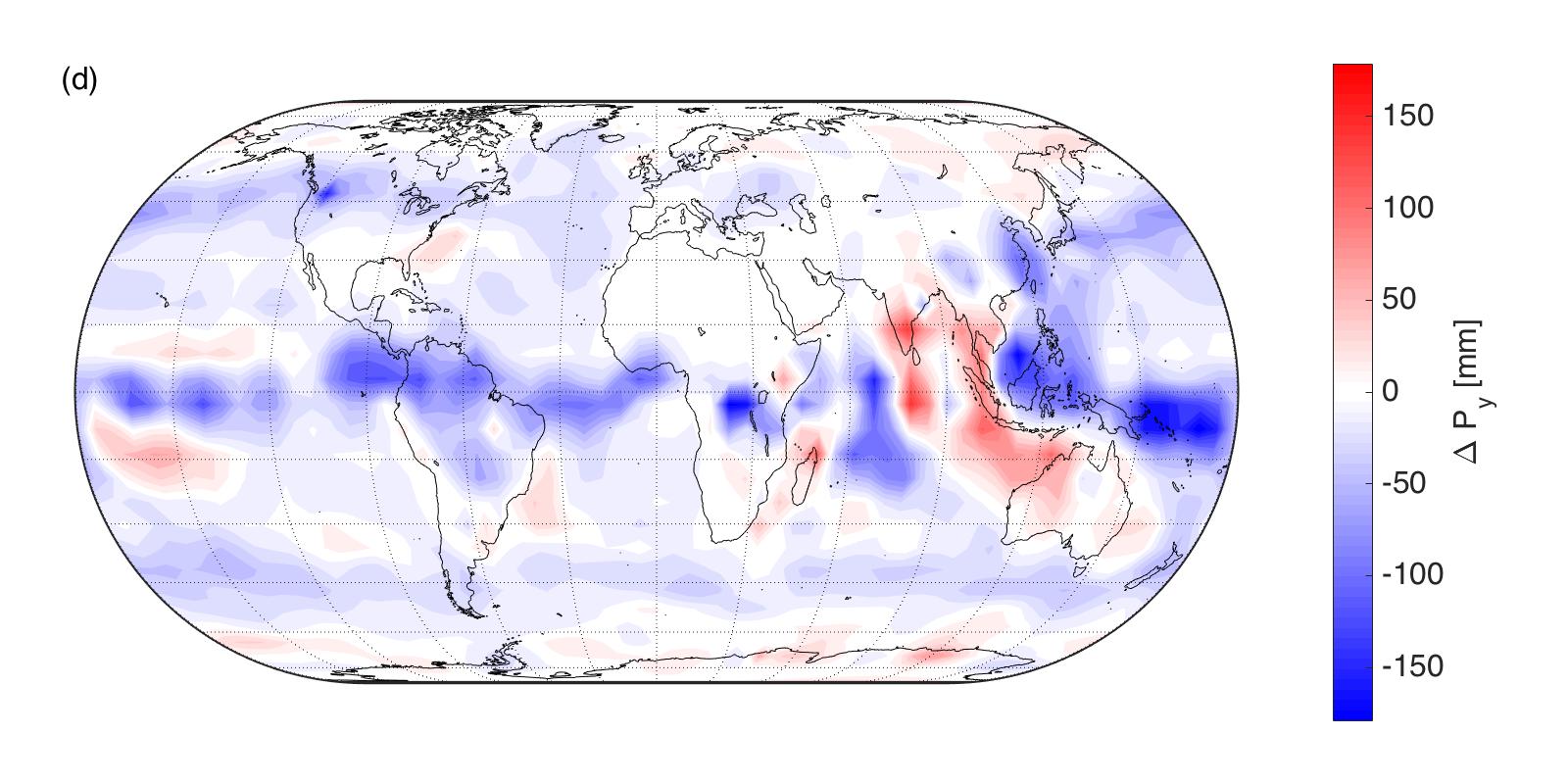} % 03.07.2018. by clim_eng_analysis_05.m
            %\hspace{0.5\linewidth}
        \end{tabular}
        \caption{\label{fig:ecs_spatpattern_cancel} Spatial variation of the stationary climate in terms of (a) the surface air temperature and (b) the annual precipitation in the BR2C experiment, when a change in the global average surface air temperature is canceled. (c){/(d)} The improved linear prediction corresponding to (a){/(b)}. 
        }
    %\end{center}
\end{figure*}

This subsection pertains to our objective (O2). Even if we managed to achieve a perfect cancellation in terms of the global averages, amounting to a success in terms of our objective (O1), it is still important to examine the total response in terms of any other observables regarding which the cancellation is not enforced, to see whether there is any unwanted residual. To this end, we look at the BR2C data. In particular, in Fig.~\ref{fig:ecs_spatpattern_cancel}, we show the spatial variations of the stationary climate in terms of (a) the surface air temperature and (b) the annual precipitation. 
The former  looks to be a result of interpolating between the maps of Figs.~\ref{fig:ecs_spatpattern}(d) and \ref{fig:ecs_spatpattern}(f), and the latter looks to have the same relationship with the maps seen in Figs.~\ref{fig:ecs_spatpattern_precip}(d) and \ref{fig:ecs_spatpattern_precip}(f). 
This implies that the variances with respect to space for BR2C (exact cancellation),  both for temperature and for precipitation, are about the same as those for BR2 (approximate cancellation employing a naive method), and are much larger than the residual total responses in terms of the respective global averages for BR2. The reason for this must be that typically the respective (not necessarily linear) local response characteristics to greenhouse and solar forcing 
are somewhat different. Furthermore, comparing the BR2 and BR2C scenarios, we see that the difference in terms of the climatic surface air temperature could be as much as 2~K, which is about 10\% of the maximum response under the corresponding greenhouse forcing alone. 
This justifies very well  the application of the improved methodology as described in Sec.~\ref{sec:improved_I}. 
{We note that the cooling tropics and warming arctic under global average surface temperature cancellation are in agreement with the findings of other studies using state-of-the-art models, including the first study examining the side-effects of geoengineering by Govindasamy and Caldeira.~\cite{doi:10.1029/1999GL006086}}

{The improved methodology to estimate susceptibilities applies of course to regional averages too. What remains to be seen now is whether linear response theory can predict the residual total responses seen in Figs.~\ref{fig:ecs_spatpattern_cancel} (a) and \ref{fig:ecs_spatpattern_cancel}(b) (O2). The corresponding linear predictions are shown in Figs.~\ref{fig:ecs_spatpattern_cancel}(c) and \ref{fig:ecs_spatpattern_cancel}(d), respectively. These predictions show a dramatic improvement on the first results shown in Figs.~\ref{fig:ecs_spatpattern}(d) and Fig.~\ref{fig:ecs_spatpattern_precip}(d), respectively. Quantitatively, however, the prediction is not perfect. We can quantify this by, for example, the Pearson correlation coefficient $C$ between the {reference} $\Delta\langle\Psi\rangle$ and the linear prediction $\langle\Psi\rangle^{(1)}$, the results for which are shown in Table~\ref{tab:nonlinearity}. (Note that no weighting of the data points with the area represented by gridpoints is done.) This shows that the prediction skill is better for the temperature than the precipitation.

\begin{table}
\caption{{
% Calculations done in clim_eng_analysis_04.m
Measures of overall nonlinearity of the response in terms of the local temperature and precipitation. $C$ is the Pearson correlation coefficient between the {reference} $\Delta\langle\Psi\rangle$ and the linear prediction $\langle\Psi\rangle^{(1)}$, and $\rho$ is defined by Eq.~(\ref{eq:rho_geoeng}). Note that to calculate std($\rho$), values of $\rho$ larger in modulus than 5 are discarded. The last column is devoted to the global averages. %[I SHOULD ALSO CALCULATE THE AVERAGE $\bar{e}_2$] TO COMPLEMENT PEARSON!!! ... perhaps just leave it because it wouldn't be an independent measure, so no surprise there
}}\label{tab:nonlinearity}
  %\begin{center}
    \begin{ruledtabular}\begin{tabular}{lccc}
      %\toprule
             & Pearson correlation coefficient & std($\rho$) & $\rho$ \tabularnewline \hline%\midrule
       $T_s$ & 0.78  & 0.26 & 0.73  \tabularnewline
       $P_y$ & 0.53  & 1.01 & 0.70
      %\bottomrule
    \end{tabular}\end{ruledtabular}
  %\end{center}
\end{table}

Whether the imperfection of the linear prediction is due to nonlinearity---as a small error $E=\Delta\langle\Psi\rangle-\langle\Psi\rangle^{(1)}$ would normally suggest---is not clear, because it is possible that the response $\Delta\langle\Psi\rangle$ is linear but errors in the susceptibility estimates determining $\langle\Psi\rangle^{(1)}$ (or rather its estimator) remain. We should therefore find a way to check linearity without relying on the linear prediction. This can be done in a naive way similarly to what we did  for single-forcing scenarios, evaluating $\rho$ as defined by Eq.~(\ref{eq:simple_nonlin}). However, this time, we have not one but two forcings present. Because of this, it turns out that a check of linearity requires not two but three data points at least. In fact, we are readily endowed by three data set candidates resulting from the BR1, BR2, and BR2C experiments. In each scenario, if the response is linear, the asymptotic climate would be given by an equation like
\begin{equation}\label{eq:lin_pred_geoeng}
 \Delta\langle \Psi_i\rangle = \chi_{\Psi,g}f_{i,g} + \chi_{\Psi,s}f_{i,s},\qquad i=1,2,3,
\end{equation}
where $i=1,2,3$ stand for, say, BR1, BR2, BR2C, in that order. One can express $\chi_{\Psi,s}$ from the equation for $i=3$, substitute into the equations for $i=1,2$, and from these latter express $\chi_{\Psi,g}$. Under linearity, the ratio of these expressions,
\begin{equation}\label{eq:rho_geoeng}
 \rho = \frac{\quad  \dfrac{\Delta\langle \Psi_2\rangle - \Delta\langle \Psi_3\rangle\frac{f_{2,s}}{f_{3,s}}}{f_{2,g} - f_{3,g}\frac{f_{2,s}}{f_{3,s}}}\quad }{\quad  \dfrac{\Delta\langle \Psi_1\rangle - \Delta\langle \Psi_3\rangle\frac{f_{1,s}}{f_{3,s}}}{f_{1,g} - f_{3,g}\frac{f_{1,s}}{f_{3,s}}}\quad },
\end{equation}
would be %of course 
unity, meaning that Eqs.~(\ref{eq:lin_pred_geoeng}) are in fact satisfied. We have evaluated $\rho$ for all gridpoints and display the results in Fig.~\ref{fig:rho}. This suggests that we do have nonlinearity  for both the temperature and  precipitation. However, this conclusion can be called into question by noticing that the three data points could be too close to one another, resulting possibly in an inaccurate estimation of the ratio $\rho$, thereby falsely suggesting nonlinearity. {\bff Instead of evaluating $\rho$, faced with finite data set size to evaluate climatic means, a test statistic would be a proper way of indicating nonlinearity at individual gridpoints.~\cite{GOTTWALD201689} Nevertheless, based on the available data still, o}ne idea to indicate that deviations from unity of both the correlation coefficient $C$ and $\rho$ are due to nonlinearity would be to demonstrate a correlation between the local errors $E$ of the linear prediction and $\rho$. We have checked %clim_eng_analysis_04.m
the scatterplots of these quantities for both the temperature and  precipitation and found no sign of correlations. This, however, does not mean that the response is linear: some unidentified effect could destroy the correlation. Our final idea is that if two situations feature different levels of nonlinearity, even if the two gridpoint-wise quantifiers of nonlinearity, $E$ and $\rho$, have random errors, typically they should indicate in a coordinated way a stronger deviation from linearity in the case when nonlinearity is actually stronger. We propose to capture this typical behavior by the correlation coefficient $C$ on the one hand, and the standard deviation std($\rho$) over the gridpoints on the other. We note that even if linearity is typical, a smaller std($\rho$) would indicate that it is more typical. We have already given the correlation coefficient in Table~\ref{tab:nonlinearity}, where we also display std($\rho$). We do indeed see that  both quantities suggest that the response of precipitation is more nonlinear.

\begin{figure*}  %[t!]
    %\begin{center}
        \begin{tabular}{cc}
            \includegraphics[width=0.5\linewidth]{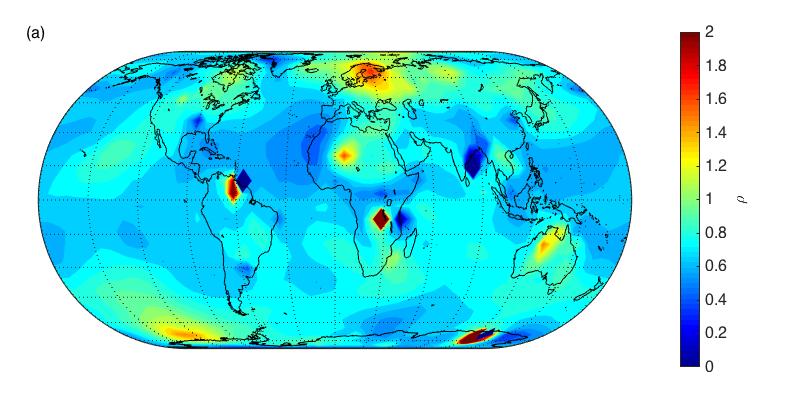} 
            \includegraphics[width=0.5\linewidth]{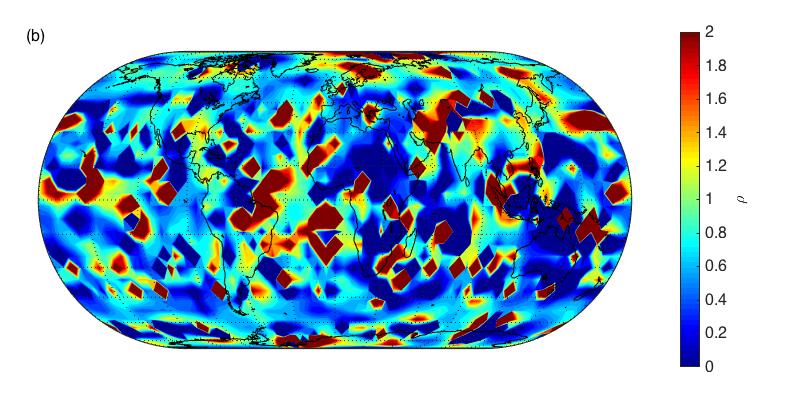} 
        \end{tabular}
        \caption{\label{fig:rho} {Non/linearity of the response in terms of (a) temperature and (b) precipitation, measured by $\rho$ given by the expression (\ref{eq:rho_geoeng}). Any values of $\rho$ lying outside  the range of the color bar are represented by the limiting red and blue colors.}
        }
   % \end{center}
\end{figure*}

In the last column of Table~\ref{tab:nonlinearity}, we show $\rho$ for the global averages $[T_s]$ and $[P_y]$ [not the average of the gridpoint-wise $\rho$'s, but having, e.g., $\Psi=[T_s]$ in Eq.~(\ref{eq:rho_geoeng})]. The steady-state values are estimated by taking the temporal mean of the ensemble means in the last 80 years. These values could be somewhat inaccurate because of the drift seen in Figs.~\ref{fig:resp2ramp}(b) and \ref{fig:resp2ramp_combine_globave_precip}(b) for the BR1 simulation. But considering the possible maximum values of $\rho$ for both $\Psi=[T_s]$ and $[P_y]$, a degree of nonlinearity still seems very likely. The figures indicate that the response of the global average precipitation, unlike the local values/regional averages, is not significantly more nonlinear than the response of temperature under geoengineering. These results caution us about the reliability of linear predictions of side-effects as part of an assessment exercise:
\begin{itemize}
 \item linear predictions of regional responses are less reliable than the global response; 
 \item some quantities can respond more nonlinearly than others.
\end{itemize}
}

\section{Summary and Outlook}\label{sec:summary}

As our objective (O1), we defined and solved an inverse problem for finding a solar forcing that can cancel global warming when applied in conjunction with a given greenhouse forcing. Our novel inverse problem approach is generic in two respects. First, we can allow for different choices of the scalar observable {to keep under control}---different either with respect to the physical quantity, or considering, for example, local variables. Second, we can prescribe an arbitrary time evolution of the chosen observable, assessing perhaps less stringent requirements than total cancellation.\cite{esd-10-453-2019} %The inverse problem constitutes thereby a generic framework for analysing or assessing geoengineering scenarios. 
The inverse problem itself was derived in the framework of linear response theory. 
% VL
We have then further generalized the problem to the case when we want to control $N$ climatic observables by including $N$ auxiliary forcings.

Because of a degree of nonlinearity of the response characteristics of interest, the degree of approximation of the solution of the inverse problem, specifically for the cancellation of global average surface air temperature, depended on the accuracy of determining the linear susceptibilities (or Green's functions) belonging to the different forcings. The inaccurate determination of the linear susceptibilities stems from the fact that for the estimation of the Green's functions we used \textit{finite}-magnitude external system identification forcings (see Appendix~\ref{sec:obtain_greens}), in which case the nonlinearity of the response is already felt, while for the cancellation, i.e., \textit{zero} total response, we would need the linear susceptibilities \textit{exactly}{---assuming the response is linear under combined forcing}. An inaccurately predicted required solar forcing leads to a nonzero residual true total response.
By a simple method, however, also used by Gritsun and Lucarini~\cite{GRITSUN201762} and independently by Liu \textit{et al.},\cite{doi:10.1175/JCLI-D-17-0462.1} here, for determining the susceptibilities, we eliminate even-order nonlinearities from the response in the system identification experiments. The price of this is having to run twice as many ensemble simulations for system identification. In the scenario of doubling CO$_2$ concentration, by this method we were able to achieve a fivefold reduction of the unwanted actual total response arising instead of cancellation. Furthermore, the linear prediction of spatial patterns using the improved \textit{local} susceptibilities improved dramatically. 

{Nevertheless, the prediction is not perfect, and, pursuing our objective (O2), we indicated that the response under combined forcing should be somewhat nonlinear, and the degree of nonlinearity could be typically stronger for some quantities. In particular, we {have seen evidence suggesting} that in PlaSim the response of precipitation is more nonlinear than that of the surface temperature. This casts a shadow over the use of linear response theory for an efficient assessment. Perhaps there would  still be value in this method, since larger-scale quantities are expected to be better predictable. {Results reported by Cao \textit{et al.}~\cite{doi:10.1002/2015JD023901} seem to indicate that in complex models this nonlinearity is not more modest.} {Therefore,} it would be desirable in the future to work out a method of predicting the nonlinear response in geoengineering scenarios.

{We note that the \textit{nonlinearity seen} in the system identification experiments \textit{is not a local property} of the unforced system. The nonlinearity starts to be manifested at a certain level of the response. This is the case only for the positive radiative forcing, but not for the negative one. This is  why---despite the nonlocality of nonlinearity---our improved methodology in Sec.~\ref{sec:improved_I}  still worked so well: the wrong susceptibility estimated from the positive forcing was averaged by the more correct susceptibility estimated from the negative forcing. That is, the error was mitigated.}

Ours is the first such analysis of the linearity of regional response under geoengineering; {previous studies~\cite{acp-16-15789-2016,doi:10.1002/2015JD023901}  only compared the linear prediction and the true model response, and, as we have argued,  such a comparison has to be considered inconclusive regarding the nonlinearity of the response}. There  is a question as to whether our findings {on the predictive power of linear response theory} in PlaSim carry over to state-of-the-art Earth System Models, because these do respond more weakly. The question certainly seems valid, however, since  CMIP5 models also feature nonlinear regional responses under [CO$_2$] forcing alone.\cite{Good2015,doi:10.1029/180GM09} The responses of global average surface air temperature and precipitation have been {reported by MacMartin and Kravitz~\cite{acp-16-15789-2016} to be} approximately linear in some CMIP5 models, seemingly more so than in PlaSim, but weaker forcing than [CO$_2$]-doubling was considered, and the linearity of regional responses was not analyzed in detail. In apparent contradiction, Cao \textit{et al.}~\cite{doi:10.1002/2015JD023901} reported a clear and not insignificant mismatch of the linear prediction and model simulation for local responses. However, only a single state-of-the-art model was considered, the HadCM3L model, and so it was not indicated whether this is typical behavior of state-of-the-art models. Furthermore, {to reiterate, regarding the novel contribution (O2) of our work, we claim that, based on these previous results reported}, it was not correct to conclude that the mismatch was due to nonlinearity.} {\bff As the latest result on the linearity of the local precipitation response, however, Ref.~\onlinecite{doi:10.1029/2019JD031093}, analyzing the Geoengineering Large Ensemble of the CESM1 model, states that ``While a rather extreme geoengineering scenario has been considered, many, but not all, of the precipitation features scale linearly with the offset global warming.'' In particular, see their Figs. 16 (c), (g), (k). Furthermore, we learn from Figs. 9 and 10 of Ref.~\onlinecite{esd-2019-48} that different ESMs feature quite different patterns of local temperature and precipitation response under geoengineering.}

We pointed out also that instead of stepwise system identification forcing, it is better to use a Kronecker delta forcing to achieve a better signal-to-noise ratio. As another gain from using a Kronecker delta forcing, the response would be much more modest in magnitude, and hence it would stay farther away from regimes with more significant contributions of nonlinear terms, and so the linear susceptibilities could be estimated more accurately, even by the naive method [see Appendix~\ref{sec:obtain_greens} and Eq.~(\ref{eq:imp_resp_by_step})].

We note that the presented method for predicting a required solar forcing is based on Green's functions that are determined by externally forcing the system of interest. This is clearly not a method that could be put in practice in the case of the Earth system, and so we are restricted to using climate models. Therefore, this is another reason, beside the unpredictability of twenty-first century greenhouse forcing, why the method is suitable only for scenario analyses. However, it might be possible to estimate the Green's functions without externally forcing the system, just from an observation of unforced fluctuations. The crucial question in this regard is whether the fluctuation--dissipation theorem~\cite{Kubo:1966,Leith:1975} is applicable.

\begin{acknowledgments}
We would like to express our gratitude to an anonymous referee of a previous submission of an earlier version of our manuscript for their very thorough feedback and many suggestions to improve the quality of our work, and also for providing references to the relevant literature. {We would like to acknowledge also Ken Caldeira for drawing our attention to his paper \cite{doi:10.1002/2015JD023901} relevant to our contribution (O2), and the then editor Ben Kravitz for informing us about his publication.\cite{acp-17-2525-2017}} {\bff We would also like to thank an anonymous referee and the editor Georg Gottwald for their helpful suggestions to improve this paper. We also acknowledge an anonymous contractor commissioned by AIP's Author Services who edited this manuscript.} TB would like to thank Jian Lu for inspiring discussions on geoengineering, and sharing their manuscript.\cite{Lu_etal:2019} %In this regard, VL has not seen this manuscript and proposed a generalisation for the multiple-objective-multiple-control-knob situation, eq. (\ref{eq:needed_forcing_freq_gen}), independently, and not necessarily identifying the different control knobs with geoengineering forcings in different spatial locations. 
This work was part of the EU Horizon 2020 project CRESCENDO (under Grant No. 641816); the financial support is gratefully acknowledged. It also received support from the EU Blue Action project (under Grant No. 727852) and from the Institute for Basic Science (IBS), Republic of Korea, under IBS-R028-D1. V.L. acknowledges support from the DFG Sfb/Transregio TRR181 project.
\end{acknowledgments}

\appendix

\section{Computing the response in time and frequency domains}\label{sec:discrete_time}

To be able to carry out (approximate) calculations involving spectral transforms, we need to clarify the formulae and algorithms applicable to \textit{discrete-time} and \textit{finite-size} data. We can approximate the time-continuous forcing $f(t)$ [appearing in Eq.~(\ref{eq:first_order_term_time})] by a \textit{staircase-like} forcing that is defined by a uniform \textit{sampling} of $f(t)$, called a \textit{sample-and-hold} approximation. It can be represented by a discrete sequence $f[n]=f(t=nT)$, $n=\dots,-1,0,1,\dots$, with $T$ being the uniform sampling interval, in which sequence the data points provide the levels of the ``stairs.'' That is, for an actual staircase-like forcing signal, $f(t=(n+\nu)T)=f[n]$ for all $\nu\in[0,1]$, where the noninteger $\nu$ can be viewed as a phase variable---the phase where the sample is taken within the interval where the forcing is constant. For such staircase-like forcings, sample values of the response with the sampling $\Psi[n]=\Psi(t=(n+\nu)T)$ of the time-continuous response at any phase $\nu\in[0,1]$ obey
\begin{equation}\label{eq:first_order_term_time_discrete}
 \langle\hat{\Psi}\rangle^{(1)}[n] = \sum^{\infty}_{k=-\infty}h_{\Psi}[k]f[n-k] = h_{\Psi}[n]\ast f[n],
\end{equation} 
where the discrete-time (DT) impulse response or DT Green's function $h_{\Psi}[n]$ is, clearly, the response $\langle\hat{\Psi}_{\perp}\rangle^{(1)}$ to a Kronecker delta function forcing: $f[n]=\delta[n]=1$ if $n = 0$ and 0 otherwise.\cite{Hespanha:2009} 
%https://en.wikipedia.org/wiki/Linear_time-invariant_theory#Impulse_response_and_convolution_2
%Note that $\langle\hat{\Psi}\rangle^{(1)}[n]\neq\langle\Psi\rangle^{(1)}(t=(n+\nu)T)$ in general, but only in the case of a staircase-like forcing, hence the distinction of the special case in the notation.
Note that we make a distinction in our notation with regard to the special forcing such that we distinguish $\hat{\Psi}$ from $\Psi$; however, for simplicity, we did not subscript $\hat{\Psi}$ by $\nu$, despite its dependence on the phase.
Note also that the DT impulse response function cannot be obtained by a straightforward sampling of the Green’s function; that is, in general, $h_{\Psi}[n]\neq G_{\Psi}^{(1)}[n]=G_{\Psi}^{(1)}(t=(n+\nu)T)$ with the same $\nu$ as $\Psi[n]$ (or $\hat{\Psi}[n]$) is defined with, or with any other fixed $\nu$, for all $n$. 
If the sampling frequency is not adequate regarding some typical time scales of the forcing, then the  calculated discrete response will be also an inadequate approximation. We note further that---unlike the Dirac delta in the continuous-time case---the Kronecker delta \textit{can} be realized for numerical purposes. It is equivalent to applying a step forcing and taking the difference: 
\begin{equation}\label{eq:imp_resp_by_step}
  h_{\Psi}[n]=\Delta\langle\hat{\Psi}_{\ulcorner}\rangle[n]-\Delta\langle\hat{\Psi}_{\ulcorner}\rangle[n-1].
\end{equation} 
This method was used by Lucarini \textit{et al.}~\cite{Lucarini2017} Such external forcings we will refer to as (system) identification forcing.

When faced with the practical situation of having \textit{finite} time series, $f[l]$ and $h_{\Psi}[l]$, $l=0,\dots,L-1$, Eq.~(\ref{eq:circ_conv_theor}) of Appendix~\ref{sec:circ_conv} 
can be used to determine the response $h_{\Psi}\ast f[l]$, $l=0,\dots,L-1$. The usefulness of this formula is due to the existence of efficient algorithms for evaluating the discrete Fourier transform, DFT; we evaluate the DFT using  {\tt fft} from MATLAB. To use the formula, one can \textit{pad} $f[l]$ and $h_{\Psi}[l]$ by $L-1$ zeros in \textit{front} (or rather $L$ zeros\footnote{This results in an odd sequence length, which has an adverse effect on the common fft algorithm performance. Therefore, in actual practice, one can produce time series data of length $L$  some power of 2, and pad by an equal number of zeros.});
%see footnote \ref{foot:efficient_fft} of Appendix~\ref{sec:circ_conv}); 
%%\ref{sec:circ_conv}); 
we will denote these padded sequences by e.g. $\tilde{f}[l]$, $l=0,\dots,2(L-1)$. The first useful ``half'' ($l=0,\dots,L-2$) of the circular convolution ${\rm DFT}^{-1}\{{\rm DFT}\{\tilde{h}_{\Psi}\}{\rm DFT}\{\tilde{f}\}\}$ resulting from Eq.~(\ref{eq:circ_conv_theor}) will then match the linear convolution $h_{\Psi}\ast f[l]$, $l=0,\dots,L-1$. Unlike this calculation in the frequency domain, the calculation in the time domain using Eq.~(\ref{eq:first_order_term_time_discrete}) is straightforward. 

% \begin{aq}
% See the comment regarding this endnote in Appendix~\ref{sec:circ_conv}.
% \end{aq}

\section{Obtaining the Green's function}\label{sec:obtain_greens}

First, to predict the response (to first order), we need to obtain for example the (first-order) Green's function. As Eq.~(\ref{eq:Greens_fun_def}) suggests, it is fully determined by the autonomous system. A direct evaluation of this formula is, however, prone to failure.\cite{Lucarini2017} Second, we note that in practice we can study only a discrete-time version of the system. This suggests that for a direct way of determining the Green's function, instead of Eq.~(\ref{eq:first_order_term_time}), we have to use Eq.~(\ref{eq:first_order_term_time_discrete}) [leading to Eq.~(\ref{eq:imp_resp_by_step})]. It also means that we cannot infer the response of the system just by observing its autonomous dynamics, but we need to force it externally in a suitable way. Third, an ensemble of experiments (appropriately initialized) is needed to obtain the expected value $\langle\hat{\Psi}\rangle$ [with the notation introduced in Appendix~\ref{sec:discrete_time}, first appearing in Eq.~(\ref{eq:first_order_term_time_discrete})]. {This was acknowledged also by MacMartin and Kravitz.~\cite{acp-16-15789-2016}} However, it is feasible  to run only a finite number of experiments, so we obtain an approximation of $\langle\hat{\Psi}\rangle$, where the error is some correlated noise process. This correlation can be made negligible by using a suitably infrequent sampling, allowed by, say, the application of a slow forcing. We use the same data as Lucarini \textit{et al.},~\cite{Lucarini2017} which consist of some ensembles of 200 members, and we have produced new data belonging to new forcing scenarios, as described in Sec.~\ref{sec:forcing_scenarios}, that consist of ensembles of 20 members.

\begin{figure*}  %[t!]
    %\begin{center}
        \begin{tabular}{cc}
            \includegraphics[width=0.5\linewidth]{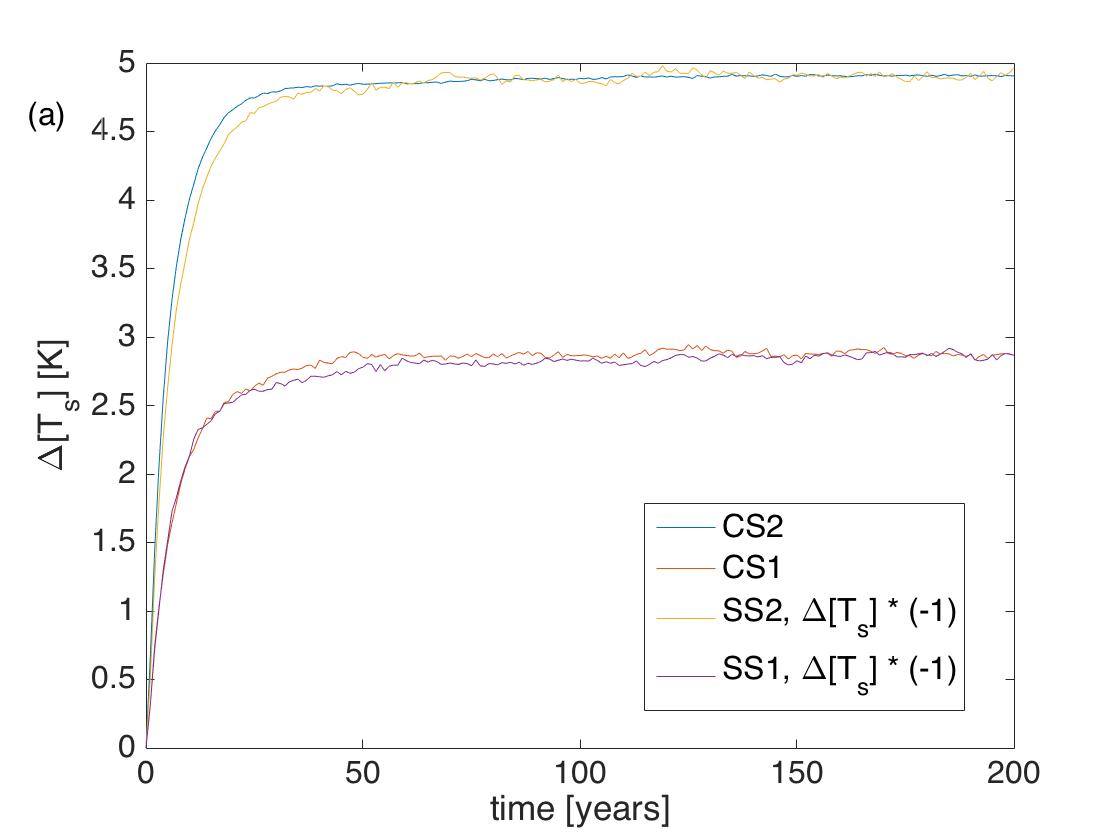} 
            \includegraphics[width=0.5\linewidth]{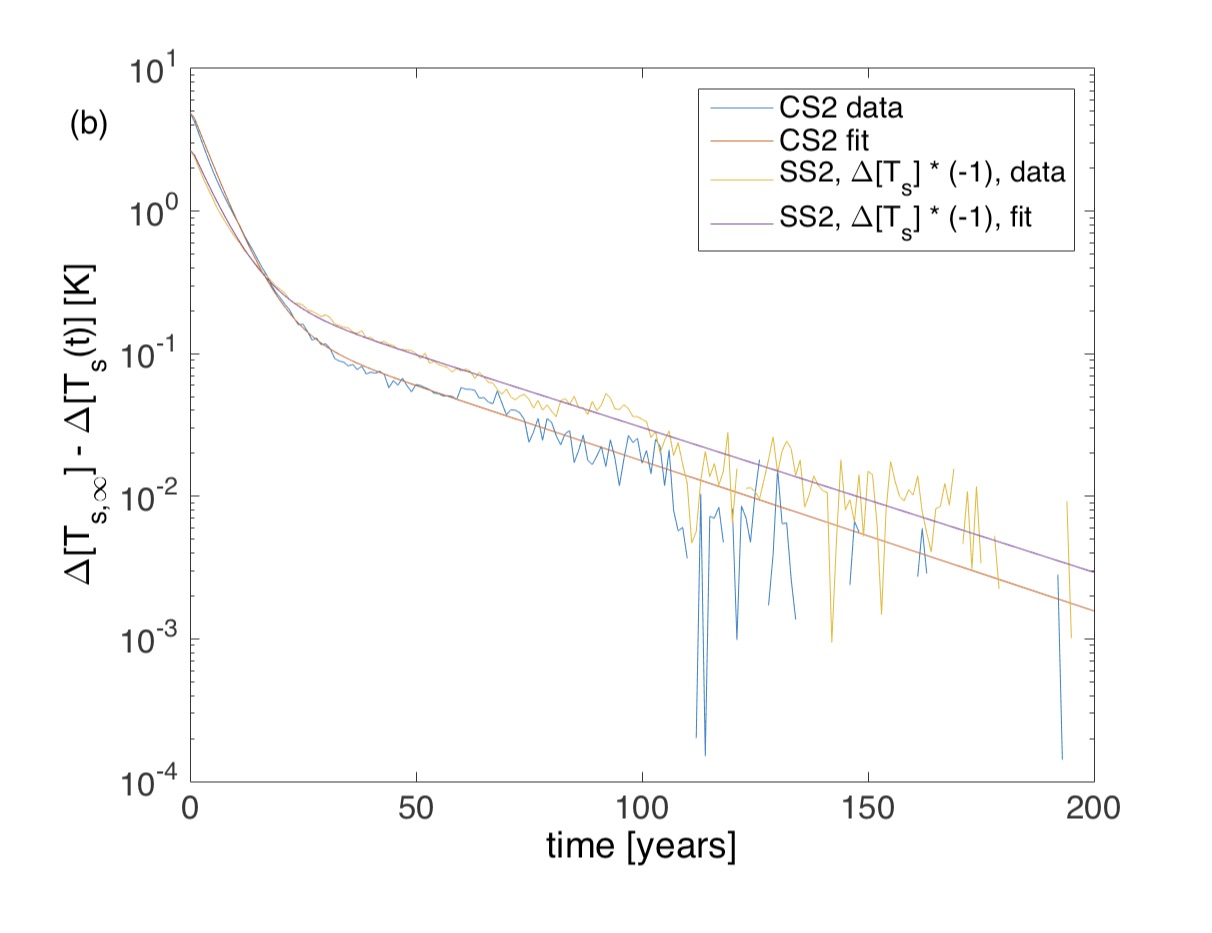}
        \end{tabular}
        \caption{\label{fig:resp2step} Simulated response to step forcings. The chosen observable is the global average surface air temperature $[T_s]$. The identification forcing scenarios are those of CS2, CS1, SS2, and SS1 from Table~\ref{tab:forcing_scenarios}. 
        After a subtraction of the limit value and displaying the response on linear--log scales in  (b), it is revealed that the high-dimensional system behaves very much like a noise-driven linear two-box model, also called a vector autoregressive (VAR) model, in view of the considered global scale variable, {as also recognized by MacMynowski (MacMartin) \textit{et al.}~\cite{doi:10.1029/2011GL048623} and Caldeira and Myhrvold.\cite{1748-9326-8-3-034039}} The two time scales of the VAR models fitted to the CS2 and SS2 data are about 5 and 40 years. {The second time scale is in disagreement with MacMynowski (MacMartin) \textit{et al.},\cite{doi:10.1029/2011GL048623} and it is not clear whether a more complex model is more reliable in this respect.} Note: the angle brackets denoting ensemble average are dropped from diagram annotations throughout this paper.}
    %\end{center}
\end{figure*}

As already spelled out in Appendix~\ref{sec:discrete_time}, two identification forcing types are particularly suitable to determine the Green's function: one is a step forcing, and the other is the Kronecker delta. When a random statistical error is present due to the finite ensemble size, represented say by a Gaussian random variable $\xi$, it is actually \textit{better to use a Kronecker delta} forcing for the following reason. Using the step forcing instead, one needs to take the difference of consecutive values---what is sometimes called ``differencing''---of the response sequence (\ref{eq:imp_resp_by_step}). This way, at any time, the variance of the error is that of the \textit{difference} of two random variables, $\xi_1$ and $\xi_2$, both distributed identically to the original random variable $\xi$. For Gaussian variables, it is straightforward to show that ${\rm Var}[\xi_1-\xi_2]= 2{\rm Var}[\xi]$. Note that we assume that $\xi$ is the same random variable to a good approximation under the delta and step forcings. {Despite  the advantage of the Kronecker delta forcing}, we apply a step forcing also in our new experiments, so that we are able to make use of data produced for the study by Lucarini \textit{et al.}~\cite{Lucarini2017} in a consistent manner. Examples of the response to step forcings are displayed in Fig.~\ref{fig:resp2step}. {The similarity of the responses to greenhouse and solar forcings here, and so the Green's functions, is consistent with the findings of others~\cite{doi:10.1175/JCLI-D-14-00214.1,1748-9326-8-3-034039,doi:10.1029/2011GL048623} {and the design of the G2 GeoMIP experiment.\cite{doi:10.1002/asl.316}}}% doi:10.1029/2005JD005776 deals only with spatial patterns but not dynamics

It is important to appreciate the following trade-off. For a better signal-to-noise ratio (SNR), one can apply a more powerful identification forcing. However, in the presence of nonlinearities, the more powerful the forcing signal, the larger is the error in estimating the Green's function \textit{belonging to the base state} $\langle\Psi\rangle_0$ (even without noise). 
{MacMartin and Kravitz~\cite{acp-16-15789-2016} applied a [CO$_2$]-quadrupling (and this is a standard forcing level for geoengineering studies~\cite{1748-9326-9-1-014001,doi:10.1002/asl.316}); however, they did not determine the solar forcing for cancellation  via Green's functions (Appendix~\ref{sec:inverse_problem}), and they checked the linearity of the response only up to a forcing level lower than [CO$_2$]-doubling. Their motivation for applying the high forcing level seems to have been only to be able to determine the Green's function with a better SNR given that no ensemble data are available from the GeoMIP experiments.}

We make here two more comments about the issue with noise. First, instead of instantaneous samples of the observable $\Psi$ and the corresponding Green's function, we will consider, like Lucarini \textit{et al.},\cite{Lucarini2017} \textit{annual averages}, $\bar{\Psi}[n]= \int_0^1\diff\nu \Psi((n+\nu)T)$. This is sensible given the slow rate of change that the applied forcing represents, and it also greatly reduces the noise level. In this regard, we point out that annual averages too obey Eq.~(\ref{eq:first_order_term_time_discrete}) \textit{exactly} if the forcing is constant over a year, because the order of summations can be interchanged, whereby a well-defined DT Green's function belonging to the annual average emerges. We  will use only annually constant staircase-like forcings in our experiments (Sec.~\ref{sec:forcing_scenarios}), so that it will be clear that a linear prediction of the response has an error not because Eq.~(\ref{eq:first_order_term_time_discrete}) does not apply exactly, but because of the missing higher-order perturbative terms appearing in the expression  (\ref{eq:perturbative}). Second, the said enhancement of noise by differencing in Eq.~(\ref{eq:imp_resp_by_step}) cannot be overcome by working in the frequency domain. The Green's function, via the frequency domain and applying Eq.~(\ref{eq:first_order_term_freq_discrete}) of Appendix~\ref{sec:circ_conv}, 
is expressed as $h_{\Psi}={\rm DTFT}^{-1}\{{\rm DTFT}\{\Delta\langle\hat{\Psi}_{\ulcorner}\rangle\}/{\rm DTFT}\{f_{\ulcorner}\}\}$, where $1/{\rm DTFT}\{f_{\ulcorner}\}=1-e^{-i\omega}$. The latter is the very factor arising in the DTFT of a differenced sequence. As far as we are aware, the only way to avoid the differencing and thereby reduce the noise is by using a Kronecker delta identification forcing as argued above.\footnote{There exist filtering techniques, but they introduce some assumptions either about the functional form of the Green's function (parametric techniques) or about the goodness of fit (nonparametric techniques) of their estimate to, say, one of the described straightforward (noisy) estimates (such as a minimal root-mean-square error). One can use, e.g., {\tt impulseest} from MATLAB.}

{We also note that the application of a stepwise identification forcing implies a specific frequency dependence of the variance of the estimator of the susceptibility. See Ref.~\onlinecite{acp-17-2525-2017} for other system identification techniques that imply different such frequency dependences.}

\section{The inverse problem}\label{sec:inverse_problem}
 
\begin{figure*}  %[t!]
    %\begin{center}
        \begin{tabular}{cc}
            \includegraphics[width=0.5\linewidth]{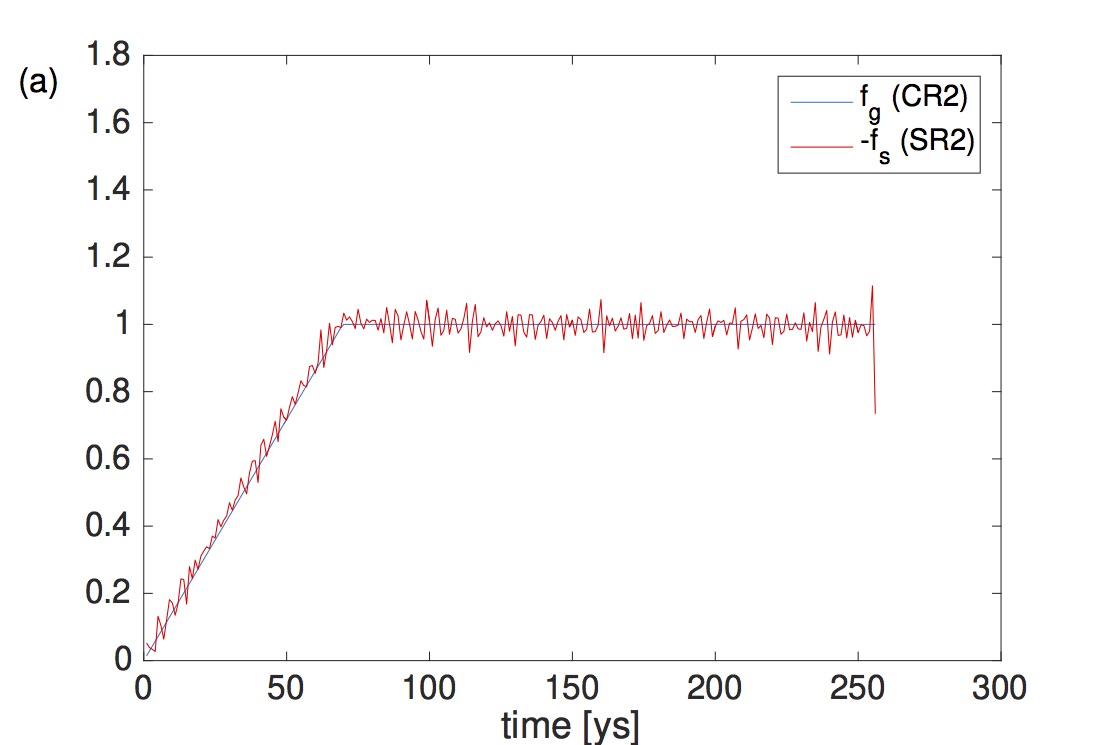} 
            \includegraphics[width=0.5\linewidth]{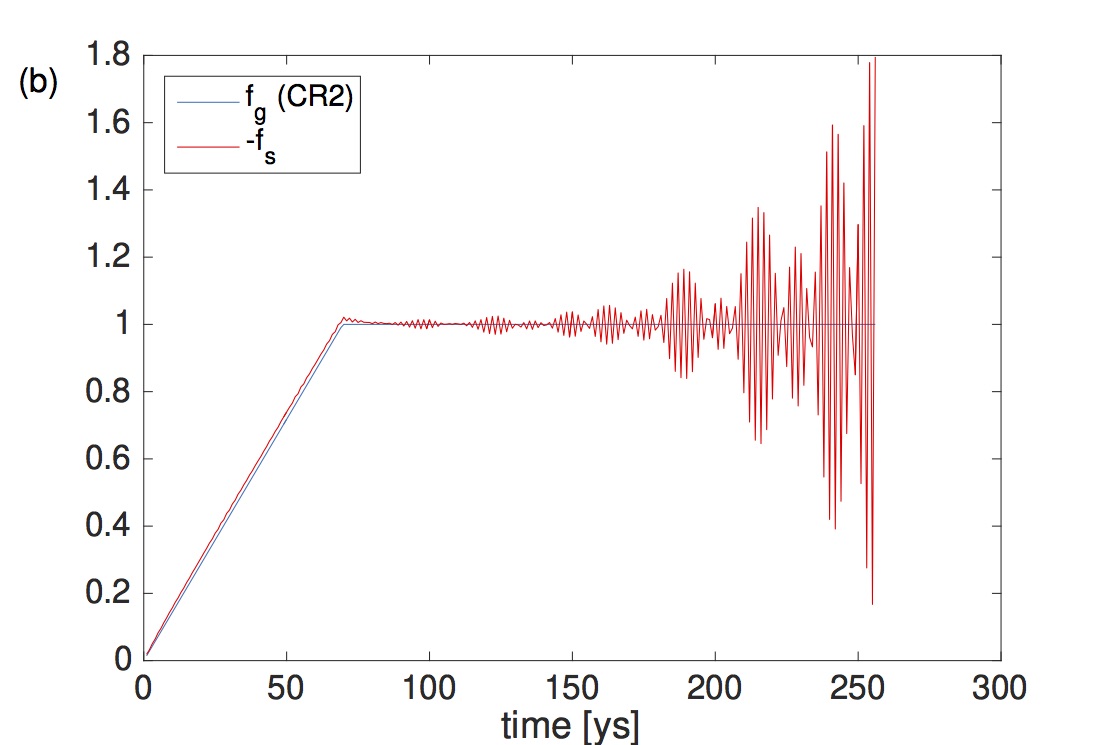}
        \end{tabular}
        \caption{\label{fig:f_sol_req_2} Imposed $[{\rm CO}_2]$ or greenhouse forcing and required solar forcing that cancels out global average surface air temperature change. Values of the time series are \textit{normalized} for the purpose of display, so that they have a unit plateau level. The required solar forcing is determined in both the frequency domain (a) and the time domain (b). We indicate in the keys the data sets from Table~\ref{tab:forcing_scenarios} to which the forcings belong. We note that in either case, we \textit{neglected the iteration}, skipping stages 1 and 2 and setting $\Delta\langle\check{\Psi}_{\varSigma}\rangle[l]=\Delta\langle\check{[T_s]}_{\varSigma}\rangle[l]=0$ for \textit{all} $l$ straightaway in stage 3, the validity of which is suggested by the very similar Green's functions $h_{[T_s],{g}}$ and $h_{[T_s],s}$, as indicated by Fig.~\ref{fig:resp2step}. Correspondingly, the required $f_s$ is very similar to the given $f_{g}$. A small gap between the red and blue ramps that can be resolved only with a smooth estimate [i.e., in  (b) but not in (a)] informs us that the system responds slightly faster to the greenhouse forcing. This characteristic  was already suggested by Fig.~\ref{fig:resp2step}(b) and the exact results of the parameter estimation by fitting. Results presented in Sec.~\ref{sec:improved_II} suggest that it is {likely} to do with nonlinearity, which makes the responses to negative and positive anomalies ``asymmetric,'' {resulting also in different spatial patterns, while the time scales associated with different locales are quite varied (results not shown)}.
        }
    %\end{center}
\end{figure*}

When different forcings act at the same time, their first-order contributions to the response---as discussed in Sec.~\ref{sec:elements}---can be \textit{superimposed}. Hence, when we desire a certain {total response} $\Delta\langle\Psi_{\varSigma}\rangle(t)$ to a \textit{combined forcing} when all forcings are given but one, there is a \textit{unique} form of that one required forcing to fulfill our desire. In terms of the geoengineering problem of  interest (Sec.~\ref{sec:geoeng_problem}), the required solar forcing $f_s$  to achieve a total response $\Delta\langle\Psi_{\varSigma}\rangle$ 
given a greenhouse forcing $f_{g}$ can be \textit{expressed}, to a first-order approximation, in the frequency domain as stated in Eq.~(\ref{eq:needed_forcing_freq}). With the most obvious choice of \textit{cancellation}, $\Delta\langle\Psi_{\varSigma}\rangle=0$, Eq.~(\ref{eq:needed_forcing_freq}) simplifies to 
\begin{equation}\label{eq:needed_forcing_freq_cancel}
 f_s(\omega) \approx -\frac{\chi_{\Psi,{g}}(\omega)}{\chi_{\Psi,s}(\omega)}f_{g}(\omega).
\end{equation} 
However, in practice, when finite time series are available, the simplification is not so trivial. As described at the end of Appendix~\ref{sec:discrete_time}, in place of the three FTs in Eq.~(\ref{eq:needed_forcing_freq}), we have to evaluate DFT$\{\tilde{f}_{g}\}$, DFT$\{\tilde{h}_{\Psi,s}\}$ and DFT$\{\tilde{h}_{\Psi,{g}}\}$. \textit{Furthermore}, the DFT in place of $\Delta\langle\Psi_{\varSigma}\rangle(\omega)$ is that of a sequence $\Delta\langle\check{\Psi}_{\varSigma}\rangle[l]$, only the first useful ``half'' ($l=0,\dots,L-2$) of which is zero, as dictated by our requirements, but its second half ($l=L-1,\dots,2(L-1)$) has nonzero values in general. That is, what we have is
\begin{align}\label{eq:needed_forcing_freq_DFT}
  \tilde{f}_s ={\rm DFT}^{-1}\{(&{\rm DFT}\{\Delta\langle\check{\Psi}_{\varSigma}\rangle\}\nonumber\\&-{\rm DFT}\{\tilde{h}_{\Psi,{g}}\}{\rm DFT}\{\tilde{f}_{{g}}\})/{\rm DFT}\{\tilde{h}_{\Psi,s}\}\},
\end{align} 
It can be shown that the said nonzero values characterize the total response to combined \textit{step} forcings (to do with the ``gap'' mentioned in the caption of Fig.~\ref{fig:f_sol_req_2}), but also depend to a certain extent on the particular finite $f_{g}[l]$ presented. The reason for this is that the Green's function is given only up to a finite time, which becomes clear upon inspection of the workings of the convolution of finite time series. Simply by rearranging Eq.~(\ref{eq:needed_forcing_freq_DFT}), the said nonzero values are given 
by 
%
% \begin{equation}\label{eq:tot_resp_iter}
%   \Delta\langle\check{\Psi}_{\varSigma}\rangle={\rm DFT}^{-1}\{{\rm DFT}\{\tilde{h}_{\Psi,{g}}\}{\rm DFT}\{\tilde{f}_{{g}}\}+{\rm DFT}\{\tilde{h}_{\Psi,s}\}{\rm DFT}\{\tilde{f}_s\}\},
% \end{equation} 
\begin{align}\label{eq:tot_resp_iter}
  \Delta\langle\check{\Psi}_{\varSigma}\rangle={\rm DFT}^{-1}\{&{\rm DFT}\{\tilde{h}_{\Psi,{g}}\}{\rm DFT}\{\tilde{f}_{{g}}\}\nonumber\\&+{\rm DFT}\{\tilde{h}_{\Psi,s}\}{\rm DFT}\{\tilde{f}_s\}\},
\end{align} 
where, however, $\tilde{f}_s$ is not known,  being the sought-for object. The idea is that we can look for $\tilde{f}_s$ by an \textit{iterative} procedure, which is initialized, say, by $\tilde{f}_s=\tilde{f}_{g}$. Note that if $h_{\Psi,{g}}$ and $h_{\Psi,s}$ are not dissimilar, then nor are $f_{g}$ and $f_s$; that is, the initial value is not far from the solution, which gives hope that it is within the basin of attraction of the solution. In each iterate, we do the following:
\begin{enumerate}
 \item Evaluate Eq.~(\ref{eq:tot_resp_iter}) using the current estimate of $\tilde{f}_s$, but instead of simply substituting this value, first we replace any nonzeros in its first half  by zeros.
 \item In the resulting $\Delta\langle\check{\Psi}_{\varSigma}\rangle[l]$, we replace any nonzeros in its \textit{first} half by zeros in order to have it in the right form.
 \item We then get a new estimate for $\tilde{f}_s$ using Eq.~(\ref{eq:needed_forcing_freq_DFT}). 
\end{enumerate}
Ideally, the first half of the $\tilde{f}_s$ estimates in stage 3 converge to zero, and the second half to some nontrivial form that is the solution. In our experience (results not shown), this is the case for systems with fairly simple and smoothly varying Green's functions. However, when the same Green's functions are corrupted by noise (which arises in practice from the finiteness of the ensemble size; see Appendix~\ref{sec:obtain_greens}), our experience is that the procedure does not necessarily converge, but iterates of $\tilde{f}_s$ can develop increasingly large 
oscillatory features. It is possible to achieve convergence for some smaller but nonzero noise level. However, even then, the limit function retains small oscillatory features over the full length of $\tilde{f}_s$. % See RESULT on 19.03.17. in resp_theo_exercise_06.m 

We emphasize that the iterative procedure was needed because we could not predict the second non-useful half of $\Delta\langle\check{\Psi}_{\varSigma}\rangle[l]$, since we do not have the Green's functions in full but only with a cutoff in time. This means that by running longer and longer \textit{ensemble} simulations, by which we can determine the Green's functions further and further in time, the solution can be approximated by a \textit{non}iterative procedure better and better. This is a numerically more expensive solution. 

As an alternative, working in the time domain, the inverse problem leads to the performance of a \textit{deconvolution}:
\begin{equation}\label{eq:needed_forcing_time}
 f_s = (\Delta\langle\check{\Psi}_{\varSigma}\rangle - h_{\Psi,{g}}\ast f_{g})\ast^{-1}h_{\Psi,s}.
\end{equation} 
Note that we have written $\Delta\langle\check{\Psi}_{\varSigma}\rangle[l]$ in the above, which should  correspond exactly to the appropriately defined circular convolution in Eq.~(\ref{eq:tot_resp_iter}) for $l=0,\dots,2(L-1)$. It is straightforward to show that in the time domain too, $f_s[l]$, $l=0,\dots,L-1$, is obtained \textit{iteratively} in three stages in a similar manner as outlined above in the frequency domain (except that in stage 1, there is no need to replace any values by zeros, only in stage 2). One can use  {\tt deconv} in MATLAB to perform the deconvolution. We find in simple examples studied (results not shown) that without noise the procedure in the time domain leads to the very same solution as the procedure in the frequency domain. However, this is not the case with additive noise, which means that the deconvolution/inverse problem is \textit{ill-posed} in this case.
% I don't have a proper citation just these two web pages:
% https://dsp.stackexchange.com/questions/2969/deconvolution-of-1d-signals/38325#38325
% http://epsc.wustl.edu/seismology/michael/CIG/workshop06/Aster/Deconvolution_Demo/deconvolution_demo.m
Nevertheless, the weaker the noise, the closer the outcome is to the true solution, either in the time or the frequency domain. We find that in the time domain, the procedure always converges to some solution; however, with increasing noise strength, this solution features oscillations of increasing amplitude as time advances. Nevertheless, for a certain noise strength when the frequency-domain procedure also converges, we find that the solution in the time domain is smoother and so closer to the true solution \textit{earlier} in time. This is also what we find considering the PlaSim data, as shown in Fig.~\ref{fig:f_sol_req_2}. We conclude, therefore, that \textit{it is preferable to work in the time domain} using Eq.~(\ref{eq:needed_forcing_time}) to produce numerical results. % 20.11.19. Since it's now an appendix, i include what was earlier a footnote as a last paragraph here.
Nevertheless, we will carry out our calculations in the frequency domain, using for example the forcing signals shown in Fig.~\ref{fig:f_sol_req_2}(a), in order to make the point that even a rough forcing signal convolved with a rough Green's function produces a not so rough response, as we see in Sec.~\ref{sec:res_global}.
% I sent the forcing signals f_sol_req_1_02.txt and f_sol_req_2_01.txt to Frank on 25.07.2016 for simulations SR1 and SR2. which was obviously obtained by f_sol_req_01.m in freq domain. I implmented the noniterated! calculation in time domain in f_sol_req_02.m only later on 21.09.16. but never saved a txt file to be sent to Frank.
% 02.07.18.

We emphasize that, in our case, multiple iterations are not needed, because $h_{[T_s],{g}}$ and $h_{[T_s],s}$ are very similar, as indicated by Fig.~\ref{fig:resp2step}. The straightforward application of Eq.~(\ref{eq:needed_forcing_freq_DFT}) is sufficient, substituting an all-zero sequence for $\Delta\langle\check{\Psi}_{\varSigma}\rangle$. Ours is obviously a special case, however, and the generic iterative procedure might be needed for other geoengineering scenarios of practical relevance.

{We note that an anonymous referee of a previous submission of this paper had suggested that, in the time domain, a simpler alternative way of obtaining the solution by a time-marching procedure should exist (not relying on deconvolution). Indeed, one can break down the convolution sum (\ref{eq:first_order_term_time_discrete}) as $\langle\hat{\Psi}\rangle^{(1)}[n] = \sum^{n}_{k=2}h_{\Psi,s}[k]f_s[n-k] + h_{\Psi,s}[1]f_s[n-1]$, which can be expressed for $f_s[n-1]$ and consider that $\langle\hat{\Psi}\rangle^{(1)}[n] = \sum^{n}_{k=1}h_{\Psi,g}[k]f_g[n-k]$ is given for all $n$. Suppose $f_g[0]=0$; then the procedure for finding $f_s[n-1>0]$ can be initialized by $f_s[0]=0$ for $n=1$. We have checked (results not shown) 
% resp_theo_excercise_11.m
that it gives exactly the same result as our iterative procedure, reproducing the time series pattern due to a particular noise realization in a simple example system.}

\section{The circular convolution theorem and its application}\label{sec:circ_conv}

Taking the discrete-time Fourier transform (DTFT) of Eq.~(\ref{eq:first_order_term_time_discrete}), we have, via the convolution theorem for discrete sequences~\cite{Katznelson:1976}, a formally analogous version of Eq.~(\ref{eq:first_order_term_freq}), with the individual Fourier transforms approximated by Fourier series:
\begin{equation}\label{eq:first_order_term_freq_discrete}
 \langle\hat{\Psi}\rangle^{(1)}_{2\pi}(\omega) = \hat{\chi}_{\Psi,2\pi}(\omega)f_{2\pi}(\omega),
\end{equation} 
where, for example, $f_{2\pi}(\omega)={\rm DTFT}\{Tf[n]\}=\sum_{n=-\infty}^{\infty}Tf[n]e^{-i\omega n}$ and $f[n]={\rm DTFT}^{-1}\{T^{-1}f_{2\pi}(\omega)\}=\frac{1}{2\pi T}\int_{-\pi}^{\pi}d\omega f_{2\pi}(\omega)e^{i\omega n}$ with a normalized nondimensional angular frequency $\omega$. Featuring instead the dimensional frequency $f$ measured in hertz (1~ Hz = 1~s$^{-1}$), the forward and inverse transform pairs are symmetrical:
$f_{1/T}(f)=f_{2\pi}(2\pi fT)=\sum_{n=-\infty}^{\infty}Tf[n]e^{-i2\pi fT n}$ and $f[n]=T\int_{1/T}df f_{1/T}(f)e^{i2\pi fTn}$.
The DTFT, a continuous function of the frequency $f$, is often sampled at $f=k/(NT)$, $k=0,\dots,N-1$:
%
% \begin{equation}\label{eq:sampled_DTFT}
%  f_{1/T}(k/(NT)) = T\sum_{n=-\infty}^{\infty}f[n]e^{-i2\pi kn/N} = T\sum_{n=n_0}^{n_0+N}f_N[n]e^{-i2\pi kn/N}=T\times{\rm DFT}\{f_N[n=n_0,\dots,n_0+N]\},
% \end{equation} 
\begin{align}\label{eq:sampled_DTFT}
 f_{1/T}(k/(NT)) &= T\sum_{n=-\infty}^{\infty}f[n]e^{-i2\pi kn/N}  \nonumber\\ &=T\sum_{n=n_0}^{n_0+N}f_N[n]e^{-i2\pi kn/N}\nonumber\\&=T\times{\rm DFT}\{f_N[n=n_0,\dots,n_0+N]\},
\end{align}
for any $n_0$, which yields the discrete Fourier transform (DFT) of the \textit{finite} sequence $f_N[n]$, $n=n_0,\dots,n_0+N$, where the full infinite sequence $f_N[n]$, $n\in\mathbb{R}$, turns out to be $N$-periodic, since it has to be, for the equivalence of the two sums in Eq.~(\ref{eq:sampled_DTFT}), in the so-called periodic summation form 
\begin{equation}\label{eq:f_N}
 f_N[n] = \sum_{m=-\infty}^{\infty}f[n-mN].
\end{equation} 
Therefore, when $f[n]$ is actually $N$-periodic, its DTFT is nonzero only at $f=k/(NT)$, $k\in\mathbb{R}$, and also periodic, such that the DFT of a single cycle of $f[n]$ is able to represent its DTFT. For such periodic sequences, to be denoted distinctively using a subscript as $f_N[n]$, it can be proved~\cite{Katznelson:1976} that 
%https://en.wikipedia.org/wiki/Discrete-time_Fourier_transform#Convolution
%https://en.wikipedia.org/wiki/Convolution_theorem#Functions_of_discrete_variable_sequences
%
% \begin{equation}\label{eq:circ_conv_eq}
%  y \ast f_N = {\rm DTFT}^{-1}\{{\rm DTFT}\{y\}{\rm DTFT}\{f_N\}\} = {\rm DFT}^{-1}\{{\rm DFT}\{y_N\}{\rm DFT}\{f_N\}\},
% \end{equation} 
\begin{align}\label{eq:circ_conv_eq}
 y \ast f_N  &={\rm DTFT}^{-1}\{{\rm DTFT}\{y\}{\rm DTFT}\{f_N\}\}  \nonumber\\ &={\rm DFT}^{-1}\{{\rm DFT}\{y_N\}{\rm DFT}\{f_N\}\},
\end{align} 
for any nonperiodic sequence $y[n]$. Note that $y \ast f_N$ is referred to as the \textit{circular convolution} of the sequences $y[n]$ and $f[n]$. When the sequences $y[n]$ and $f[n]$  are of finite length, $n=0,\dots,N-1$ with any $N\geq1$, so that, for example, $f_N[n]=f[{\rm mod}(n,N)]$, their circular convolution can be shown to be~\cite{Katznelson:1976} (see also \url{https://uk.mathworks.com/help/signal/ug/linear-and-circular-convolution.html}) % ~\cite{Katznelson:1976,circ_conv_theor} 
%
% \begin{equation}\label{eq:circ_conv_theor}
%  (y \ast f_N)[n=0,\dots,N-1] = \sum_{k=0}^{N-1} y[k]f_N[n-k] = {\rm DFT}^{-1}\{{\rm DFT}\{y\}{\rm DFT}\{f\}\},
% \end{equation} 
\begin{align}\label{eq:circ_conv_theor}
 (y \ast f_N)[n=0,\dots,N-1] &= \sum_{k=0}^{N-1} y[k]f_N[n-k]  \nonumber\\ &={\rm DFT}^{-1}\{{\rm DFT}\{y\}{\rm DFT}\{f\}\}.
\end{align}
This equality is called the \textit{circular convolution theorem}. It follows that when  $y[n]=0$ and $f[n]=0$ for $n=0,\dots,N_f-1$ and $N_y-1$, respectively, then $(y \ast f_N)[{\rm mod}(n-1,N)]=(y \ast f)[n]$ for $n=N,\dots,N+{\rm min}(N_f+N_y,N-1)$.
%https://en.wikipedia.org/wiki/Convolution#Fast_convolution_algorithms
%https://en.wikipedia.org/wiki/Discrete_Fourier_transform#Circular_convolution_theorem_and_cross-correlation_theorem
Furthermore, $(y \ast f)[n]$, $n=1+N_f+N_y,\dots,N+1+N_y$, is the segment that represents the part of the linear convolution that can be considered useful in the sense that it coincides with the occurrence of the finite values of $f$ in a finite time interval of length $N-N_f$. Therefore, the circular convolution $(y \ast f_N)[n]$ captures the useful part of the linear convolution over $n=\max(1+N_f+N_y,N),\dots,N+\min(1+N_y,N_f+N_y,N-1)$.

Therefore, when faced with the practical situation of having \textit{finite} time series, $f[l]$ and $h_{\Psi}[l]$, $l=0,\dots,L-1$, Eq.~(\ref{eq:circ_conv_theor}) can be used to determine the response $h_{\Psi}\ast f[l]$, $l=0,\dots,L-1$ (whose usefulness  comes from an efficient algorithm for evaluating the DFT, called the fast Fourier transform algorithm, FFT). In particular, if the two sequences are to be \textit{padded} in front by a number $N_f=N_h=N_0$ of zeros equally (so that the circular convolution (\ref{eq:circ_conv_theor}) is well defined), then the reconstructed length of the linear convolution $h_{\Psi}\ast f$ (the response of a causal system coinciding with the forcing) is $1+N_0-\max(N_0-L+1,0)$. This is a linear function of $N_0$ saturating at $N_0=L-1$ reaching the full length $L$. Therefore, for simplicity one can pad by $N_0=L-1$ zeros,\cite{Note5} 
%\footnote{This results in an odd sequence length, which has an adverse effect on the common fft algorithm performance. Therefore, in actual practice, one can produce time series data of length $L$ some power of 2, and pad by an equal number of zeros.}, 
and we will \textit{denote these padded sequences} by, for example, $\tilde{f}[l]$, $l=0,\dots,2(L-1)$. Note that padding with fewer or no zeros results in a circular convolution that better approximates either the useful or the not useful part of the linear convolution, which approximation is  better the more zeros are used. In the extreme case of no padding, very little of the useful part can be approximated well. %[***IF THIS GETS TO THE APPENDIX, I SHOULD INCLUDE 3 FIGURES WITH $N_0=0,L/2,L-1$, BOTH WHAT IS FIG. 5 by {\tt resp\_theo\_excercise\_05.m} NOW IN MATLAB WITH 2 CYCLES AND A CORRESPONDING DIFFERENCE.] 
The key to the applicability of Eq.~(\ref{eq:circ_conv_eq}) is that it does not matter how the forcing $f[n]$---and with it the response $\langle\hat{\Psi}\rangle[n]$---continue after our experiment, and so they can be thought of as periodic. 

% \begin{aq}
% Note that the use of \textbackslash label  and \textbackslash ref commands to cross-reference this endnote does not work: as you can see from the original manuscript, it produced a reference to ``footnote E''! There does not seem to be a straightforward way of repeatedly citing an endnote in Revtex, other than perhaps producing it in the form of an entry in the .bib file and citing it in the same way as an ordinary reference. Instead, I've treated  this note in the usual way with \textbackslash footnote in Appendix~\ref{sec:discrete_time} and then used the label generated by BibTeX in the .bbl file (``Note5'') to cross-reference it here using \textbackslash cite. Of course, if any  changes are made to  earlier endnotes, this label will have to be amended. 
% \end{aq}

\section{Zonal average surface temperature}\label{sec:res_zonal}

\begin{figure*}  %[t!]
    %\begin{center}
        \begin{tabular}{cc}
            \includegraphics[width=0.5\linewidth]{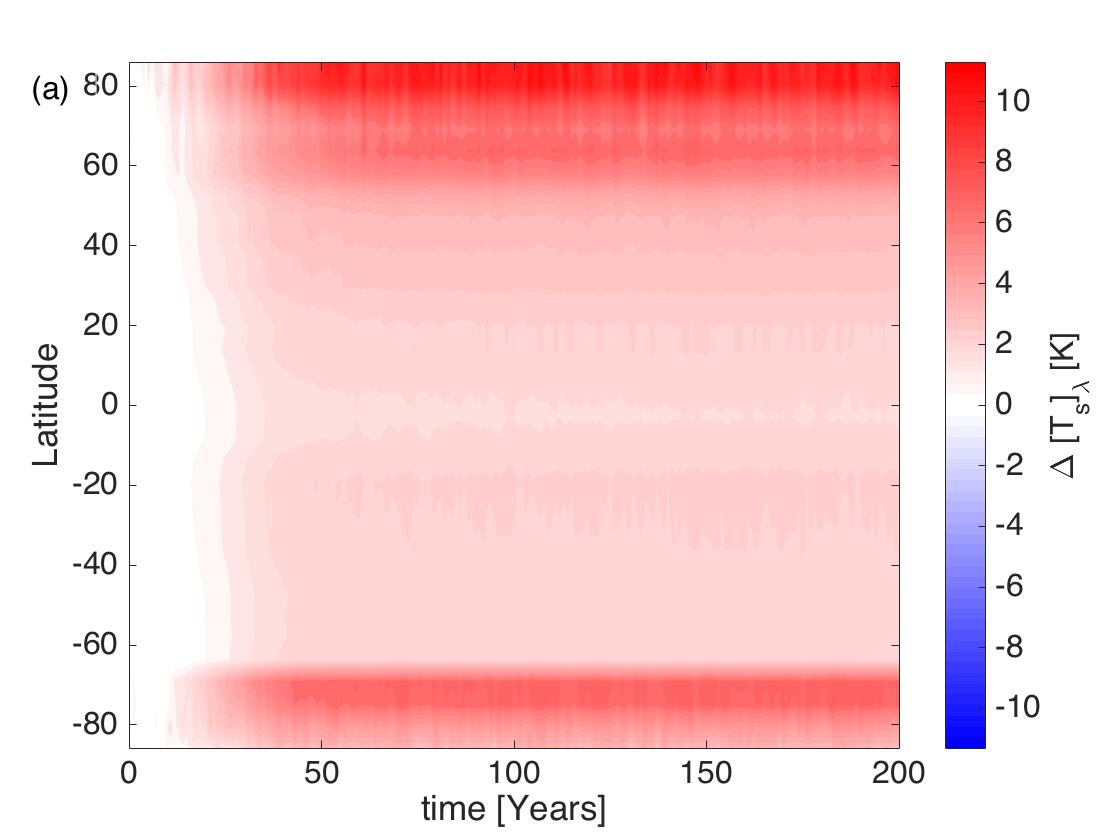} 
            \includegraphics[width=0.5\linewidth]{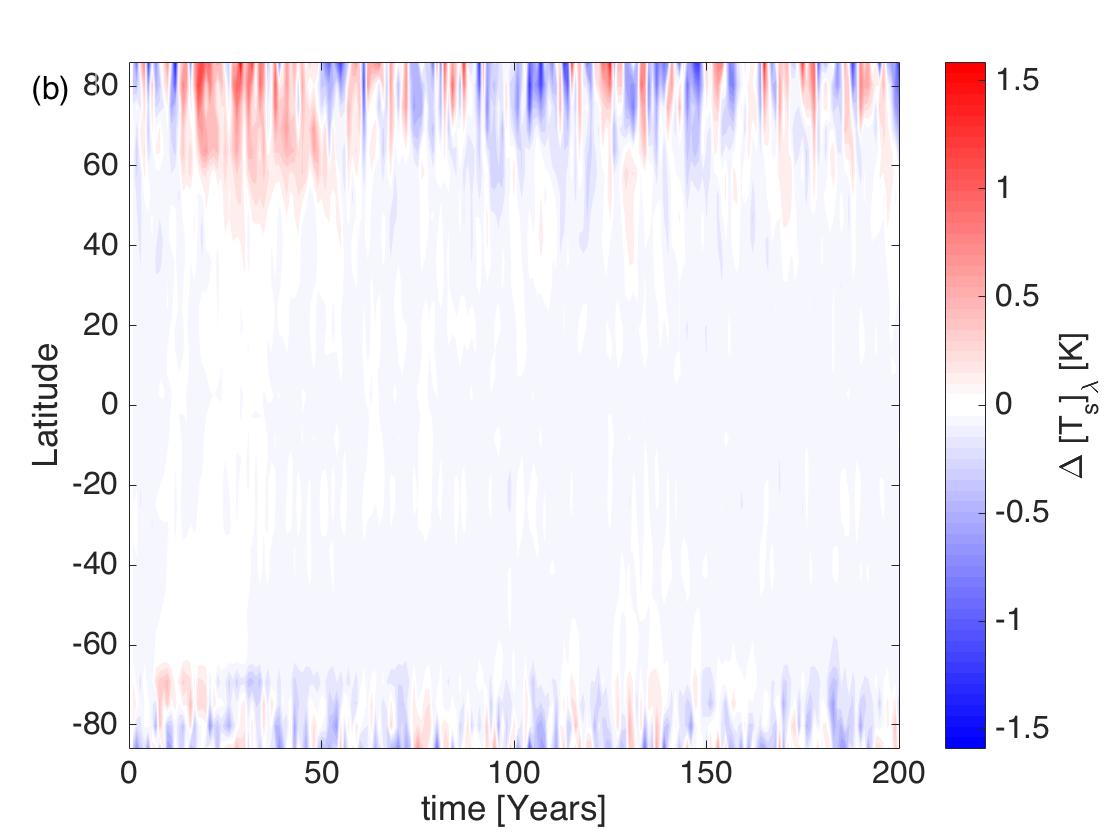} \\
            \includegraphics[width=0.5\linewidth]{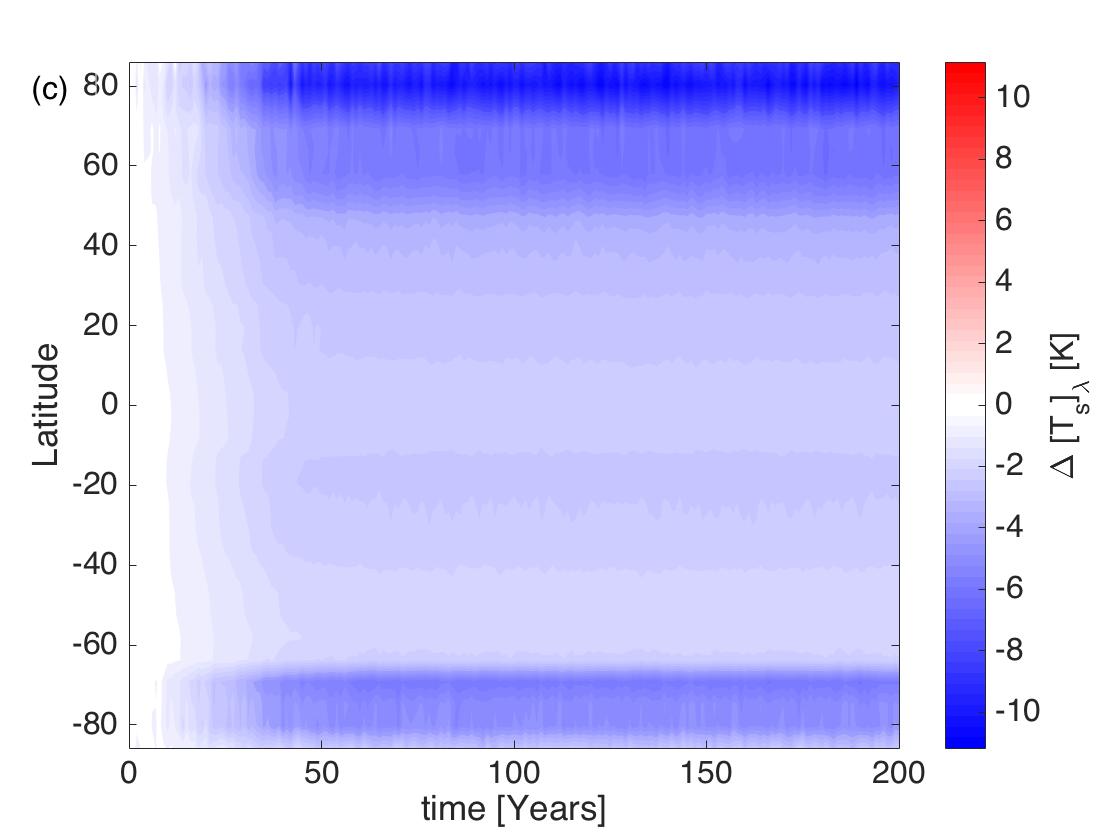} 
            \includegraphics[width=0.5\linewidth]{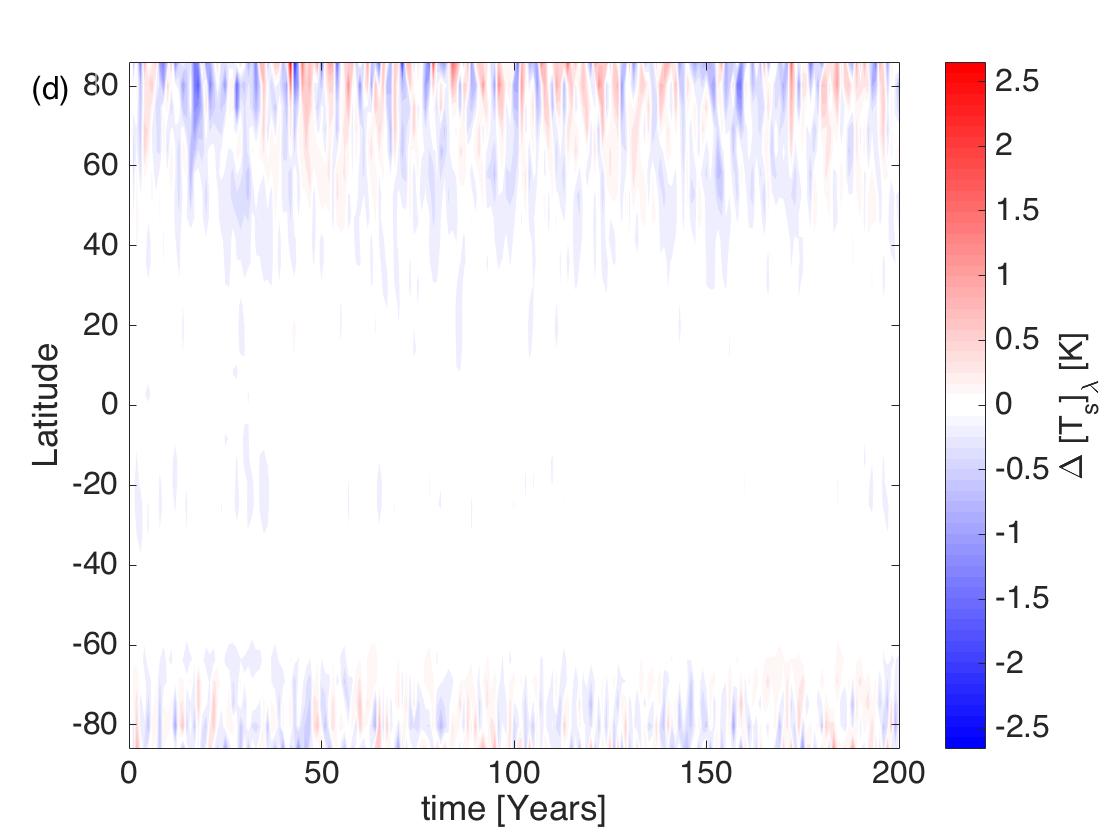}
        \end{tabular}
        \caption{\label{fig:resp2ramp_zonave} Response of the zonally averaged surface air temperature to ramp forcings. The first column shows the true responses in the model and the second  the errors in the linear predictions. The first and second rows belong to the CR1 and SR1 forcing scenarios, respectively. Similar diagrams as in the first row but for CR2 are shown in Fig.~6 of Ref.~\onlinecite{Lucarini2017}. 
        }
   % \end{center}
\end{figure*}

\begin{figure*}  %[t!]
   % \begin{center}
        \begin{tabular}{cc}
            \includegraphics[width=0.5\linewidth]{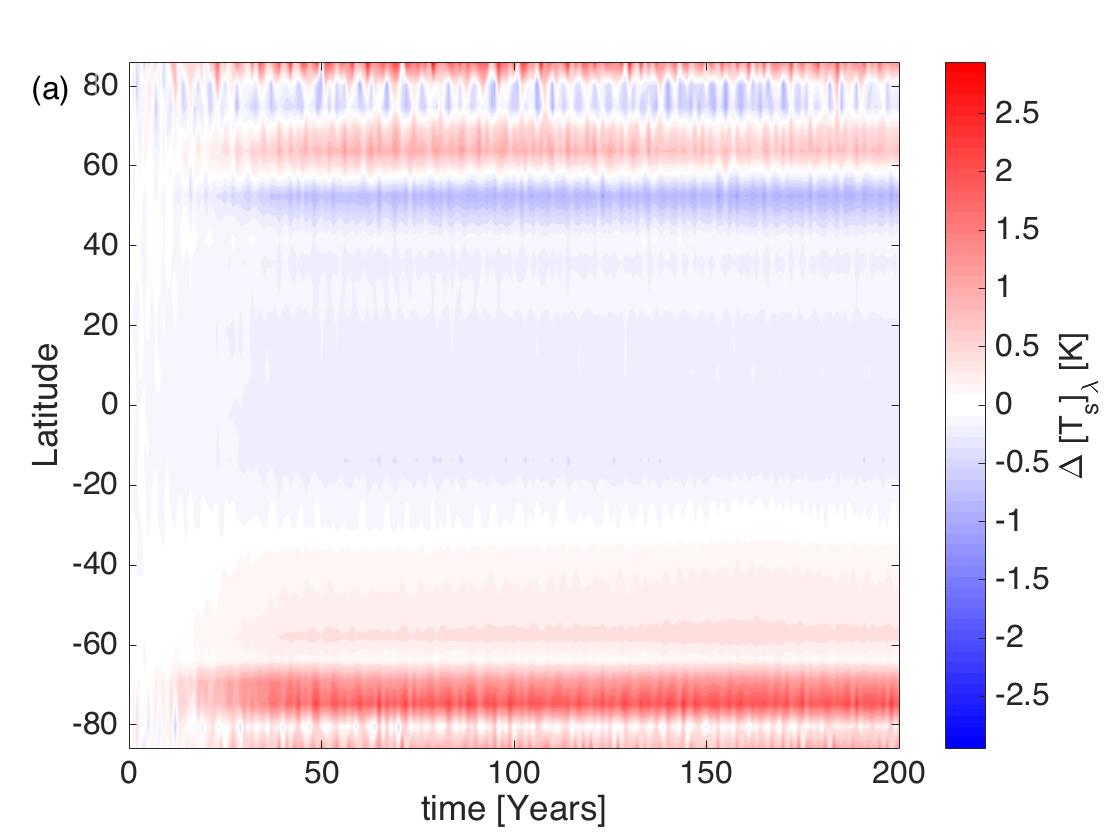} 
            \includegraphics[width=0.5\linewidth]{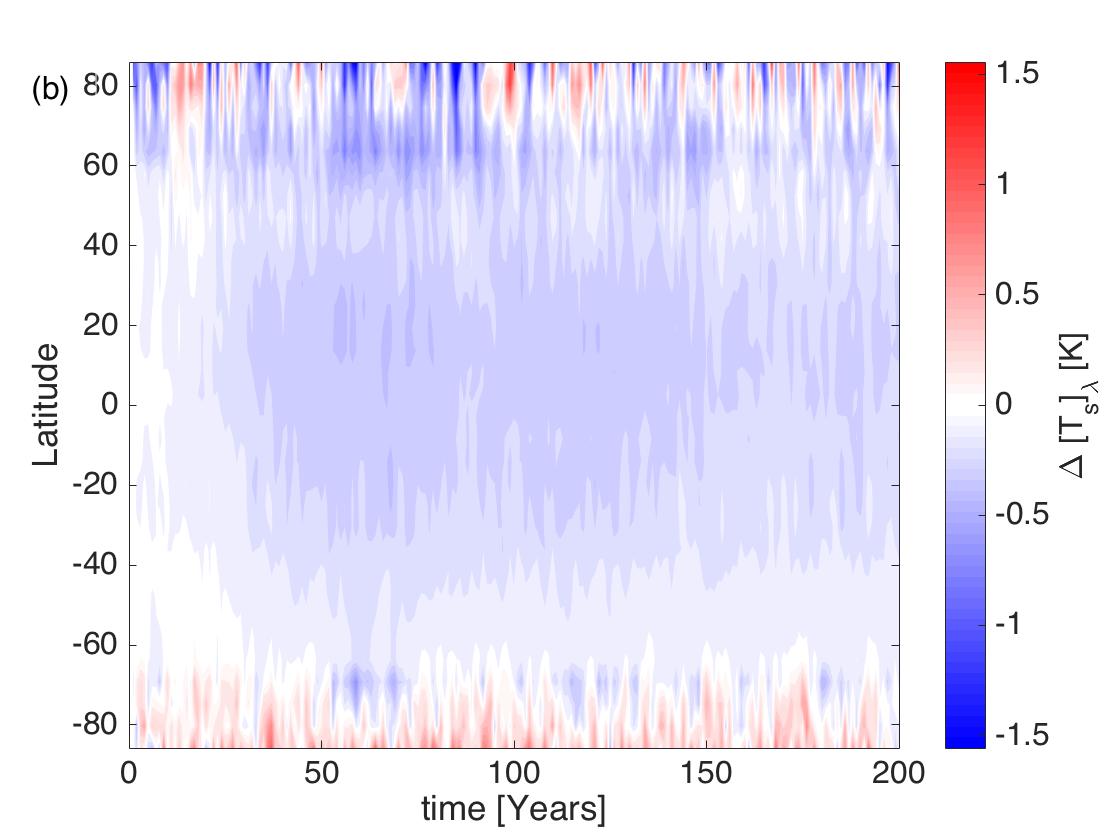} 
        \end{tabular}
        \caption{\label{fig:resp2ramp_combine_zonave} Predicted (a) and true (b) total responses of the zonally averaged surface air temperature to combined ramp forcings (BR1). 
        }
  %  \end{center}
\end{figure*}

For the diagnostics of any residual total response, a good starting point is to look at the zonally averaged fields of the surface air temperature. As a reminder, any observable other than the one for which a desired evolution has been enforced (the global average surface air temperature in our case; see Sec.~\ref{sec:res_global}) can evolve in an uncontrolled fashion. This is something that is not intended and should be checked. 
First, we show results with the
$\sqrt{2}$-fold $[{\rm CO}_2]$ increase (CR1, BR1).  Following Lucarini \textit{et al.}~\cite{Lucarini2017} (where only the case of $[{\rm CO}_2]$-doubling was treated), treating zonal means in a similar fashion to global means informs us
that the response to either greenhouse or solar forcing is  strongest at
high-latitude/polar regions; see Figs.~\ref{fig:resp2ramp_zonave}(a) and \ref{fig:resp2ramp_zonave}(c). This is where the response is most nonlinear, as indicated by Figs.~\ref{fig:resp2ramp_zonave}(b) and \ref{fig:resp2ramp_zonave}(d), showing the difference between {reference} and prediction. This nonlinearity should be due to albedo saturation and/or nonlinear characteristics of radiation physics, as discussed by others.\cite{doi:10.1029/180GM09,Good2015,Lucarini2017,Fukai_etal:2019} We note that in Figs.~\ref{fig:resp2ramp_zonave}(b) and \ref{fig:resp2ramp_zonave}(d), nonzero values under the stationary climate (after a transient) represent finite-ensemble-size statistical errors only. As a consequence of the said nonlinearities, in the high-latitude regions, linear response theory performs  very poorly in predicting the total response to combined forcing, as it does also in the regime of stationary climate; compare Figs.~\ref{fig:resp2ramp_combine_zonave}(a) and \ref{fig:resp2ramp_combine_zonave}(b) showing the prediction and {reference}, respectively. 

\begin{figure*}  %[t!]
    %\begin{center}
        \begin{tabular}{cc}
            \includegraphics[width=0.5\linewidth]{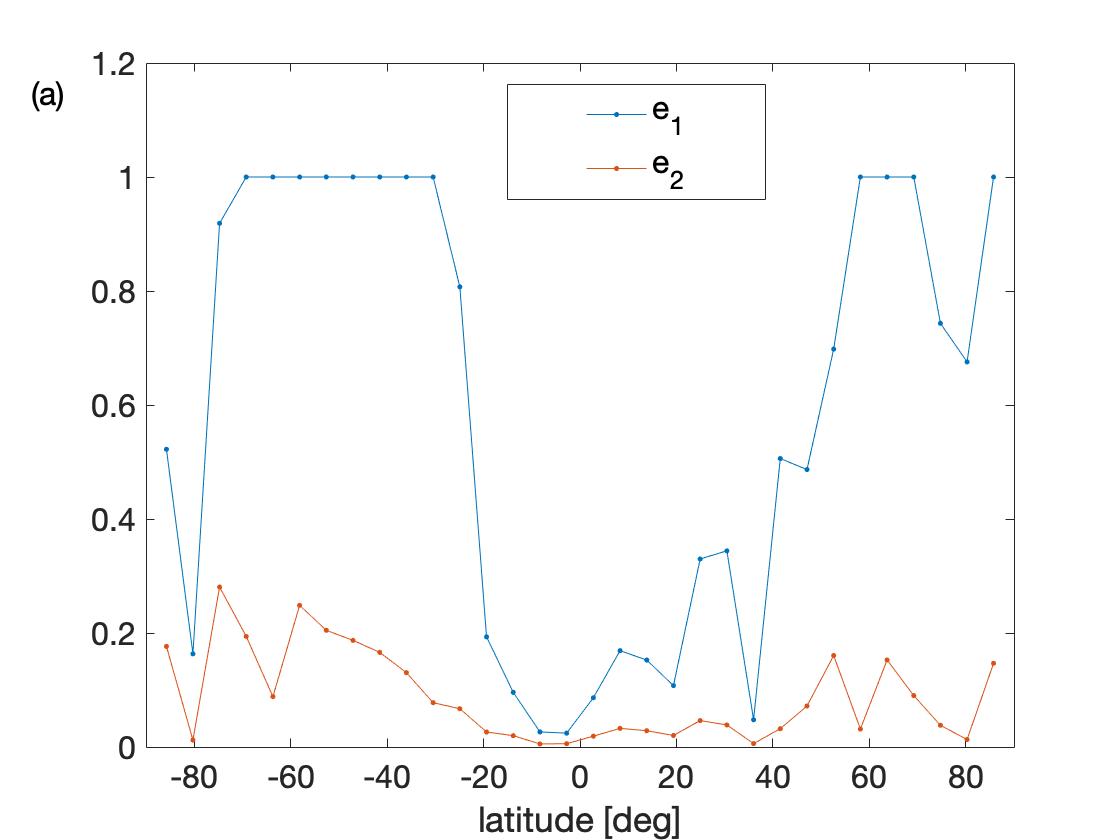} 
            \includegraphics[width=0.5\linewidth]{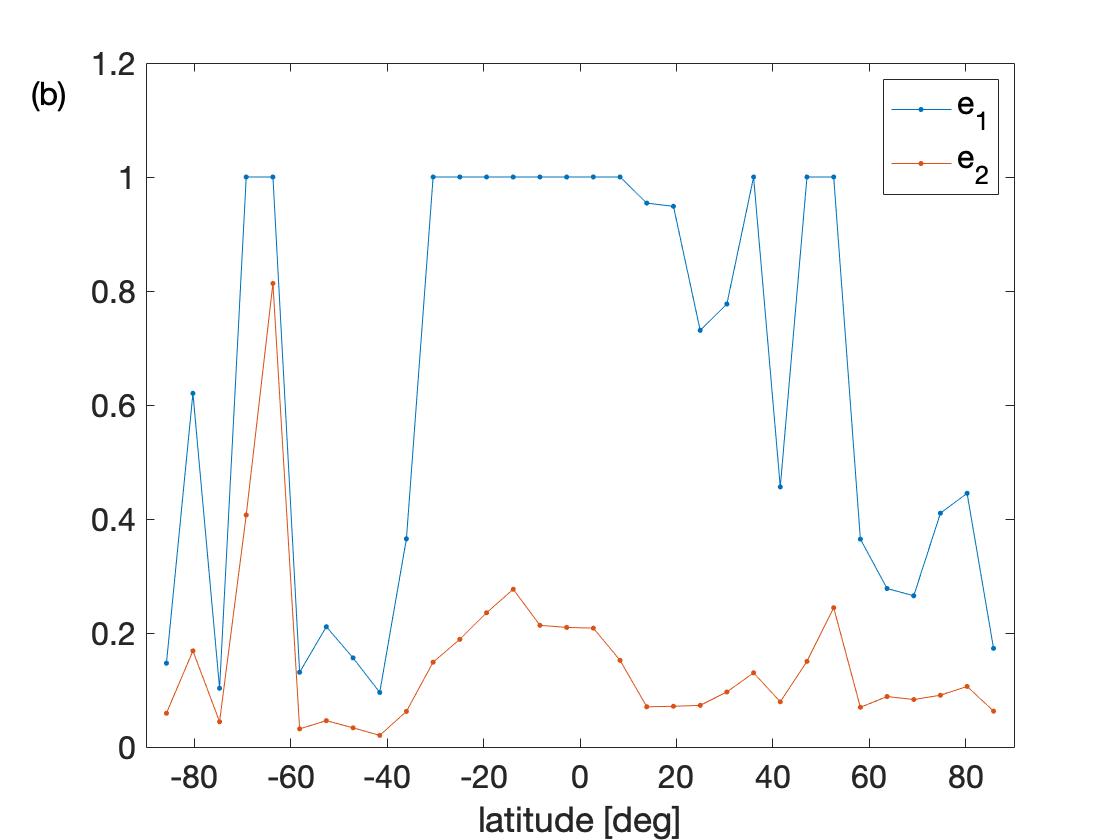} 
        \end{tabular}
        \caption{\label{fig:resp2ramp_combine_zonave_relerr} Relative errors $e_1$ and $e_2$ defined respectively by Eqs.~(\ref{eq:e1}) and (\ref{eq:e2}) for the predicted total responses of the zonally averaged surface air temperature to combined ramp forcings. 
        (a) is a companion diagram to those in Figs.~\ref{fig:resp2ramp_zonave} and \ref{fig:resp2ramp_combine_zonave} belonging to the weak-forcing scenarios (CR1, BR1), whereas (b) shows the same for the stronger-forcing scenarios (CR2, BR2).  Discrete data points are connected by lines to aid reading the diagrams.
        }
    %\end{center}
\end{figure*}

In addition to such a visual comparison, it is customary to quantify the discrepancy by measuring the error of the prediction \textit{relative} to the true value. However, the true value can be zero at certain latitudes, and this naive relative error measure then lacks an obvious meaning. In these situations, it is customary~\cite{relative_err} to analyze the following relative error:
\begin{equation}\label{eq:e1}
 e_1 = \frac{|\Delta\langle\Psi\rangle_{BRX}-\langle\Psi\rangle^{(1)}_{BRX}|}{|\Delta\langle\Psi\rangle_{BRX}|+|\langle\Psi\rangle^{(1)}_{BRX}|}.
\end{equation} 
This takes  values in the interval [0,1] for all values of $\Delta\langle\Psi\rangle_{BRX}$ and $\langle\Psi\rangle^{(1)}_{BRX}$; and a larger value should be considered worse. {Clearly, $e_1(\mu)$ as a function of latitude would facilitate the comparison of the predictive skill of linear response theory at different latitudes.} We note that in Eq.~(\ref{eq:e1}), $\langle\Psi\rangle^{(1)}_{BRX}$ is meant to be an estimator of the actual quantity, and this estimator is biased, but to keep things simple, we do not introduce a separate symbol for the estimator. Another possibility in our situation is that we measure the error of prediction of the response to combined forcing relative to the response to one of the forcings: 
\begin{equation}\label{eq:e2}
 e_2 = \frac{|\Delta\langle\Psi\rangle_{BRX}-\langle\Psi\rangle^{(1)}_{BRX}|}{\Delta\langle\Psi\rangle_{CRX}}.
\end{equation} 
We evaluate $e_1$ and $e_2$ only with respect to the stationary climate, in which case the estimation is very accurate since we can take an average also with respect to time. Figure~\ref{fig:resp2ramp_combine_zonave_relerr}(a) shows the result in the case of the weaker  
forcing (CR1, BR1). Both $e_1$ and $e_2$ indicate {with good agreement} that the prediction is  poorest at some high-latitude regions. 

With $[{\rm CO}_2]$-doubling (CR2, BR2), as shown by the results  in Fig.~\ref{fig:resp2ramp_combine_zonave_relerr}(b), the performance has different characteristics in comparison with the case of  weak forcing. {Both} $e_1$ and $e_2$ are  highest at both equatorial and some high-latitude regions, and somewhat less at polar and some Southern Hemisphere midlatitude regions. 

%\nocite{*} % no idea why it was active by default!!
%\bibliography{resp_theory_geoeng.bib}% Produces the bibliography via BibTeX.
%merlin.mbs aipnum4-1.bst 2010-07-25 4.21a (PWD, AO, DPC) hacked
%Control: key (0)
%Control: author (8) initials jnrlst
%Control: editor formatted (1) identically to author
%Control: production of article title (0) allowed
%Control: page (1) range
%Control: year (1) truncated
%Control: production of eprint (0) enabled
%

\end{document}